\documentclass[pdflatex,sn-nature]{sn-jnl}%

\usepackage{float}
\usepackage{adjustbox}%
\usepackage{longtable}%
\usepackage{amsmath,amssymb,amsfonts}%
\usepackage{cancel}%
\usepackage{mathtools}%
\usepackage{xcolor}%
\usepackage{colortbl}%
\usepackage{booktabs}%
\usepackage{algorithm}%
\usepackage{enumitem}
\definecolor{red}{RGB}{249, 65, 68}
\definecolor{blue}{RGB}{25, 130, 196}

\DeclareUnicodeCharacter{2032}{\ensuremath{{}^{\prime}}}

\title{A concentration-independent paradigm rendering weak interactions inherently quantifiable}

\begin{document}

\author[1]{\fnm{Masahiko} \sur{Yoshimura}}\equalcont{These authors contributed equally to this work.}
\author[1]{\fnm{Fuyuki} \sur{Matsuda}}\equalcont{These authors contributed equally to this work.}
\author[1]{\fnm{Yoshiki} \sur{Ikeda}}\equalcont{These authors contributed equally to this work.}
\author[1]{\fnm{Chihiro} \sur{Mori}}
\author[1]{\fnm{Tomoko} \sur{Yoneda}}
\author[1]{\fnm{Minako} \sur{Kikukawa}}
\author[1]{\fnm{Rie} \sur{Murakami}}
\author[2]{\fnm{Chiharu} \sur{Nogami}}
\author[3,4,5]{\fnm{Yukihiko} \sur{Sugita}}
\author[1]{\fnm{Yoshiko} \sur{Nakada-Nakura}}
\author[1]{\fnm{Masahiko} \sur{Tsujimoto}}
\author*[1,6]{\fnm{Daishi} \sur{Fujita}}\email{dfujita@icems.kyoto-u.ac.jp}

\affil[1]{\orgdiv{Institute for Integrated Cell-Material Sciences (iCeMS), Institute for Advanced Study}, \orgname{Kyoto University}, \orgaddress{\street{Yoshida, Sakyo-ku}, \city{Kyoto}, \postcode{606-8501}, \country{Japan}}}
\affil[2]{\orgdiv{Faculty of Science}, \orgname{Kyoto University}, \orgaddress{\street{Sakyo-ku}, \city{Kyoto}, \postcode{606-8502}, \country{Japan}}}
\affil[3]{\orgdiv{Institute for Life and Medical Sciences}, \orgname{Kyoto University}, \orgaddress{\street{53 Shogoin Kawahara-cho, Sakyo-ku}, \city{Kyoto}, \postcode{606-8507}, \country{Japan}}}
\affil[4]{\orgdiv{Graduate School of Biostudies}, \orgname{Kyoto University}, \orgaddress{\street{Yoshida-Konoecho, Sakyo-ku}, \city{Kyoto}, \postcode{606-8501}, \country{Japan}}}
\affil[5]{\orgdiv{Hakubi Center for Advanced Research}, \orgname{Kyoto University}, \orgaddress{\city{Kyoto}, \country{Japan}}}
\affil[6]{\orgname{Inamori Research Institute for Science}, \orgaddress{\city{Kyoto}, \country{Japan}}}

\presentaddress{Chiharu Nogami: Institute for Life and Medical Sciences, Kyoto University, 53 Shogoin Kawahara-cho, Sakyo-ku, Kyoto 606-8507, Japan; Graduate School of Biostudies, Kyoto University, Yoshida-Konoecho, Sakyo-ku, Kyoto 606-8501, Japan.}

\abstract{Measuring weak interactions essential to cellular function often requires bulk concentrations beyond the limits set by protein solubility and sample availability. Conventional biochemical experiments, including titrations, typically vary the number of molecules, $N$, at fixed volume, $V$, to control molecular concentration, $N/V$. We establish a parallel framework that holds $N$ fixed and controls $V$, the space within each partner's reach in an isolated pair. Tether lengths and attachment positions inside DNA nanocavities set each partner's reach and where the pair can interact. Theory relates these spatial design choices to binding, enabling quantitative measurement and modulation. Standard competition assays quantified a protein--peptide dissociation constant on the order of 10~mM using only 25~fmol of tethered material per well and less than 1~pmol per titration. A standard plate reader gave a signal-to-noise ratio near $10^{3}$. We used the same approach to screen for protein--protein interaction modulators based on their ability to enhance weak interactions rather than their affinity for an isolated target. Optimization of a hit from an integrin--talin screen yielded a compound that increased fibronectin-dependent cell adhesion despite no detectable binding to an isolated talin domain. Thus, the framework extends biochemical measurement and modulation beyond the practical concentration limits of $N$-controlled experiments.}

\maketitle

\section*{Introduction}

Cellular function depends on weak molecular interactions that current biochemical methods can fail to capture \cite{Protein_interactome_1,Protein_interactome_2,Protein_interactome_3}. The thermodynamic basis of high-affinity binding, characterised by equilibrium dissociation constants ($K_{\mathrm{d}}$) in the nanomolar to micromolar range, is well understood. In contrast, many functionally critical interactions exhibit weak affinity ($K_{\mathrm{d}}$ in the millimolar range), such that most binding partners remain unbound at typical intracellular concentrations \cite{Weak_protein_complexes}. These low-occupancy interactions are essential for cellular homeostasis, mediating dynamic protein assemblies \cite{LLPS_1,LLPS_2,LLPS_3,Weak_interaction_cell_adhesion}, metabolic channelling \cite{The_role_of_dynamic_enzyme_assemblies,Metabolon_review1,Metabolon_sorghum,Metabolon_Flavonoid,Metabolon_review2}, and plasticity within signalling networks \cite{Dynamic_Protein_Interaction_Networks,Rewiring_cell_signaling}. However, their low occupancy renders them largely invisible to conventional ensemble assays. Indeed, a comprehensive database analysis highlights the scarcity of quantitative data on millimolar-affinity protein--protein interactions (PPIs; Fig. 1). Chien and Gierasch described their quantitative detection and characterization as `dream experiments' \cite{DREAM_EXPERIMENTS}, reflecting both their biological importance and persistent experimental elusiveness. Why, then, have these interactions remained so difficult to quantify?

    \begin{figure}[H]
    \centering
    \includegraphics{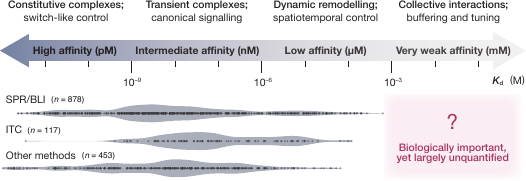}
    \caption{
Millimolar-affinity protein--protein interactions (PPIs) are largely unquantifiable by routine binding assays. Reported dissociation constants ($K_{\mathrm{d}}$) for selected wild-type PPIs in the PPB-Affinity dataset \cite{Ye2024PPBAffinity} (N = 1,460, 1990--2018). The $K_{\mathrm{d}}$ axis is logarithmic, and larger values indicate weaker binding. Displayed points denote individual reported PPIs (1,448 shown, excluding 12 below the axis minimum of $10^{-12}$ M), grouped by method: surface plasmon resonance or biolayer interferometry (SPR/BLI, n = 878), isothermal titration calorimetry (ITC, n = 117), and other (n = 453). Grey violin density envelopes summarize the method-specific distributions. Upper annotations show illustrative biological contexts, not definitive categories. Coverage is dense from picomolar to micromolar and thins sharply at high micromolar $K_{\mathrm{d}}$. All 1,460 records fall below 1 mM, with the highest reported $K_{\mathrm{d}}$ being 0.635 mM. Therefore, the millimolar range highlighted in pale pink describes the weak-interaction frontier that we target in this study.}
    \label{fig:1}
    \end{figure}

    \begin{figure}[!htbp]
    \centering
    \includegraphics[width=\textwidth]{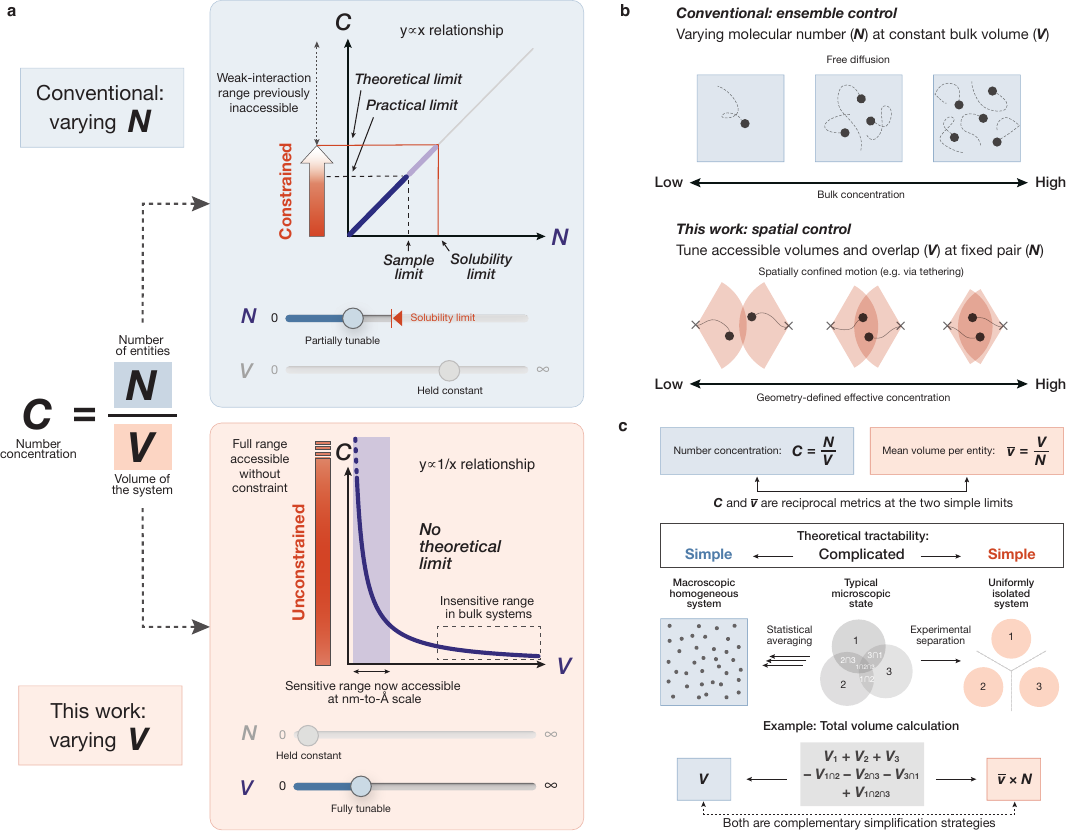}
    \caption{
The denominator side of concentration control has no theoretical ceiling.
a. Conventional titration changes the number concentration (C) by varying the number of molecules (N) at constant bulk volume (V). Although this titration follows C $\propto$ N, sample solubility and availability limit the attainable concentration, making weak-binding measurements unattainable. The V-controlled path instead varies the denominator. As C $\propto$ 1/V, the ideal curve rises without bound as the partners are confined more tightly, and it falls to zero when they are fully separated. At macroscopic length scales, practical changes in V barely alter 1/V for a single pair in absolute terms. Modern nanoscale architectures make the sensitive nm-to-\AA{} range experimentally accessible.
b. Physical interpretation of the two perturbations. Conventional ensemble assays perturb a shared bulk system by adding molecules. In the present approach, a fixed molecular pair is perturbed by changing the overlap of the regions its partners can explore, such as by tethering each partner to a fixed anchor. Moving the same pair closer together or farther apart tunes a geometry-defined effective concentration, emulating a concentration titration and extending its range without increasing bulk abundance.
c. Complementary simplification limits. Number concentration (C = N/V) is natural for macroscopic homogeneous systems, where statistical averaging simplifies many-particle behaviour. Mean volume per entity ($\overline v$ = V/N) becomes natural at the opposite limit, where entities are experimentally isolated in addressable regions. Typical microscopic states lie between these limits, where overlapping volumes generate cross terms. Three components already require seven volume terms: three individual volumes, three pairwise overlaps, and one triple overlap. Thus, macroscopic pooling and spatial isolation are complementary simplification strategies. The full formal treatment is given in our companion paper \cite{CompanionPaper}.}
    \label{fig:2}
    \end{figure}

The answer lies not in instrumentation alone, but in the molar concentration metric through which binding is conventionally measured. For over a century, the $K_{\mathrm{d}}$ metric has dominated because its molar units are simple and intuitive. For a 1:1 interaction, it describes the free-ligand concentration at half occupancy. Molar concentration is determined by molecular number per volume (N/V), and routine titrations address it almost exclusively by increasing N while keeping V fixed. This practice works well for high-affinity interactions, whose affinity range is reached at modest concentrations. However, weak-affinity interactions expose its limitation: measuring millimolar $K_{\mathrm{d}}$ values requires free-ligand concentrations that routinely exceed solubility limits or sample availability. For example, even well-behaved proteins suitable for crystallization rarely reach concentrations above 10 mg/mL ($\sim$200 \textmu{}M for a 50 kDa protein) \cite{McPherson2014,McPherson2014_2,Cagney2003}, which makes millimolar measurements impractical in concentration-based assays, regardless of the method employed (Fig. 1). Requiring higher bulk concentrations to measure weaker interactions, the concentration paradigm turns solubility limits and sample availability into a measurement ceiling. Protocols and instruments built around this requirement inherit this limitation, which is commonly mistaken as inevitable. Here, we name this conventional framework the concentration-based (N-controlled) paradigm and challenge its foundations.

A complementary experimental approach would be to control the denominator of N/V: instead of adding molecules, control the space available to them. At the molecular scale, we can implement this by controlling each interaction partner's accessible volume, thus presenting an alternative experimental axis. We have named this the volume-based (V-controlled) paradigm. In our model, each partner's range of motion is constrained by a flexible linker, and the overlap of these ranges determines how readily the binding pair can come into contact (Fig. 2a, b). Thus, what biochemists call `local concentration' and `proximity effects' can be viewed as consequences of the volume available to each partner \cite{High_Local_Concentration,Proximity,Scaffold_Proteins}. DNA-based nanocavities, connected to the surrounding solution through openings, can model this scheme; the range of motion of proteins tethered inside the cavity can be controlled within the nanoscale solution domains. These engineered zones form a minimal single-pair platform, defining cavities that act as `molecular interrogation rooms' where pairwise behaviour can be perturbed and examined (Fig. 3a).

    \begin{figure}[!htbp]
    \centering
    \includegraphics[width=\textwidth]{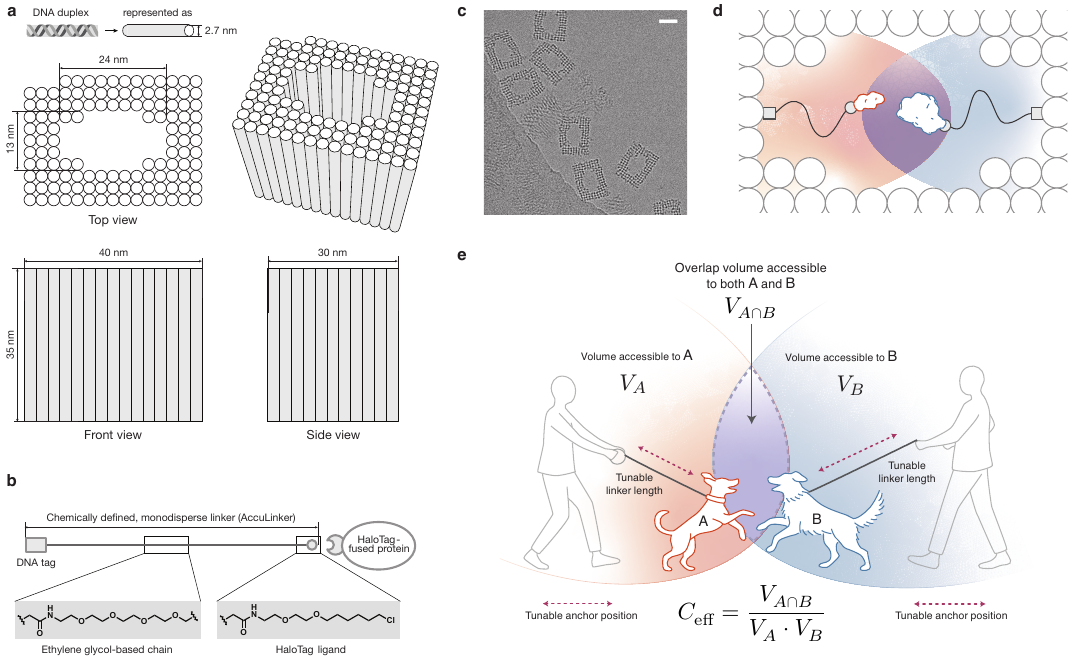}
    \caption{
DNA nanocavities define the accessible volumes that convert pairwise geometry into molar effective concentration.
a. DNA-nanocavity scaffold. DNA duplexes are represented as 2.7-nm cylinders; orthogonal and perspective views show the designed dimensions.
b. Stepwise-synthesised, defined-length AccuLinker for tethering proteins to DNA tags. The ethylene glycol-based chain can be functionalised with different terminal ligands; here, a HaloTag ligand \cite{HaloTag} is used to covalently tether a HaloTag-fused protein.
c. Representative cryo-EM image of assembled DNA nanocavities embedded in thin vitreous ice. Scale bar, 25 nm.
d. Schematic of tethered pair ranges. Two tethered partners explore partially overlapping regions inside a nanocavity connected to the surrounding solution. The red- and blue-shaded regions represent the volumes accessible to each partner. The purple region represents the shared region in which encounters can occur.
e. Analogy for panel d. Leash lengths and owner positions determine where two dogs can meet. For this single pair, linker length and anchor position set the accessible volumes $V_{A}$ and $V_{B}$, equivalent to $\overline v_{A}$ and $\overline v_{B}$, and their overlap, $V_{A\cap B}$. In the idealised tethered geometry shown, the overlap represents the complex's accessible volume $\overline v_{AB}$. For independently and uniformly distributed unbound partners, the number-based effective concentration is then $C_{\mathrm{eff}} = V_{A\cap B}/(V_A\cdot V_B)$. Dividing by the Avogadro constant gives $c_{\mathrm{eff}}$, a first-principles molar effective concentration that places local pair geometry on the same scale as molarity. The full framework extends this relation to non-uniform spatial distributions \cite{CompanionPaper}. The analogy shows how changing range overlap, rather than molecular number, tunes encounter likelihood.}
    \label{fig:3}
    \end{figure}

We show that controlling V, rather than N, presents an experimental opportunity to overcome the bulk-concentration barrier, enabling the measurement and modulation of weak-affinity interactions. Weak-affinity-interaction analysis is a practical application of the broader framework developed in our companion paper \cite{CompanionPaper}. Here, we present an intuitive physical explanation of the underlying principles, while the companion paper provides the full mathematical formulation and derivation. Our results reveal that controlled spatial arrangement can invert a bulk solution-phase binding hierarchy. The same control enabled an absolute protein--peptide $K_{\mathrm{d}}$ of order 10 mM to be quantified using less than a picomole of tethered material per titration. A final screen identified protein--protein interaction modulators that enhance weak associations by reweighting local encounters rather than by possessing high intrinsic affinity in isolation. These results shift the V-controlled paradigm from a theoretical coordinate change towards a practical path for measuring and manipulating weak-affinity interactions beyond the reach of conventional methods.

\section*{A V-centric theoretical framework}

We begin with the elementary equilibrium $\mathrm{A}+\mathrm{B}\rightleftharpoons\mathrm{AB}$. We write the concentrations as number concentrations, [X] = $N_{X}$/$V_{X}$, so that the Avogadro factors do not obscure this argument. The volume $V_{X}$ is specific to each interacting species. The usual molar $K_{\mathrm{d}}$ differs only in the constant unit conversion. Lowercase c denotes molar concentration throughout.

The key step is to view the concentration axis from the perspective of its denominator. Number concentration (C) is N/V, so its reciprocal, V/N, is a mean volume per molecule. In ordinary bulk solution, this quantity is only notional, not the physical range explored by a freely diffusing molecule. However, suppose that the volume available to species X is divided into $N_{X}$ mutually isolated compartments, one for each X molecule (Fig. 2c). Each compartment has a mean accessible volume $\overline v_{X}$, so the partition satisfies $V_{X}$ = $N_{X}$ $\cdot$ $\overline v_{X}$. Substituting this relation into the equilibrium expression allows the key cancellation in which $N_{X}$ terms disappear and only accessible volumes remain (Eq. 1):
\begin{equation}\label{eq:mutual_isolation}
    K_{\mathrm{d}}=\frac{[\mathrm{A}][\mathrm{B}]}{[\mathrm{AB}]}
    =\frac{\dfrac{N_A}{V_A}\cdot\dfrac{N_B}{V_B}}{\dfrac{N_{AB}}{V_{AB}}}
    \quad
    \xRightarrow[\;V_X=N_X\cdot\overline v_X\;]{\text{mutual isolation}}
    \quad
    \frac{\dfrac{\cancel{N_A}}{\cancel{N_A}\cdot\overline v_A}\cdot\dfrac{\cancel{N_B}}{\cancel{N_B}\cdot\overline v_B}}{\dfrac{\cancel{N_{AB}}}{\cancel{N_{AB}}\cdot\overline v_{AB}}}
    =\frac{\overline v_{AB}}{\overline v_A\cdot\overline v_B}
    \end{equation}

This cancellation is the essence of the V-controlled paradigm. Under this construction, equilibrium depends on accessible volumes, not on pooled molecule numbers. Ordinary bulk solution contains no such compartments, so the construction remains hypothetical until nanoscale architecture imposes mutual isolation. Under such conditions, V/N ceases to be a formal reciprocal and becomes an experimental variable.

Viewed from the denominator, the formal concentration axis has no ceiling. Shrinking co-localized accessible regions towards a shared point drives the effective concentration upward without bound; separating those regions completely drives it to zero. Thus, in the limiting case, $0\leq c<\infty$ becomes accessible. At macroscopic scales, changes in practical volume barely alter 1/V for a single interacting pair in absolute terms. However, nanoscale architectures, in which cavity geometry, tether length, and steric exclusion can directly tune accessible volume, enable the experiment to access the sensitive range of this reciprocal curve (Fig. 2a, b).

\section*{Experimental platform: linker-mediated volume control}

To realize denominator control experimentally, a single molecular pair must be confined and isolated from all other pairs, without isolating that pair from the solution. With this objective, we envisaged DNA nanocavities that separate neighbouring linker-tethered pairs and remain open to the surrounding solution (Fig. 3a, d). The walls of these nanocavities would prevent tethered partners in different cavities from meeting, thus removing the corresponding cross-pair terms from the total population output. We have termed this architecture the Zonal Engineered Nano (ZEN) platform. In each ZEN cavity, anchor position and linker length define each partner's accessible range of motion and the overlap between these ranges (Fig. 3d, e).

To achieve the design freedom to place anchors at defined internal coordinates, we constructed the nanocavities using DNA bricks \cite{DNA_Bricks_Nature,DNA_Bricks_Science}, a modular variant of DNA origami that enables sub-nanometre positioning precision and the facile repositioning of tether anchors. The self-assembled cuboid architecture contains an elliptical cylindrical pore, with single-stranded DNA (ssDNA) overhangs at defined coordinates that can position tether anchors across a range of diagonal distances (Fig. 3a; Supplementary Figs.~\ref{fig:Extended Data Fig. Bricks design}, \ref{fig:Extended Data Fig. Detail Bricks design} and \ref{fig:Extended Data Fig. Detail Bricks design2}). Cryo-electron microscopy (EM) and negative-stain transmission EM (nsTEM) confirmed the structural integrity of the assembly (Fig. 3c; Supplementary Fig.~\ref{fig:Extended Data Fig. TEM.}). Gel electrophoresis indicated efficient assembly, with a maximum isolated-product yield of 48\% (Supplementary Fig.~\ref{fig:Extended Data Fig. DB01}).

In the ZEN architecture, accessible volume is controlled by linkers of defined length, and their reactivity can be modified with modular end chemistry. To realize this architecture, we synthesised AccuLinkers, ethylene glycol-based monodisperse spacers made by stepwise synthesis rather than polymerization, so that each tether allows a predetermined exploration radius (Fig. 3b; Supplementary Figs.~\ref{fig:Extended Data Fig. Detail Bricks design2}, \ref{fig:Extended Data Fig. AccuLinker and protein tethering}, \ref{fig:Extended Data Fig. Synthetic plan}, \ref{fig:Extended Data Fig. AccuLinker Evi}, and \ref{fig:Extended Data Fig. AccuLinker Pro}). To create the anchor terminus of the tether, we attached ssDNA complementary to a specific sequence inside the DNA cavity wall. The other, reactive, terminus of the tether can be functionalised with different terminal ligands; we attached a HaloTag ligand \cite{HaloTag} for the covalent linkage of HaloTag-fused proteins. Together with anchor position and cavity geometry, the linker length defines each partner's accessible range, and the cavity walls prevent unintended encounters. Fluorescence monitoring showed that a 10-fold excess of protein--ssDNA conjugate drove cavity loading to a plateau within 2 h. Removal of excess conjugate gave the tether-saturated material, which exhibited the expected site-specific tether localization (Supplementary Figs.~\ref{fig:Extended Data Fig. Protein conju1}, \ref{fig:Extended Data Fig. Protein conju2}, \ref{fig:Extended Data Fig. Protein conju3}, and \ref{fig:Extended Data Fig. Protein tethering}). Therefore, varying the linker length and anchor position across ZEN populations yielded fraction-bound measurements that reflected overlap-dependent binding.

    \begin{figure}[!htbp]
    \centering
    \includegraphics[width=\textwidth]{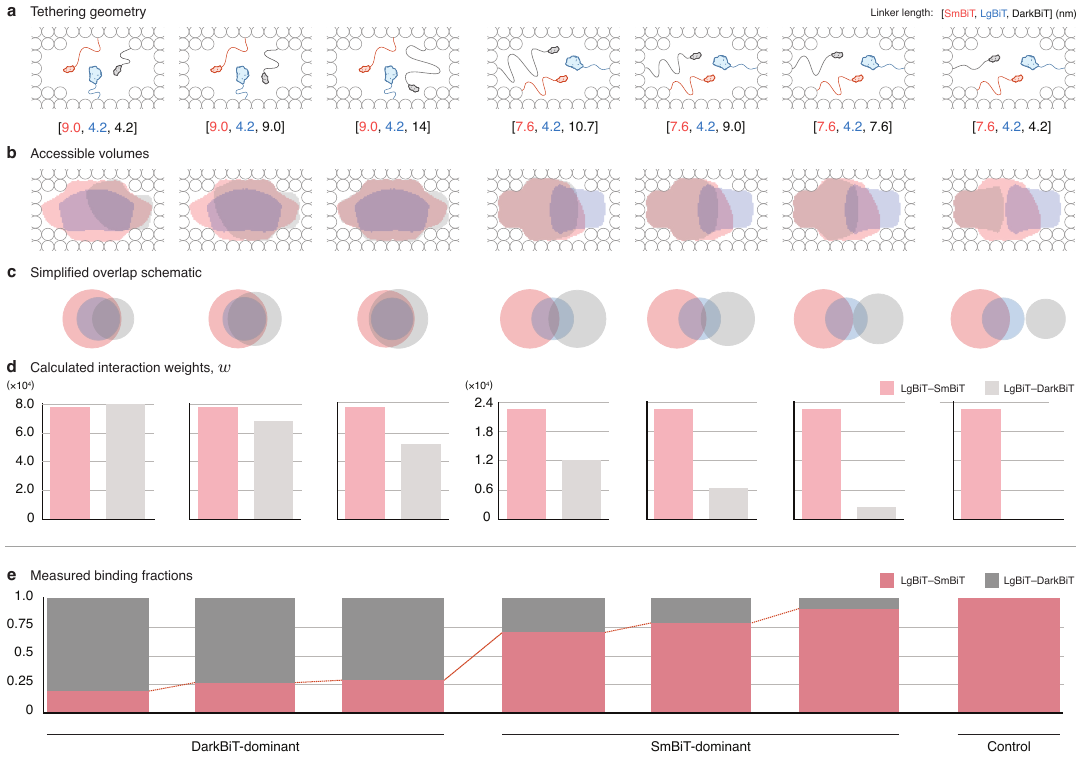}
    \caption{
Designed spatial overlap inverts the binding hierarchy in a single nanocavity. In Zonal Engineered Nano (ZEN) cavities, defined tethering geometries position LgBiT, the large luciferase fragment, and two peptide partners (all tethered through HaloTag fusions): SmBiT114 (labelled SmBiT) restores luciferase activity when bound to LgBiT, and DarkBiT~\textbf{1} (labelled DarkBiT) has higher solution-phase affinity for LgBiT but does not restore activity. Bold numerals denote compound names. Unbound LgBiT and LgBiT--DarkBiT are dark, so luminescence reports the LgBiT--SmBiT binding fraction.
a. Tethering geometries, specifying DNA-helix attachment positions and linker lengths in nm ([SmBiT, LgBiT, DarkBiT]).
b. Accessible volumes calculated for each of the seven designs based on a simple tethered-sphere Monte Carlo model. Red, blue, and grey indicate the regions explored by SmBiT, LgBiT and DarkBiT, respectively.
c. Simplified overlap schematics preserving the relative fractions from panel b while regularising the shapes for visual comparison. Because SmBiT and DarkBiT both compete for LgBiT, comparing the red--blue and grey--blue overlaps indicates which interaction is geometrically favoured.
d. Calculated interaction weights, w, for LgBiT--SmBiT and LgBiT--DarkBiT, which quantify the spatial overlap underlying $c_{\mathrm{eff}}$ (Fig. 3e), with constants common to the two competitors omitted. Geometry progressively favours SmBiT despite DarkBiT's stronger solution-phase affinity.
e. Measured binding fractions for the same nanocavity designs. The dominant partner switches from DarkBiT to SmBiT in the order predicted by the calculated weights (panels b--d). In the control design, DarkBiT is tethered but cannot reach LgBiT. The companion paper analysis reports quantitative agreement of the w-based model across tethering geometries using one shared calibration factor (Pearson's r = 0.994, root-mean-square error = 1.09-fold) \cite{CompanionPaper}.}
    \label{fig:4}
    \end{figure}

\section*{Spatial control reshapes binding preferences}

Figs. 3d and 3e illustrate the central premise that only the shared accessible region contributes to molecular encounters between the binding pair. We therefore assessed whether this overlap principle can condition partner choice in a three-component ZEN system. In the NanoLuc\textsuperscript{\textregistered} Binary Technology (NanoBiT) two-subunit luciferase system, LgBiT is the large fragment, SmBiT is the luminescent peptide partner that restores signal upon binding LgBiT, and DarkBiT is a non-luminescent SmBiT analogue that competes for the same LgBiT site \cite{LgBiT-SmBiT,DarkBiT}. Here, we used the HaloTag-fused variants SmBiT114 and DarkBiT~\textbf{1}, implementing the system in seven ZEN designs, each containing one copy of each component in the same cavity volume. Only the linker length and anchor position were varied (Fig. 4a). Therefore, a copy-number-per-cavity-volume view of local concentration would assign all designs the same value. We estimated the regions accessible to each component by simple Monte Carlo sampling of tethered motion (Supplementary Fig.~\ref{fig:Extended Data Fig. Size of LgBiT and SmBiT}), then redrew the same overlap patterns as simplified schematics that show whether LgBiT overlaps more with SmBiT114 or DarkBiT~\textbf{1} (Fig. 4b, c). For the two possible LgBiT complexes, LgBiT--SmBiT114 and LgBiT--DarkBiT~\textbf{1}, we calculated an interaction weight, w, a pair-specific overlap measure that provides the geometric basis for $c_{\mathrm{eff}}$. In the uniform-independent limit, $w_{XY}\propto V_{X\cap Y}/(V_X\cdot V_Y)$, with the general probability-distribution form given in the companion paper framework \cite{CompanionPaper} (Fig. 4d).

We then tested these geometries with the three-component NanoBiT competition assay to determine whether the designed overlap can override the solution-phase binding hierarchy. DarkBiT~\textbf{1} binds LgBiT more strongly in solution ($K_{\mathrm{d}}$ $\approx$ 10 \textmu{}M; Supplementary Fig.~\ref{fig:Extended Data Fig. Trp analysis}) than SmBiT114 ($K_{\mathrm{d}}$ $\approx$ 0.3 mM; Fig. 5b) \cite{LgBiT-SmBiT,DarkBiT}. As linker length and anchor position are designed variables, w can in principle be tuned continuously. The seven designs sampled this range, from configurations favouring LgBiT--DarkBiT~\textbf{1} to those favouring LgBiT--SmBiT114 (Fig. 4a--d). Despite this tens-of-fold affinity advantage in solution, the measured LgBiT--SmBiT114 binding fractions followed the geometry-defined order, with the favoured partner switching from DarkBiT~\textbf{1} to SmBiT114 across the designed geometries (Fig. 4e). In our companion paper analysis, the w-based model reproduced the measured fractions, with Pearson's $r > 0.99$ using one shared calibration factor, supporting the use of w as a quantitative measure of designed overlap \cite{CompanionPaper}.

    \begin{figure}[!htbp]
    \centering
    \includegraphics[width=0.94\textwidth]{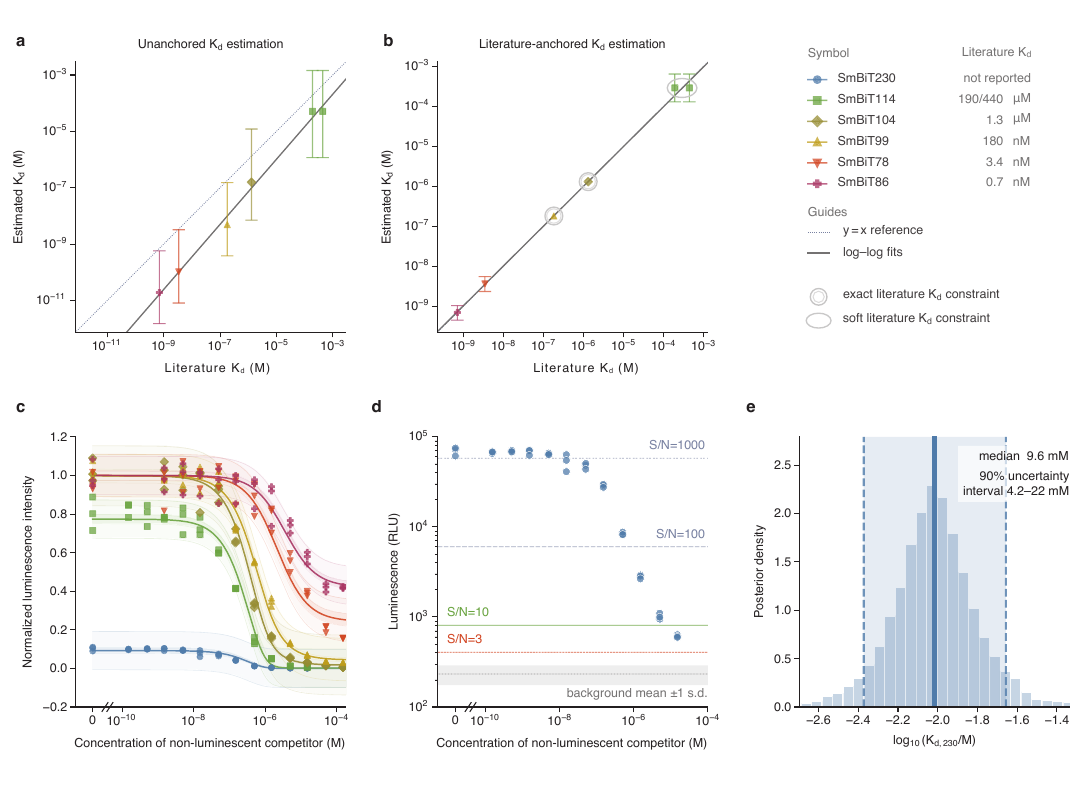}
    \caption{
The V-controlled paradigm and the ZEN system enable $K_{\mathrm{d}}$ measurements for ultra-weak interactions. SmBiT variants and LgBiT shared a ZEN geometry that fixed accessible volume. DarkBiT~\textbf{2} (the free DarkBiT) was titrated from bulk, and luminescence reported SmBiT-bound LgBiT.
a. Unanchored Bayesian fitting. For variants with literature affinities, unanchored $K_{\mathrm{d}}$ estimates are plotted against literature values. The unanchored estimates lie near published values across five orders of magnitude, with a tight log-linear relation and a systematic offset. The titrations alone quantify every affinity ratio, so the series can be converted into absolute $K_{\mathrm{d}}$ values using either the effective concentration set by the cavity geometry or a published affinity. The anchors in b are published affinities and leave the ratios almost unchanged.
b. Literature-anchored calibration. SmBiT99 and SmBiT104 were fixed at literature $K_{\mathrm{d}}$ values (grey halos, no error bars). SmBiT114 was constrained loosely by its two reported $K_{\mathrm{d}}$ values. Other estimates fall on the calibrated line with narrow uncertainty intervals, showing that this approach can predict $K_{\mathrm{d}}$ values with high confidence across the series. In a and b, dotted lines represent the identity line, grey lines represent log--log fits, points/error bars represent median estimates/90\% uncertainty intervals, and duplicate x positions represent two literature values.
c. Normalised DarkBiT~\textbf{2} titrations and anchored fits. Luminescence was normalised to matched intact-luciferase activity and to the fitted response scale, removing reporter-brightness differences while retaining reconstitution differences. Symbols denote replicates (n = 3 per point), curves represent median fits, dark bands represent 95\% fitted-response uncertainty intervals, and pale bands represent 95\% prediction intervals. Low-affinity curves start below full signal and retain high-dose luminescence. Both features follow quantitatively from the chemistry--geometry crossover and finite local capacity (each ZEN cavity holds only a limited number of molecules) \cite{CompanionPaper}.
d. Raw SmBiT230 luminescence in relative light units (RLUs), plotted logarithmically without fits to show unused S/N capacity. The competitor-free plateau reaches an S/N of $10^{3}$. The grey band represents background $\pm$ 1 standard deviation (SD), guides represent background plus a multiple of SDs, and S/N = 3 and 10 denote detection and quantification guides, respectively \cite{ICH_Q2R2}. Extrapolating the curve places S/N =10 near a $K_{\mathrm{d}}$ of $\sim$1 M.
e. Inferred distribution of $\log_{10}$($K_{\mathrm{d},230}$/M) using the anchored model. Median $K_{\mathrm{d},230}$ = 9.6 mM; 90\% uncertainty interval, 4.2--22 mM. Literature values are from NanoBiT characterization \cite{LgBiT-SmBiT}; none has been reported for SmBiT230. Fitting used the companion finite-capacity distribution-averaged inhibition model in its closed-form hypergeometric (${}_2F_1$) expression \cite{CompanionPaper}.}
    \label{fig:5}
    \end{figure}

These results show that nanostructured space can render geometry the primary control axis for binding preference; this axis is invisible in an N/V view of concentration. The same set of components can form different local complexes when linker-defined accessible regions are rearranged. In some geometries, this spatial arrangement alone can invert the solution-phase $K_{\mathrm{d}}$ hierarchy. Pairwise $K_{\mathrm{d}}$ values still define intrinsic affinities, but local preference follows the geometry-defined encounter landscape. Programmable geometry thereby becomes the operational control axis of the V-controlled paradigm, in which the effect of an experimentally set variable, analogous to concentration in conventional assays, is quantified by w.

\section*{Detection of millimolar-affinity interactions}

Geometry-controlled inversion of affinity hierarchy reveals a practical strength of the V-controlled paradigm: competition is no longer governed by solution $K_{\mathrm{d}}$ alone, as accessible-volume geometry provides a second, programmable axis. We addressed this axis in an open titration format. The SmBiT variant and LgBiT were tethered in the same ZEN geometry, thus setting the local encounter domain for the luminescent pair, while the free DarkBiT peptide was added from bulk solution as a competitor for the same LgBiT site (Fig. 5a--c). This ultra-weak tethered interaction thus becomes a standard-format inhibition assay: ZEN samples are aliquoted into a microtitre plate, mixed with a DarkBiT concentration series, and read by luminescence on a standard plate reader. We applied this strategy to a community-validated NanoBiT toolkit of six SmBiT variants (SmBiT78, SmBiT86, SmBiT99, SmBiT104, SmBiT114, and SmBiT230), chosen because it spans orders of magnitude in reported affinity and includes SmBiT230, the weakest member, for which no conventional $K_{\mathrm{d}}$ has been assigned, along with the DarkBiT~\textbf{2} peptide \cite{LgBiT-SmBiT,DarkBiT}.

The resulting inhibition curves reveal quantitative affinity information across the SmBiT variants. Without literature anchors, fitting the companion paper response model already placed the estimates near the published affinities, with a systematic offset (Fig. 5a) \cite{CompanionPaper}. The response model uses a shared $c_{\mathrm{eff}}$ for all six SmBiT variants, so the titrations determine their relative affinities. We can then convert the whole series to absolute $K_{\mathrm{d}}$ values using either an independently determined effective concentration or a published affinity (Supplementary Section~\ref{subsec:calibration_free_identifiability}). As the geometric calculation of $c_{\mathrm{eff}}$ is still being validated, we repeated the fitting using literature anchors to ensure the reported $K_{\mathrm{d}}$ values remained comparable to published ones. SmBiT99 and SmBiT104 served as anchors in the micromolar-to-submicromolar regime, where standard concentration-based assays are most reliable (Fig. 5b). SmBiT114 lies just outside that regime, and its two reported $K_{\mathrm{d}}$ values are different, so we used them only as a loose constraint. In this anchored fit, SmBiT86 and SmBiT78, whose reported $K_{\mathrm{d}}$ values were withheld, both fell close to the identity line, with narrow uncertainty intervals, and the $K_{\mathrm{d}}$ value for SmBiT114 fell between its two reported values. Both results demonstrate accurate $K_{\mathrm{d}}$ prediction across the SmBiT variants. Anchoring changed the SmBiT230/SmBiT114 ratio by only 0.15\%.

With the literature-anchored model, SmBiT230 yielded a median $K_{\mathrm{d}}$ of 9.6 mM (Fig. 5e), far above the affinity range reported in Fig. 1. Through control of the accessible volume, the intrinsic ceiling of the N-controlled paradigm is bypassed. Cavity geometry offers another route: combining the titration ratios with the computed $c_{\mathrm{eff}}$ of 1.20 mM yields 11.8 mM for SmBiT230, with no literature affinity reference. The geometric and literature-based $K_{\mathrm{d}}$ estimates differ by 23\% (Supplementary Sections~\ref{subsec:calibration_free_identifiability} and \ref{subsec:geometric_agreement}). Therefore, more accurate $c_{\mathrm{eff}}$ simulation could enable the assay to measure absolute affinities without literature calibration.

With a single ZEN setup defined by fixed cavity size, linker lengths, and anchor positions, and the same soluble DarkBiT~\textbf{2} titration format, the assay could address more than seven orders of magnitude of affinity, from picomolar to millimolar. Part of this range enters the chemistry--geometry crossover regime described in our companion paper, where local geometry can dominate apparent inhibition so strongly that a $\sim$$10^{4}$-fold difference in solution $K_{\mathrm{d}}$ shifts the inhibition midpoint by only $\sim$1\% \cite{CompanionPaper}. Geometry thereby supplies the matching parameter that conventional bulk competition lacks, allowing interaction affinities differing by orders of magnitude in solution to be evaluated in one measurable local-competition window.

The accessible $K_{\mathrm{d}}$ range still has headroom beyond 10 mM. On a conventional, legacy-generation plate reader, SmBiT230 produced luminescence at a signal-to-noise (S/N) ratio of $\sim$$10^{3}$ from 25 fmol of tethered material per well, corresponding to under 1 pmol for a complete titration (Fig. 5d). With the same cavity design and sample loading, the fitted response model and measured background noise project a signal above the S/N =10 threshold up to a $K_{\mathrm{d}}$ of $\sim$1 M (Fig. 5d). The cavity size, linker lengths, and anchor positions used here were chosen to address a broad picomolar-to-millimolar panel, not to maximize sensitivity at the weakest-binding limit. Therefore, the accessible range could be further extended by higher sample loading, more sensitive detection hardware, or ZEN designs tuned for weak binding.

\section*{Geometry-guided screening of PPI modulators}

The ZEN platform enables applications inaccessible to affinity-based bulk assays; for example, the selection of PPI modulators according to how they reshape local molecular encounters. Affinity-ranked screens are poorly suited for this class of modulators because their primary output is how tightly a compound binds, not how it changes the local encounter balance between weak partners \cite{Glue_1,Glue_2,PPI,kinetic_stabilizer,Polymer_inducer,PPI_stabilizer_review,PPI_stabilizer_review_2}. The ZEN assay makes this encounter balance the readout by measuring occupancy within a local response landscape jointly determined by binding chemistry ($K_{\mathrm{d}}$) and encounter geometry (w) \cite{CompanionPaper}. Therefore, candidate compounds can be selected for their ability to shift occupancy (potentially without even forming a stable ternary complex), even when isolated bulk binding is weak or not detectable under standard assay conditions.

    \begin{figure}[!htbp]
    \centering
    \includegraphics[width=\textwidth]{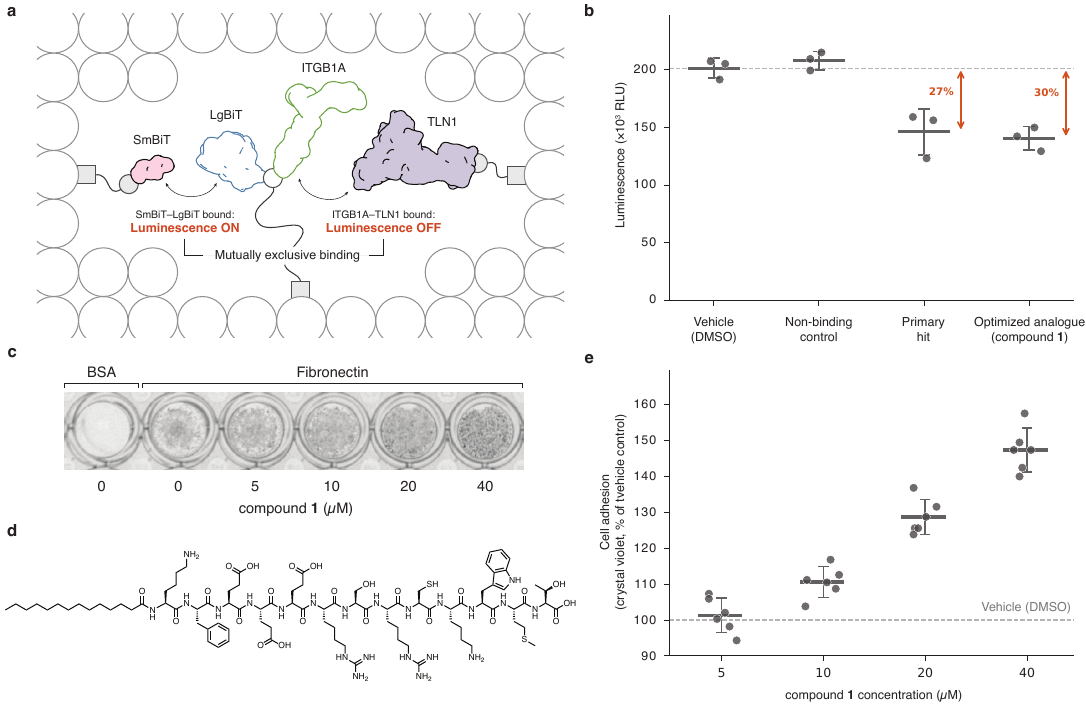}
    \caption{
The ZEN assay renders reporter and target binding mutually exclusive, enabling PPI modulator screening based on shifting local peptide binding.
a. Reporter ON/OFF architecture for the weak interaction between the cytoplasmic tail of integrin subunit beta 1 (ITGB1) isoform A (ITGB1A) and the FERM domain of talin 1 (TLN1) (NMR-based $K_{\mathrm{d}}$ = 491$\pm$10 \textmu{}M for the F3 subdomain of TLN1 \cite{Anthis_Structure_2010}). SmBiT--LgBiT binding forms the luminescence-ON reporter state, while ITGB1A--TLN1 binding forms a competing luminescence-OFF state within the same engineered encounter volume. Therefore, compounds that shift local occupancy towards the target ITGB1A--TLN1 engagement reduce reporter luminescence \cite{CompanionPaper}.
b. Screening output for a focused peptide library based on the integrin $\beta$ cytoplasmic tail. Vehicle and a non-binding control maintained luminescence, whereas a primary hit and its optimised analogue, compound~\textbf{1}, reduced the signal by 27\% and 30\% at 40 \textmu{}M, respectively, indicating shifts towards target-state occupancy. Points show replicate wells (n = 3), bars denote the mean $\pm$ standard deviation (SD), and the dashed line denotes the vehicle mean. Luminescence is reported in relative light units (RLUs). No isolated binding of compound~\textbf{1} was detected by SPR (Supplementary Fig.~\ref{fig:SPR}).
c. Enlarged images of crystal violet (CV)-stained wells containing human U-87 MG glioblastoma cells on bovine serum albumin (BSA)- or fibronectin 1 (FN1)-coated surfaces at the indicated compound~\textbf{1} concentrations. BSA-coated wells showed minimal staining.
d. Chemical structure of compound~\textbf{1}.
e. Adhesion assay after 5 days. Compound 1 dose-dependently increased blank-subtracted CV absorbance ($A_{595}$) to 147\% $\pm$ 6\% of the vehicle control at 40 \textmu{}M, whereas parallel CellTiter-Glo 3D ATP signals showed no corresponding increase (Supplementary Fig.~\ref{fig:adhesion_biomass}), linking the ZEN screening signal to an adhesion phenotype consistent with enhanced ITGB1A--TLN1 engagement. Points show replicate wells (n = 6), bars represent the mean $\pm$ SD, and the dashed line represents the vehicle mean.}
    \label{fig:6}
    \end{figure}

To implement this screen, we configured a mutually exclusive ZEN assay around the interaction between the cytoplasmic tail of integrin subunit beta 1 (ITGB1) isoform A (ITGB1A) and the FERM domain of talin 1 (TLN1) \cite{Integrin_activation_by_talin}; ITGB1A--TLN1 is a weakly binding pair with a nuclear magnetic resonance (NMR)-based $K_{\mathrm{d}}$ of 491 $\pm$ 10 \textmu{}M for the F3 subdomain of TLN1 (Supplementary Fig.~\ref{fig:Talin-Integrin}) \cite{Anthis_Structure_2010}. SmBiT--LgBiT comprised the luminescent reporter state, while ITGB1A--TLN1 presented a competing target-PPI state that reduced reporter occupancy and decreased luminescence (Fig. 6a). When a pair binds, the tethering geometry prevents the other partners from reaching each other. Therefore, a compound that shifts local occupancy towards ITGB1A--TLN1 engagement should reduce the reporter signal. In a peptide-library screen, vehicle treatment and a non-binding control preserved luminescence, whereas a primary hit reduced the signal by 27\%. Optimization yielded compound~\textbf{1}, which reduced the signal by 30\% under the same fixed-copy-number and fixed-geometry conditions (Fig. 6b, d). Compound 1 did not show detectable isolated binding by surface plasmon resonance (SPR; Supplementary Fig.~\ref{fig:SPR}). Through varying the reporter-pair affinity, the assay threshold can be independently tuned from sensitive discovery to selective refinement (Supplementary Fig.~\ref{fig:switch sensitivity}).

To confirm this result in a standard in vitro extracellular matrix adhesion assay, compound~\textbf{1} increased fibronectin 1 (FN1)-dependent cell adhesion in a dose-dependent manner, reaching approximately 150\% of the vehicle control at 40 \textmu{}M, whereas the bovine serum albumin (BSA) control surface showed minimal adhesion (Fig. 6c, e). The cellular response, with no detectable change in proliferation, is consistent with enhanced ITGB1A--TLN1 engagement \cite{Integrin_activation_by_talin}. This proof of concept establishes spatial control as a systematic strategy for PPI-modulator discovery, offering a path from affinity-ranked or stable-ternary-complex screens towards a geometry-based screen of modulators for weak PPIs.

\section*{Summary and outlook}

The control of molecular interactions is no longer confined to molecular number; molecular space provides an independent experimental axis. Exploiting the denominator V in N/V, the V-controlled paradigm circumvents the practical concentration limits of the conventional N-controlled paradigm. Modulating the spatial parameter, the ZEN architecture inverted a solution-phase binding hierarchy, quantified a $K_{\mathrm{d}}$ of order 10 mM from sub-picomole samples at an S/N of $\sim$$10^{3}$, and successfully guided the optimization of a PPI modulator.

Our results reveal a risk in the routine interpretation of molecular interactions: concentration-based frameworks can be misleading when transposed unchanged from homogeneous bulk systems onto heterogeneous microscopic ones \cite{protein_interaction}. Equation 1 illustrates the risk visually. Taken literally, and shown schematically for clarity, the cancellation it allows would make the constant $K_{\mathrm{d}}$ vary with accessible volume. Strictly, however, its right-hand side also includes a factor describing the balance between unbound and bound states. In the bulk, this balance is folded into a concentration ratio. When local states are resolved, this factor must remain explicit (Supplementary Section~\ref{SI_detailed_derivation}). As accessible volume changes, the occupancy balance adjusts to keep intrinsic $K_{\mathrm{d}}$ constant; therefore, volume controls occupancy, not affinity (the bulk concentration form represents a well-mixed limit). At the opposite single-pair limit described within the ZEN architecture, a Michaelis--Menten form emerges, with geometric effective concentration in place of free-ligand concentration \cite{CompanionPaper}. Thus, an interaction classified as weak in bulk solution can become functionally strong within a nanoscale niche, where its effective concentration can exceed its $K_{\mathrm{d}}$. This apparent paradox highlights both the risk of extrapolating from unmodified bulk solution and the advantage of a spatial control axis.

Spatial control is especially powerful at the nanoscale. Conventional bulk methods cannot control accessible volume at this scale, while modern nanoscale architectures can. At the nanoscale, small perturbations in geometry or accessible range can produce disproportionately large shifts in encounter probability and occupancy (Fig. 2a). Reshaping a local environment can require less energy than changing molecular abundance. Indeed, biology may exploit spatial control as an energy-efficient means to tune local equilibria. Familiar proximity effects need a quantitative reinterpretation: at the nanoscale, local geometry and accessible range can become more influential than intrinsic binding chemistry, as the chemistry--geometry crossover shows \cite{nanoscale_biology1,nanoscale_biology2,nanoscale_biology3,nanoscale_biology4,CompanionPaper}.

The V-controlled paradigm offers more than a second control axis: it also presents a much larger parameter space than the abundance-based N-controlled paradigm. The N-controlled paradigm provides one concentration variable per species, whereas the V-controlled paradigm can encode accessible volumes and pairwise or higher-order overlaps (for n species, n concentration variables versus up to $2^{n}-1$ volume terms, respectively; Fig. 2c). Even a two-component system gains a third spatial parameter: the overlap of the accessible regions of the two components. Therefore, rational variation in molecular placement can generate many local states from the same molecular set, whereas bulk mixing compresses these states into a few averages. From this parameter-count perspective, it is not surprising that simply mixing the right components can fail to reproduce cellular behaviour \cite{A_Cell,Biology_of_boundary_conditions,ProteinSynthesis}.

The ZEN architecture provides a minimal experimental implementation of this expanded space. Its modular design can position additional components, vary cavity geometry, and accommodate diverse analytical outputs; the plate-reader assay used here is only one example. As each implementation retains the same spatial variables, data from functional assays and microscopic observations can be integrated directly with theory and simulation, supporting the analysis of interaction networks and regulatory principles across molecular and cellular scales \cite{CompanionPaper}. This modularity enables complex molecular logic to be constituted in simplified, controllable models. A few precisely placed, balanced components can encode an interaction landscape far richer than their number alone suggests.

\clearpage
\bibliography{bio}

\begin{itemize}
 \item[{\small Acknowledgements}] The authors thank the iCeMS Analysis Center at Kyoto University for access to the ICP-AES facilities and technical support with sample identification. This work was supported by JSPS KAKENHI (grant Nos. JP23K26777 to D. F. and JP23H02776 to Y. I.), AMED (grant No. 22am0401020h0004 to D. F.), the AMED Research Support Project for Life Science and Drug Discovery (BINDS; grant Nos. JP22ama121033, JP23ama121033 and JP24ama121033 to Y. I.), JST ACT-X (grant No. JPMJAX20BK to M. Y.) and JST FOREST (grant Nos. JPMJFR203R to D. F., JPMJFR220E to M. Y. and JPMJFR255J to Y. I.). D. F. was additionally supported by JST-Mirai (grant No. JPMJMI22H5), JKA (KEIRIN RACE promotion funds), the Inamori Foundation (InaRIS Fellowship) and the Asian Young Scientist Fellowship.
 \item[{\small Author contributions}] M. Y. designed and performed the chemical synthesis and ZEN binding-affinity experiments. Y. I. designed and performed the modulator screening and cell-adhesion assays, and prepared and analysed the surface plasmon resonance experiments. F. M. and D. F. established the theoretical framework and conducted the analysis. M. Y. and D. F. designed the DNA nanostructure, and M. Y. assembled it. C. M., T. Y., M. K., R. M. and Y. N. prepared recombinant proteins. M. Y. and C. N. performed the protein conjugation with AccuLinkers. Y. S. and M. Y. performed cryo-EM analysis. M. T. and M. Y. conducted nsTEM analysis. M. Y., F. M. and D. F. wrote the manuscript. M. Y., Y. I. and D. F. acquired funding. D. F. conceived and supervised the project. All authors critically read and approved the manuscript.

 \item[{\small Competing interests}] The authors declare the following competing interests: Kyoto University is the applicant for a pending international patent application published as WO 2025/079662 A1, covering structures and methods for positioning biomolecules and analysing their interactions as implemented in the ZEN platform described in this work; D. F., M. Y., F. M. and Y. I. are named inventors. D. F. and Y. I. are co-founders of SeedFairing Institute, a Japanese non-profit general incorporated association established to facilitate dissemination of this technology. The remaining authors declare no competing interests.

 \item[{\small Data availability}] The data supporting the findings of this study are available within the paper and its Supplementary Information. Source data underlying the main and Supplementary figures are available from the corresponding author upon reasonable request. The authors plan to make reagents and technical support related to the ZEN platform available through SeedFairing Institute upon reasonable request.

 \item[{\small Code availability}] Code for the Bayesian $K_{\mathrm{d}}$ inference from the ZEN titration assay with DarkBiT~\textbf{2} (Fig.~5) is available at \url{https://github.com/FujitaG/zen-kd-estimation}, and code for the tethered-sphere Monte Carlo overlap calculator (Fig.~4b) at \url{https://github.com/FujitaG/zen-spatial-overlap}. Both are archived at Zenodo (\url{https://doi.org/10.5281/zenodo.21777953} and \url{https://doi.org/10.5281/zenodo.21777964}, respectively). The response-model implementation used by the inference is distributed with the companion paper\cite{CompanionPaper}.

 \item[{\small Use of large language models}] Large language models (Claude, Anthropic; GPT, OpenAI) were used to improve the readability and language of the manuscript, and to assist in writing the analysis code listed above.

 \item[Correspondence] Correspondence and requests for materials should be addressed to Daishi Fujita~(dfujita@icems.kyoto-u.ac.jp).
\end{itemize}

\clearpage

\noindent{\bfseries \LARGE Supplementary Information}\setlength{\parskip}{12pt}%

\setcounter{figure}{0}
\renewcommand{\theHfigure}{supp.\arabic{figure}}
\renewcommand{\thefigure}{S\arabic{figure}}  %
\setcounter{table}{0}
\renewcommand{\theHtable}{supp.\arabic{table}}
\renewcommand{\thetable}{S\arabic{table}}

\noindent Throughout, bold numerals and letters denote compound names, so DarkBiT~\textbf{1}, compound~\textbf{1} and compound~\textbf{A} are different molecules. The bold numerals in Supplementary Table 9 number the DNA strands. Numerals set close to a name identify sequence variants (for example SmBiT114).

\section{Chemicals and reagents}
\label{sec:Chemicals and reagents}

All chemical reagents were purchased from BroadPharm, Tokyo Chemical Industry Co., Ltd. (TCI), Sigma-Aldrich and BLDpharm at the highest commercially available quality, and used without further purification. Synthetic peptides were purchased from Biologica with $>$95\% purity. Synthetic oligonucleotides for DNA bricks were purchased from Eurofins Genomics as oligonucleotide purification cartridge (OPC)-purified grade. Azide- and/or fluorescent-dye-modified oligonucleotides were purchased from Integrated DNA Technologies, Inc. as high-performance liquid chromatography (HPLC)-purified grade. For cell-based assays, human plasma fibronectin (Fibronectin Neosilk, catalogue no. 54071) was purchased from Immuno-Biological Laboratories Co., Ltd. (IBL). High-glucose Dulbecco’s modified Eagle’s medium (DMEM; catalogue no. 11965118), penicillin/streptomycin (catalogue no. 15140122) and non-treated 96-well Nunc plates (catalogue no. 2674427) were purchased from Thermo Fisher Scientific. Fetal bovine serum (FBS; catalogue no. FB-1003/500; batch no. S00KK; identification no. S00KK10002) was purchased from Biosera. The Nano-Glo Luciferase Assay System (catalogue no. N1110; 10 mL) and the CellTiter-Glo 3D Cell Viability Assay (catalogue no. G9681) were purchased from Promega, and crystal violet (catalogue no. C0428) was purchased from TCI.

\section{General chemical reaction}
\label{sec:General chemical reaction}
All reactions were carried out under an argon atmosphere with dry solvents unless otherwise noted. Chromatographic purifications were performed on a Teledyne ISCO CombiFlash Rf 200i equipped with standard silica columns (SepaFlash SilicaFlash Cartridge UltraPure irregular Silica Gel, 40--63 \textmu m, 60 {\AA} from Santai Technologies). Thin-layer chromatography (TLC) was performed on precoated silica-gel 60 TLC glass plates with fluorescence indicator UV${}_{254}$. The TLC plates were stained with iodine. ${}^1$H and ${}^{13}$C nuclear magnetic resonance (NMR) spectra were recorded on a Bruker spectrometer operating at 500 and 126 MHz, respectively. Chemical shifts are given in ppm and calibrated using residual undeuterated solvent as an internal reference (in CDCl${}_3$, $\delta$ = 7.26 for ${}^1$H NMR and $\delta$ = 77.1 for ${}^{13}$C NMR). The following abbreviations describe the multiplets: s, singlet; d, doublet; dd, doublet of doublets; ddd, doublet of doublets of doublets; tdd, triplet of doublets of doublets; t, triplet; dt, doublet of triplets; q, quartet; tq, triplet of quartets; quin, quintet; sext, sextet; m, multiplet; br, broad. Mass spectra were recorded on a high-resolution mass spectrometer (timsTOF) from Bruker equipped with an electrospray source (polarity: positive, capillary: 3.4 kV, capillary temperature: 200~${}^\circ$C, sheath gas flow rate: 5.0 L/min, mass range: $m/z$ 50 to 1600).

\newpage

\section{Synthesis of molecules}
\label{sec:Synthesis of molecules}

\noindent OEG denotes oligo(ethylene glycol); BCN, bicyclo[6.1.0]non-4-yne; NHS, $N$-hydroxysuccinimide. The spacers used throughout are monodisperse oligomers obtained by stepwise synthesis, not by polymerization.

\textbf{Synthesis of compound A }
\label{sec:Synthesis of compound A}

\begin{center}
\includegraphics[width=\linewidth]{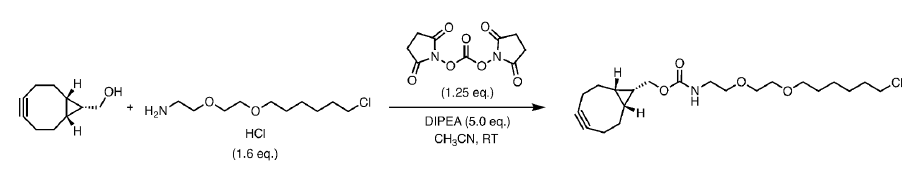}
\end{center}

\noindent To a solution of BCN-OH (25 mg, 0.16 mmol) and $N,N'$-disuccinimidyl carbonate (53 mg, 0.21 mmol, 1.25 equiv.) in acetonitrile (0.4 mL, 0.4 M) was added $N,N$-diisopropylethylamine (DIPEA; 145 \textmu L, 5.0 equiv.). The mixture was stirred at room temperature for 3.5 hours. The reaction was monitored by TLC, which showed complete consumption of BCN-OH. HaloTag ligand amine HCl salt (69 mg, 0.27 mmol, 1.6 equiv.) was added to the reaction solution at room temperature. The solution was then stirred for 16 hours. The reaction mixture was diluted with water and extracted with EtOAc three times. The combined organic phase was washed with saturated NaHCO${}_3$ aqueous solution and saturated NH${}_4$Cl aqueous solution. The resulting organic layer was dried over anhydrous MgSO${}_4$, filtered and concentrated under reduced pressure. The crude product was purified by chromatography over silica gel with EtOAc and hexane (1:1) to afford the desired carbamate \textbf{A} (25 mg, 38\% yield) as a colourless oil. ${}^1$H NMR (500 MHz, CDCl${}_3$) $\delta$ = 5.17 (br, 1H), 4.15 (d, $J$ = 8.0 Hz, 2H), 3.60--3.62 (m, 2H), 3.55--3.57 (m, 4H), 3.53 (t, $J$ = 6.5 Hz, 2H), 3.46 (t, $J$ = 6.5 Hz, 2H), 3.38 (q, $J$ = 5.0 Hz, 2H), 2.20--2.29 (m, 5H), 1.78 (quint, $J$ = 6.5 Hz, 2H), 1.59--1.65 (m, 3H), 1.43--1.48 (m, 2H), 1.33--1.41 (m, 3H), 0.93--0.96 (m, 2H), 0.83--0.88 (m, 2H). The obtained ${}^1$H NMR signals are consistent with the reported signals\cite{Gruskos2016}. HR-MS: 422.2068 (C${}_{21}$H${}_{34}$ClNNaO${}_4$${}^+$, [M+Na]${}^+$; calc. 422.2069).

\noindent\textbf{Synthesis of compound B}

\begin{center}
\includegraphics[width=\linewidth]{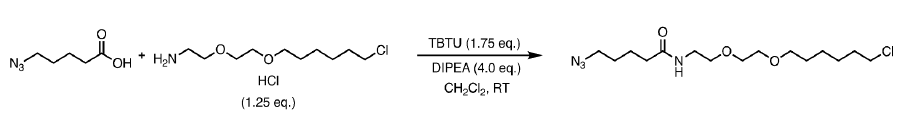}
\end{center}

\noindent To a solution of 5-azidopentanoic acid (30 mg, 0.21 mmol) in CH${}_2$Cl${}_2$ (0.7 mL, 0.3 M) were added $O$-(benzotriazol-1-yl)-$N,N,N',N'$-tetramethyluronium tetrafluoroborate (TBTU; 118 mg, 0.37 mmol, 1.75 equiv.) and DIPEA (145 \textmu L, 4.0 equiv.) at room temperature. After adding HaloTag ligand amine HCl salt (68 mg, 0.26 mmol, 1.25 equiv.), the reaction solution was stirred at room temperature for 14 hours. The reaction mixture was diluted with water and extracted with EtOAc three times. The combined organic phase was washed with saturated NaHCO${}_3$ aqueous solution and saturated NH${}_4$Cl aqueous solution. The resulting organic layer was dried over anhydrous MgSO${}_4$, filtered and concentrated under reduced pressure. The crude product was purified by chromatography over silica gel with EtOAc and hexane (17:3) to afford the desired amide \textbf{B} (9.0 mg, 12\% yield) as a colourless oil. ${}^1$H NMR (500 MHz, CDCl${}_3$) $\delta$ = 6.00 (br, 1H), 3.60--3.62 (m, 2H), 3.51--3.57 (m, 6H), 3.43--3.47 (m, 4H), 3.29 (t, $J$ = 7.0 Hz, 2H), 2.20 (t, $J$ = 7.5 Hz, 2H), 1.69--1.78 (m, 4H), 1.57--1.65 (m, 4H), 1.42--1.48 (m, 2H), 1.34--1.40 (m, 2H). ${}^{13}$C NMR (126 MHz, CDCl${}_3$) $\delta$ = 172.32, 71.27, 70.28, 70.04, 69.82, 51.20, 45.04, 39.17, 35.90, 32.53, 29.48, 28.43, 26.69, 25.43, 22.81. HR-MS: 349.2002 (C${}_{15}$H${}_{30}$ClN${}_4$O${}_3$${}^+$, [M+H]${}^+$; calc. 349.2001), 371.1821 (C${}_{15}$H${}_{29}$ClN${}_4$NaO${}_3$${}^+$, [M+Na]${}^+$; calc. 371.1820).

\noindent\textbf{Synthesis of compound C}

\begin{center}
\includegraphics[width=\linewidth]{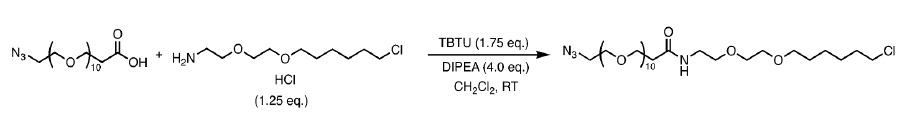}
\end{center}

\noindent To a solution of azide-OEG10-acid (30 mg, 0.05 mmol) in CH${}_2$Cl${}_2$ (0.18 mL, 0.3 M) were added TBTU (30 mg, 0.09 mmol, 1.75 equiv.) and DIPEA (38 \textmu L, 4.0 equiv.) at room temperature. After adding HaloTag ligand amine HCl salt (17.5 mg, 0.07 mmol, 1.25 equiv.), the reaction solution was stirred at room temperature for 18 hours. The reaction mixture was diluted with water and extracted with EtOAc three times. The combined organic phase was washed with saturated NaHCO${}_3$ aqueous solution and saturated NH${}_4$Cl aqueous solution. The resulting organic layer was dried over anhydrous MgSO${}_4$, filtered and concentrated under reduced pressure. The crude product was purified by chromatography over silica gel with MeOH and CHCl${}_3$ (1:9) to afford the desired amide \textbf{C} (24 mg, 58\% yield) as a colourless oil. ${}^1$H NMR (500 MHz, CDCl${}_3$) $\delta$ = 6.58 (br, 1H), 3.73 (t, $J$ = 5.0 Hz, 2H), 3.57--3.66 (m, 40H), 3.51--3.55 (m, 6H), 3.40--3.44 (m, 4H), 3.38 (t, $J$ = 5.0 Hz, 2H), 2.46 (t, $J$ = 6.3 Hz, 2H), 1.77 (dt, $J$ = 6.7 Hz, $J$ = 14.8 Hz, 2H), 1.59 (dt, $J$ = 6.5 Hz, $J$ = 15 Hz, 2H), 1.40--1.46 (m, 2H), 1.31--1.38 (m, 2H). ${}^{13}$C NMR (126 MHz, CDCl${}_3$) $\delta$ = 171.49, 77.41, 77.36, 77.16, 76.90, 71.39, 70.83, 70.80, 70.77, 70.72, 70.70, 70.53, 70.45, 70.41, 70.17, 69.98, 67.41, 50.82, 45.15, 39.25, 37.16, 32.65, 29.59, 26.81, 25.55. The carbon peaks in the repeating units of the oligoethylene glycol moiety overlap. HR-MS: 761.4308 (C${}_{33}$H${}_{66}$ClN${}_4$O${}_{13}$${}^+$, [M+H]${}^+$; calc. 761.4309), 783.4129 (C${}_{33}$H${}_{65}$ClN${}_4$NaO${}_{13}$${}^+$, [M+Na]${}^+$; calc. 783.4129).

\noindent\textbf{Synthesis of compound D}

\begin{center}
\includegraphics[width=\linewidth]{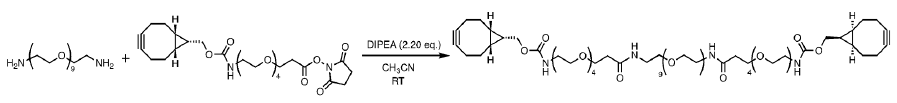}
\end{center}

\noindent To a solution of BCN-OEG4-NHS (11.0 mg, 20 \textmu mol, 2.0 equiv.) in CH${}_3$CN (0.3 mL, 0.033 M) were added amino-OEG9-amine (4.6 mg, 10.2 \textmu mol, 1.0 equiv.) and DIPEA (3.0 mg, 22 \textmu mol, 2.2 equiv.) at room temperature. The reaction solution was stirred for 22 hours. The reaction was monitored by TLC with iodine staining. The reaction solution was diluted with EtOAc, then washed twice with saturated aqueous NaHCO${}_3$ to remove $N$-hydroxysuccinimide and the hydrolysed carboxylic acid. The organic layer was subsequently washed twice with saturated aqueous NH${}_4$Cl to remove the unreacted amine starting material and the monoreacted intermediate. The solution was dried over anhydrous MgSO${}_4$ for 30 minutes, filtered and concentrated \textit{in vacuo} to afford the desired compound \textbf{D} (6.4 mg, 48\% yield) as a colourless oil that gave a single spot by TLC with iodine staining. The resulting product was used for the next click reaction with azide-modified single-stranded DNA oligonucleotides (Section~\ref{Functionalization of oligonucleotides}) without further purification.

\noindent\textbf{Synthesis of compound E}

\begin{center}
\includegraphics[width=\linewidth]{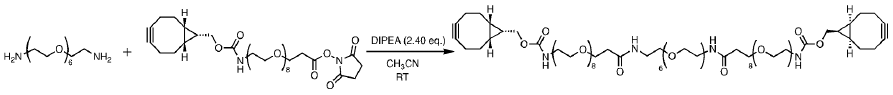}
\end{center}

\noindent To a solution of BCN-OEG8-NHS (9.5 mg, 13.5 \textmu mol, 2.2 equiv.) in CH${}_3$CN (160 \textmu L, 0.037 M) were added amino-OEG6-amine (2.0 mg, 6.2 \textmu mol, 1.0 equiv.) and DIPEA (2.0 mg, 15 \textmu mol, 2.4 equiv.) at room temperature. The reaction solution was stirred for 24 hours. The reaction was monitored by TLC with iodine staining. The reaction solution was diluted with EtOAc, then washed twice with saturated aqueous NaHCO${}_3$ to remove $N$-hydroxysuccinimide and the hydrolysed carboxylic acid. The organic layer was subsequently washed twice with saturated aqueous NH${}_4$Cl to remove the unreacted amine starting material and the monoreacted intermediate. The solution was dried over anhydrous MgSO${}_4$ for 30 minutes, filtered and concentrated \textit{in vacuo} to afford the desired compound \textbf{E} (1.2 mg, 21\% yield) as a colourless oil that gave a single spot by TLC with iodine staining. The resulting product was used for the next click reaction with azide-modified single-stranded DNA oligonucleotides (Section~\ref{Functionalization of oligonucleotides}) without further purification.

\noindent\textbf{Compound F}

\noindent Compound \textbf{F}, a commercially available BCN dimer, was purchased from BroadPharm and used without further purification.

\noindent The ${}^1$H and ${}^{13}$C NMR spectra of the synthetic small molecules are collected after the Methods sections.

\newpage

\section{Construct design and protein preparation}

\textbf{Construction of tagged recombinant cDNA plasmids}

\noindent Genes encoding human integrin $\beta$1 cytoplasmic tail (ITGB1A) and talin-1 FERM domain (TLN1) were amplified by polymerase chain reaction (PCR) using human cDNAs from the Mammalian Gene Collection\cite{MGC2002} as templates. DNA sequences of HaloTag, LgBiT and the SmBiT variants, which possess different binding affinities for LgBiT, were retrieved from the Promega website (\url{https://www.promega.jp/products/cloning-and-dna-markers/cloning-vectors-and-kits/ph6htn-and-ph6htc-his6halotag-t7-vectors/?catNum=G7971}), the Addgene website (\url{https://www.addgene.org/188319/sequences/}) and the original report\cite{LgBiT-SmBiT}, respectively. The DNA sequences were synthesized by Eurofins Genomics. Overlapping sequences were added at the 5′ and 3′ ends for seamless cloning. Amplified ITGB1A and TLN1 were subcloned into the pH6HTN (Promega, catalogue no. G7971) and pET-15b (Novagen, catalogue no. 69661) vectors using Gibson Assembly seamless cloning. The DarkBiT~\textbf{1} plasmid was derived from the SmBiT104 plasmid using an inverse-PCR point-mutation method. Detailed information on the constructs used in this study is given in \textbf{Supplementary Tables 1--4}.

\noindent\textbf{Preparation of recombinant proteins}

\noindent The recombinant His-tagged-HaloTag-LgBiT, His-tagged-HaloTag-SmBiT, His-tagged-HaloTag-LgBiT-SmBiT and His-tagged-HaloTag-DarkBiT~\textbf{1} were expressed in the BL21 Star \textit{E.~coli} strain. The bacteria were grown in LB medium to an optical density at 600 nm (OD${}_{600}$) of 0.8. Isopropyl $\beta$-D-1-thiogalactopyranoside (IPTG) was then added to a final concentration of 1 mM, and the cells were grown for a further 2 hours. After collection by centrifugation at \(8{,}000 \times g\) for 20 minutes, the bacterial pellets were resuspended in wash buffer (50 mM 4-(2-hydroxyethyl)-1-piperazineethanesulfonic acid (HEPES), pH 7.5, 150 mM NaCl, 10 mM imidazole) and sonicated. To remove the debris, the samples were centrifuged at \(10{,}000 \times g\) for 30 minutes. The supernatant was then subjected to purification using Ni-NTA agarose resin (QIAGEN, \#30210) following the manufacturer’s protocol with customized buffers (wash buffer: 50 mM HEPES (pH 7.5), 150 mM NaCl, 10 mM imidazole; elution buffer: 50 mM HEPES (pH 7.5), 150 mM NaCl, 200 mM imidazole).
The recombinant His-tagged-LgBiT-HaloTag-ITGB1A and His-tagged-HaloTag-TLN1 were prepared in the same manner from the same strain, using customized buffers (wash buffer: 50 mM HEPES (pH 7.5), 1 M NaCl, 10 mM imidazole; elution buffer: 50 mM HEPES (pH 7.5), 1 M NaCl, 200 mM imidazole).
The protein samples were divided into small portions and frozen in liquid nitrogen. The recombinant protein samples were stored at $-80$~${}^\circ$C until use.

\section{Functionalization of oligonucleotides}
\label{Functionalization of oligonucleotides}

\textbf{Synthesis of BCN-conjugated single-stranded DNAs}

\noindent A 50 \textmu L portion of a 100 \textmu M solution of azide- and/or fluorescent-dye-modified oligonucleotide in Tris-EDTA (TE) buffer (10 mM Tris-HCl, 0.5 mM ethylenediaminetetraacetic acid (EDTA), pH 8.0) was reacted with 5 \textmu L of a 10 mM solution of BCN-modified small molecule (compound \textbf{A}, \textbf{D}, \textbf{E} or \textbf{F}) in $N,N$-dimethylformamide (DMF), and the reaction solution was then heated at 60~${}^\circ$C for 2 hours. After the reaction, the solution was cooled to room temperature and subjected to 5\% native agarose gel electrophoresis at 100~V for 1 hour (gel prepared with PrimeGel agarose PCR-sieve HRS from Takara in Tris/borate/EDTA (TBE) buffer (82 mM Tris-HCl, 81 mM B(OH)${}_3$, 1.9 mM EDTA, pH 8.3)). Then, the target gel bands were excised and melted. The target oligonucleotide was purified using a DNA extraction kit (NucleoTrap) from Takara. The concentration of the purified solution containing the target oligonucleotide was determined by A${}_{260}$ absorbance.

\noindent\textbf{Synthesis of HaloTag ligand-conjugated single-stranded DNAs (AccuLinkers)}

\noindent A 30 \textmu L portion of BCN-conjugated oligonucleotide in a 60--80 \textmu M aqueous solution was reacted with 25 equivalents of azide-modified HaloTag ligand (compound \textbf{B} or \textbf{C}) as a 50 mM solution in DMF, and the reaction solution was then heated at 60~${}^\circ$C for 2 hours. After the reaction, the solution was cooled to room temperature and subjected to 5\% native agarose gel electrophoresis at 100~V for 1 hour (gel prepared with PrimeGel agarose PCR-sieve HRS from Takara in TBE buffer). Then, the target gel bands were excised and melted. The target oligonucleotide was purified using a DNA extraction kit (NucleoTrap) from Takara. The concentration of the purified solution containing the target oligonucleotide was determined by A${}_{260}$ absorbance. Representative gels are shown in Supplementary Fig.~\ref{fig:Extended Data Fig. AccuLinker Evi}.

\section{Design and synthesis of DNA brick (DB01)}
\label{Design and Synthesis of DNA brick (DB01)}

The DB01 structure was designed using the open-source software NanoBricks\cite{DNA_Bricks_Nature} (\url{https://nanobricks.software/}). To assemble the structures, DNA strands were mixed to a final concentration of 100 nM per strand (DB01 contains 705 strands) in TE (10 mM Tris-HCl, 0.5 mM EDTA, pH 8.0) supplemented with 30 mM MgCl${}_2$. The strand mixture was then annealed in a PCR thermal cycler by a linear cooling ramp from 80~${}^\circ$C to 60~${}^\circ$C over 1 hour, followed by a 100-hour ramp from 60~${}^\circ$C to 30~${}^\circ$C. Annealed samples were concentrated in an Amicon Ultra-0.5 device (100 kDa molecular-weight cut-off) from Merck by centrifugation at \(3{,}000 \times g\), then subjected to 1.5\% native agarose gel electrophoresis at 100~V for 1 hour (gel prepared in 0.5$\times$TBE buffer supplemented with 6 mM Mg(OAc)${}_2$) in a cold room at 4~${}^\circ$C. DNA was stained by incubation with SYBR Gold (Thermo Fisher Scientific). Then, the target gel bands were excised and placed into a centrifugal filter tube from Merck. The excised gel was crushed into fine pieces by centrifugation at \(10{,}000 \times g\) for 10 minutes, and target DNA bricks were extracted in the filtered solution. The buffer was exchanged for 20 mM MgCl${}_2$ solution using an Amicon Ultra-0.5 device (100 kDa) by centrifugation at \(3{,}000 \times g\), and the concentration of the resulting solution was then determined by A${}_{260}$ absorbance. The isolated yield was calculated from that concentration, the recovered volume and a DB01 molecular weight estimated from the average nucleotide mass, relative to the theoretical maximum set by the 100 nM per-strand annealing concentration. The value quoted in the main text is the highest obtained across preparations. The design and the purified product are shown in Supplementary Figs.~\ref{fig:Extended Data Fig. Bricks design} and~\ref{fig:Extended Data Fig. DB01}.

\section{HaloTag conjugation with oligonucleotides}
\label{HaloTag conjugation with oligonucleotides}

AccuLinker (5 \textmu M) was reacted with 3 equivalents of HaloTag-fused recombinant protein (15 \textmu M) in buffer (50 mM Tris-HCl, 100 mM NaCl, pH 7.5) supplemented with 20 mM MgCl${}_2$ on ice for 1 hour. The HaloTag conjugation reaction was monitored by polyacrylamide gel electrophoresis (PAGE). The starting material, AccuLinker, was quantitatively converted to the desired protein-conjugated oligonucleotide, and the resulting solution was used for the next protein tethering step without further purification. Conjugation was verified by gel electrophoresis (Supplementary Figs.~\ref{fig:Extended Data Fig. Protein conju1}--\ref{fig:Extended Data Fig. Protein conju3}).

\section{Protein tethering to DB01}
\label{Protein tethering to DB01}

DB01 (30 nM) was reacted with 10 equivalents of protein-conjugated oligonucleotide (300 nM) in buffer (50 mM Tris-HCl, 100 mM NaCl, pH 7.5) supplemented with 20 mM MgCl${}_2$ on ice for 2 hours. Excess proteins and protein-conjugated oligonucleotides were removed using an Amicon Ultra-0.5 device (100 kDa) by centrifugation at \(3{,}000 \times g\). After five rounds of centrifugation, the concentration of the resulting solution was determined by A${}_{260}$ absorbance, and the solution was diluted to 1 nM. Tethering efficiency is shown in Supplementary Fig.~\ref{fig:Extended Data Fig. Protein tethering}.

\section{Luminescence measurement of the LgBiT-SmBiT complex}
\label{Luminescence measurement of the LgBiT-SmBiT complex}

In a 96-well white plate (low-binding-surface 1/2-Area OptiPlate, PerkinElmer), 25 \textmu L of protein-immobilized DB01 solution (1 nM) was mixed with 25 \textmu L of NanoBiT substrate solution (Nano-Glo Luciferase Assay System, Promega) at room temperature. After a 3-minute incubation, the luminescence intensity was measured at room temperature on a plate reader (Infinite F500, Tecan) with an integration time of 200 ms. Background was estimated from 12 wells in plate row H measured on the same plate, and the signal-to-noise ratio was defined as \((I-\bar B)/s_B\), where \(\bar B\) and \(s_B\) are the mean and sample standard deviation of the background-well signals. The acquired data are summarized and plotted in \textbf{Figs. \ref{fig:4} and~\ref{fig:5}} and \textbf{Supplementary Fig. \ref{fig:local conc}}.

\section{Peptide regulation of protein-protein interaction in DB01}
\label{Peptide regulation of protein-protein interaction in DB01}

Protein-immobilized DB01 solution was diluted to 1 nM with an aqueous solution containing the peptide used in each experiment (DarkBiT~\textbf{2}, one of the screening peptides of Section~\ref{PPI modulator screening} or the optimized modulator compound~\textbf{1}), and the solution was incubated on ice for 1 hour. The final aqueous solution contained 50 mM Tris-HCl, 100 mM NaCl and 20 mM MgCl${}_2$ (pH 7.5), with a final dimethyl sulfoxide (DMSO) concentration below 1\%. In a 96-well white plate, 25 \textmu L of the incubated solution was mixed with 25 \textmu L of NanoBiT substrate solution (Nano-Glo Luciferase Assay System, Promega) at room temperature. After a 3-minute incubation, the luminescence intensity was measured at room temperature on a plate reader (Infinite F500, Tecan) with an integration time of 200 ms. The acquired data are summarized and plotted in \textbf{Figs. \ref{fig:5} and~\ref{fig:6}} and \textbf{Supplementary Fig. \ref{fig:switch sensitivity}}.

\section{Determination of the dissociation constant of \texorpdfstring{DarkBiT~\textbf{1}}{DarkBiT 1}}
\label{Determination of the dissociation constant of DarkBiT1}

To determine the affinity of HaloTag-DarkBiT~\textbf{1} for HaloTag-LgBiT, recombinant HaloTag-LgBiT (3 \textmu M) was incubated with HaloTag-DarkBiT~\textbf{1} at concentrations from 0 to 146 \textmu M in 30 \textmu L of reaction buffer (50 mM Tris-HCl, 100 mM NaCl, pH 7.5) in a 96-well white plate. After 30 minutes on ice, intrinsic fluorescence was recorded with excitation at 290 nm and emission from 310 to 400 nm. Tryptophan fluorescence quenching on complex formation, $I=F_a+F_b-F_c$, was obtained at 340 nm from the fluorescence of HaloTag-LgBiT alone ($F_a$), HaloTag-DarkBiT~\textbf{1} alone ($F_b$) and their mixture ($F_c$). Because $I$ is proportional to the complex concentration, plotting it against the HaloTag-DarkBiT~\textbf{1} concentration and fitting gave the dissociation constant ($K_{\mathrm{d}}$) (Supplementary Fig.~\ref{fig:Extended Data Fig. Trp analysis}).

\section{Protein--protein interaction (PPI) modulator screening}
\label{PPI modulator screening}
\textbf{Creation of a focused peptide library and improvement of PPI modulator activity}

\noindent While designing PPI modulators of integrin--talin binding, we observed that low concentrations of the integrin-$\beta$ peptide ITGB3 (Supplementary Table 5) slightly promoted integrin--talin complex formation. We therefore prepared a focused peptide library mimicking the cytoplasmic tail of integrin $\beta$. Iterative evaluation and refinement of the hit peptides, guided by simulation-based stability scoring in MolDesk Screening (IMSBIO Co., Ltd.; \url{https://www.moldesk.com}), yielded the PPI modulator compound~\textbf{1}.

\noindent\textbf{PPI modulator screening using DB01}

\noindent Selected peptides were evaluated using the DB01-based switching system (\textbf{Fig. \ref{fig:6}} and \textbf{Supplementary Fig. \ref{fig:switch sensitivity}}). The experimental protocol is described in Section \ref{Peptide regulation of protein-protein interaction in DB01}.

\section{Surface plasmon resonance}
\label{sec:SPR}

\noindent Compound~\textbf{1} was tested for direct binding to talin-1 on a Biacore S200 instrument (Cytiva). Biotinylated talin-1 FERM domain was captured on a NeutrAvidin sensor chip (Series S Sensor Chip NA, catalogue no. 29699622, Cytiva) at 3,000--5,000 resonance units (RU), the instrument's measure of the mass bound to the sensor surface. Analytes were injected for 120~s in 0.1~M Tris-HCl (pH 8.0), 0.5~M NaCl, 0.01\% Tween-20 and 5\% DMSO.

\noindent Compound~\textbf{1} was dissolved in DMSO to 2.0~mM. The running buffer tolerates no more than 5\% DMSO, so this stock could make up at most one twentieth of an injected sample, setting a ceiling of 100~\textmu M. The series was a two-fold dilution from that ceiling.

\noindent The sensitivity of the assay was checked with a peptide of known affinity, the myristoylated integrin-$\beta$3 peptide ITGB3 (Supplementary Table 5), injected over the same protein preparation. NMR reports a dissociation constant of $273\pm6.4~\mu\mathrm{M}$ for the $\beta$3 cytoplasmic tail binding the talin-1 F3 subdomain\cite{Anthis_Structure_2010}. At the 0.625--10~\textmu M injected here, this peptide therefore binds no more than 3.5\% of the immobilized talin-1. Even so, its response was clearly measurable and increased with concentration (Supplementary Fig.~\ref{fig:SPR}a). Compound~\textbf{1}, injected at ten times those concentrations (6.25--100~\textmu M), showed no such increase (Supplementary Fig.~\ref{fig:SPR}b,c). It therefore binds talin-1 far more weakly than the reference under these conditions.

\section{Biological evaluation of the PPI modulator}
\label{Biological evaluation of PPI modulator}

Because integrin--talin complex formation is essential for cell adhesion, we established an integrin-dependent cell-adhesion assay. First, a non-treated 96-well plate was coated with 100 \textmu L of a 10 \textmu g/mL fibronectin solution for 2 hours, then washed twice with 1$\times$ PBS(-) (phosphate-buffered saline lacking Ca$^{2+}$ and Mg$^{2+}$). In parallel, U87MG human glioblastoma cells were cultured at 37~${}^\circ$C in DMEM/10\% FBS/1\% penicillin/streptomycin (P/S) under 5\% CO${}_2$. The cells were trypsinized and collected by centrifugation at \(400 \times g\) for 5 minutes, and the medium was changed to DMEM/1\% FBS/1\% P/S. Then $4\times10^{3}$ cells were plated in each well, and two plates were prepared under the same conditions. After incubation at 37~${}^\circ$C for 10 minutes, the peptide compounds were added to the wells at the desired concentrations, and incubation was continued at 37~${}^\circ$C for 5 days. One plate was treated with CellTiter-Glo 3D Cell Viability Assay solution, an adenosine triphosphate (ATP)-based indicator of proliferation, according to the manufacturer’s protocol, and luminescence was measured using an EnSpire plate reader (PerkinElmer). These values were analysed independently as a control for viable cell number and were not used to normalize the crystal-violet absorbance. In the other plate, the medium was aspirated and the cells were fixed with 4\% paraformaldehyde for 30 minutes, and the wells were washed once with 1$\times$ PBS(+) (PBS containing Ca$^{2+}$ and Mg$^{2+}$). To quantify the cells that remained bound to fibronectin, a stock of 0.2\% crystal violet and 2\% EtOH was diluted five-fold in PBS(-), and 100 \textmu L of the resulting solution was added and incubated for 30 minutes. The plate was washed three times with 1$\times$ PBS(+). It was then inverted and allowed to dry completely in the dark. Finally, 100 \textmu L of 1\% sodium dodecyl sulfate (SDS) solution was added and the plate was incubated for 30 minutes at room temperature with agitation. The absorbance at 595 nm was measured using an EnSpire plate reader (PerkinElmer). Readings were blank-subtracted and then expressed as a percentage of the mean of the vehicle-treated wells on the same plate. These values were used as indicators of cell adhesion. Proliferation and biomass data are shown in Supplementary Fig.~\ref{fig:adhesion_biomass}.

\section{Negative-stain transmission electron microscopy (nsTEM)}
\label{Negative-stain transmission electron microscopy (nsTEM)}

For imaging, 3.0 \textmu L of agarose-gel-purified DB01 was adsorbed for 30 seconds onto glow-discharged, 2 nm carbon-coated transmission electron microscopy (TEM) grids. The grids were then stained twice, each time for 1 minute, with 25\% aqueous EM stainer (Nisshin EM). Imaging was performed using a JEOL Field Emission Transmission Electron Microscope (FE-TEM) JEM-2200FS operated at 200 kV, equipped with an energy filter for high-contrast imaging. Representative images are shown in Supplementary Fig.~\ref{fig:Extended Data Fig. TEM.}.

\section{Cryo-EM analysis}
\label{Cryo-EM analysis}

The DB01 sample (30 nM) was applied to glow-discharged Quantifoil R0.6 holey grids and plunge-frozen using a Vitrobot Mark IV (Thermo Fisher Scientific) at 4~${}^\circ$C and 100\% humidity. The data were acquired using the EPU software (Thermo Fisher Scientific) on a Glacios cryo-transmission electron microscope operated at 200 kV and equipped with a Falcon 4 direct electron detector, at the Institute for Life and Medical Sciences, Kyoto University. A representative image is shown in Fig.~\ref{fig:3}c of the main text.

\section{Survey of reported protein--protein binding affinities}
\label{sec:reported_affinities}

Figure~\ref{fig:1} summarizes the PPB-Affinity dataset\cite{Ye2024PPBAffinity}.
Because multiple mutant records can be associated with the same wild-type
complex, only records without annotated mutations were selected. Records were
further required to have an annotated affinity-measurement method and a
parseable affinity-release date. The resulting dataset comprised 1,460
records: 879 measured by surface plasmon resonance or biolayer interferometry
(SPR/BLI), 118 by isothermal titration calorimetry (ITC), and 463 by other
methods, including records annotated as Unknown. Twelve records with
$K_{\mathrm{d}} < 10^{-12}\,\mathrm{M}$ fell outside the plotted range.
The final plot contains 1,448 records: 878 SPR/BLI, 117 ITC, and 453
other-method records.

\newpage
\noindent\textbf{${}^1$H and ${}^{13}$C NMR of synthetic small molecules}

\noindent Compound \textbf{A} is a known compound and is shown by ${}^1$H NMR only\cite{Gruskos2016}. Compounds \textbf{D} and \textbf{E} were carried into the next step without further purification and are not included.

\noindent ${}^1$H NMR of compound \textbf{A}

\begin{center}
\includegraphics[width=\linewidth]{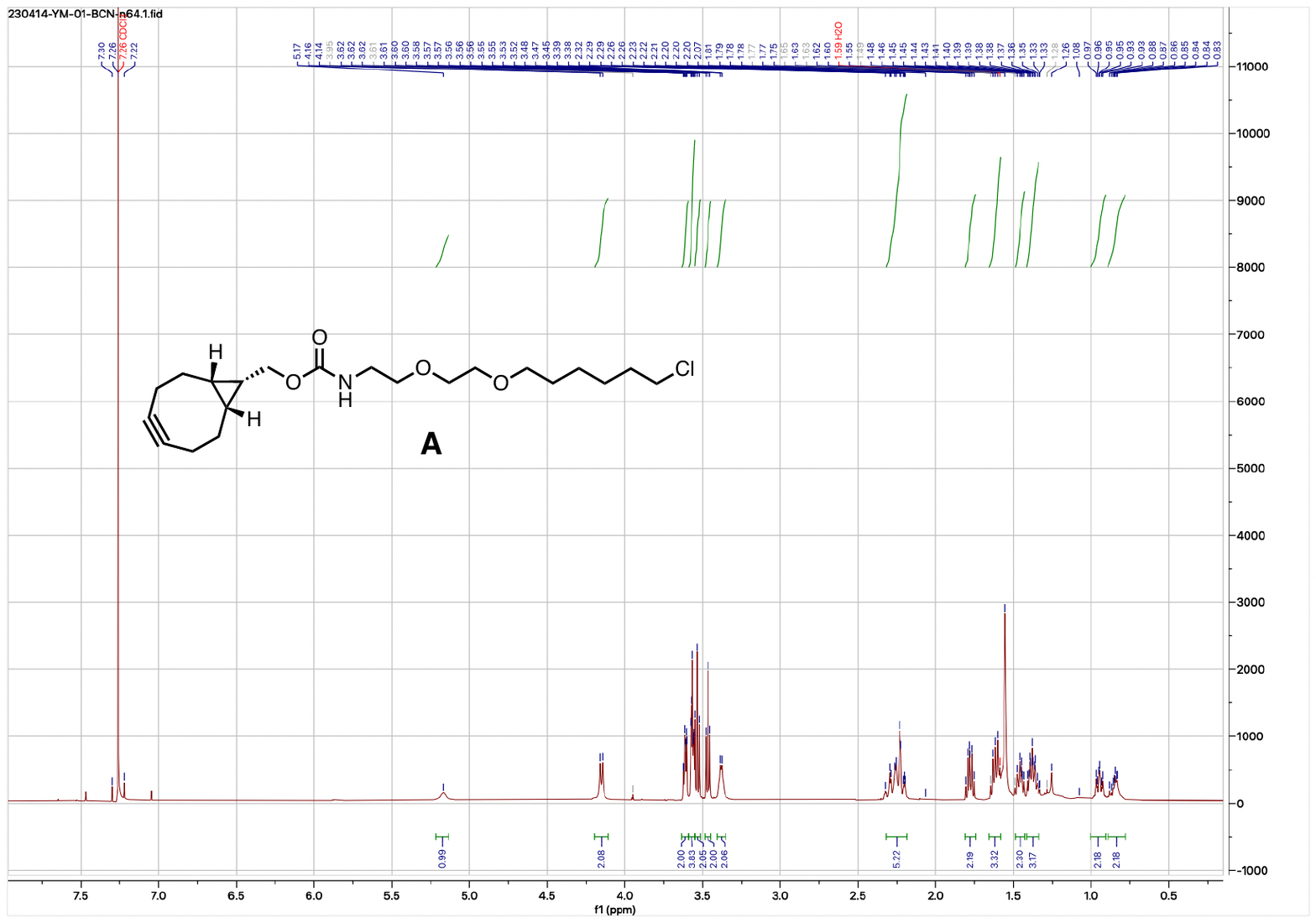}
\end{center}

\newpage
\noindent ${}^1$H NMR of compound \textbf{B}

\begin{center}
\includegraphics[width=\linewidth]{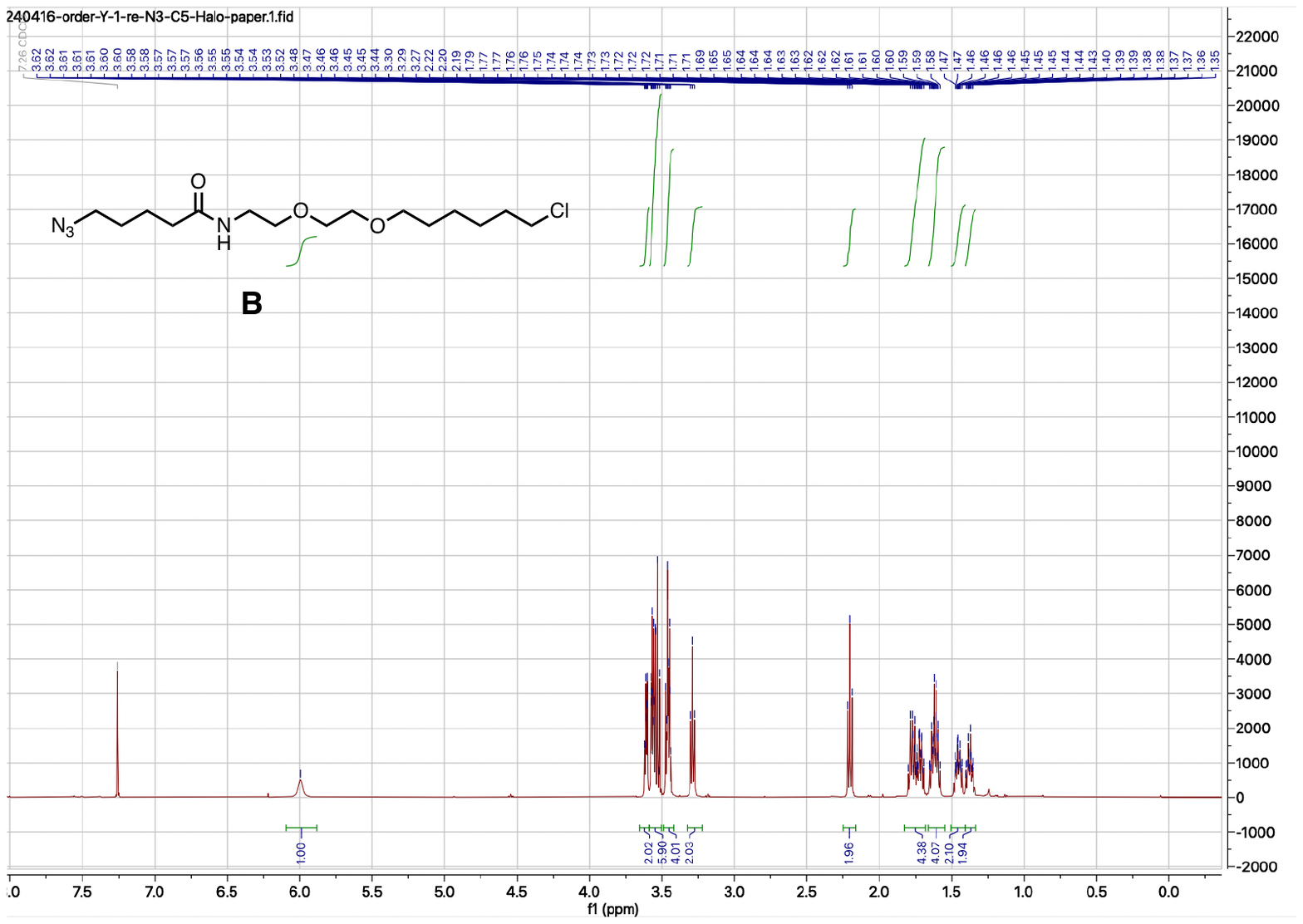}
\end{center}

\newpage
\noindent ${}^{13}$C NMR of compound \textbf{B}

\begin{center}
\includegraphics[width=\linewidth]{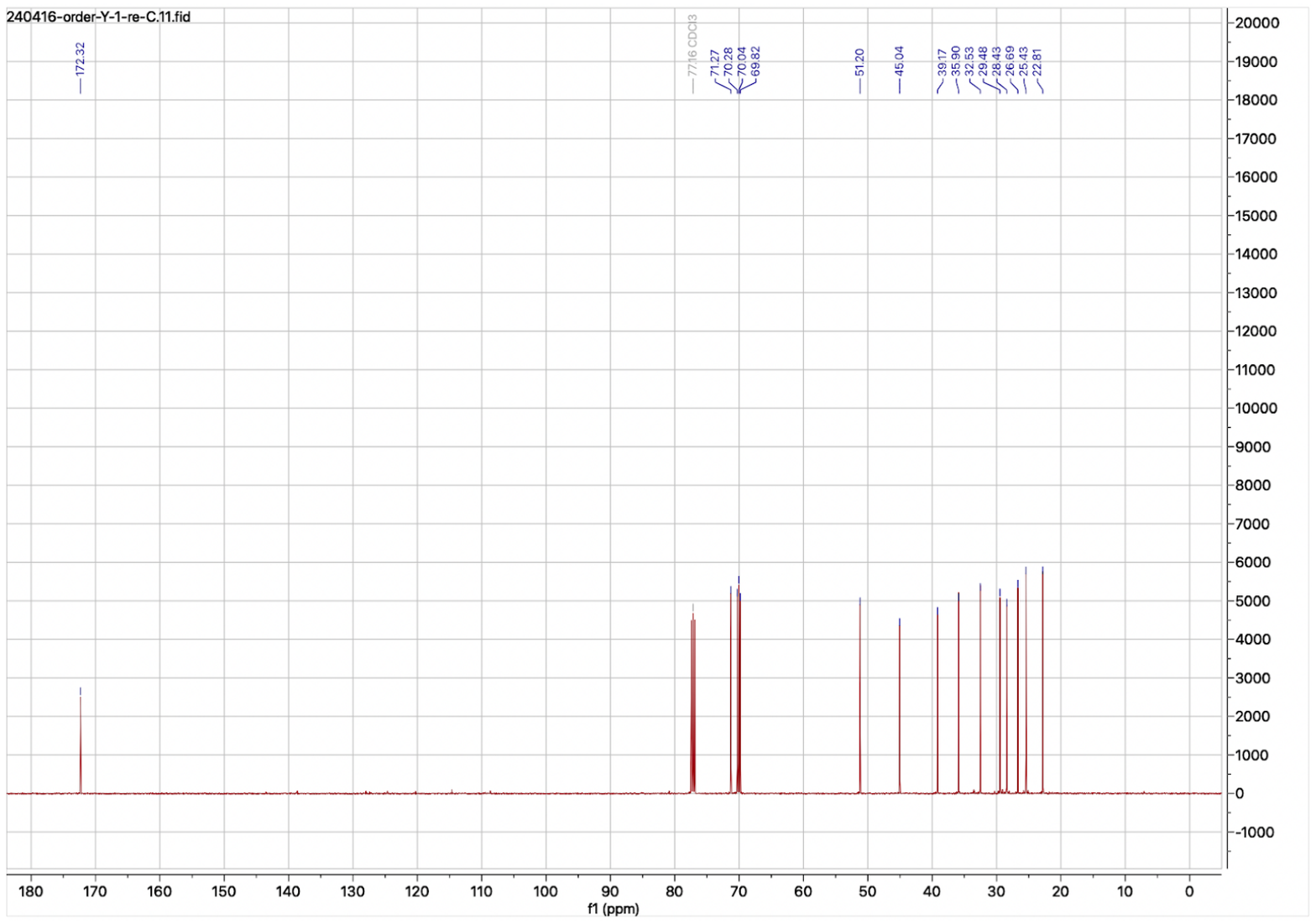}
\end{center}

\newpage
\noindent ${}^1$H NMR of compound \textbf{C}

\begin{center}
\includegraphics[width=\linewidth]{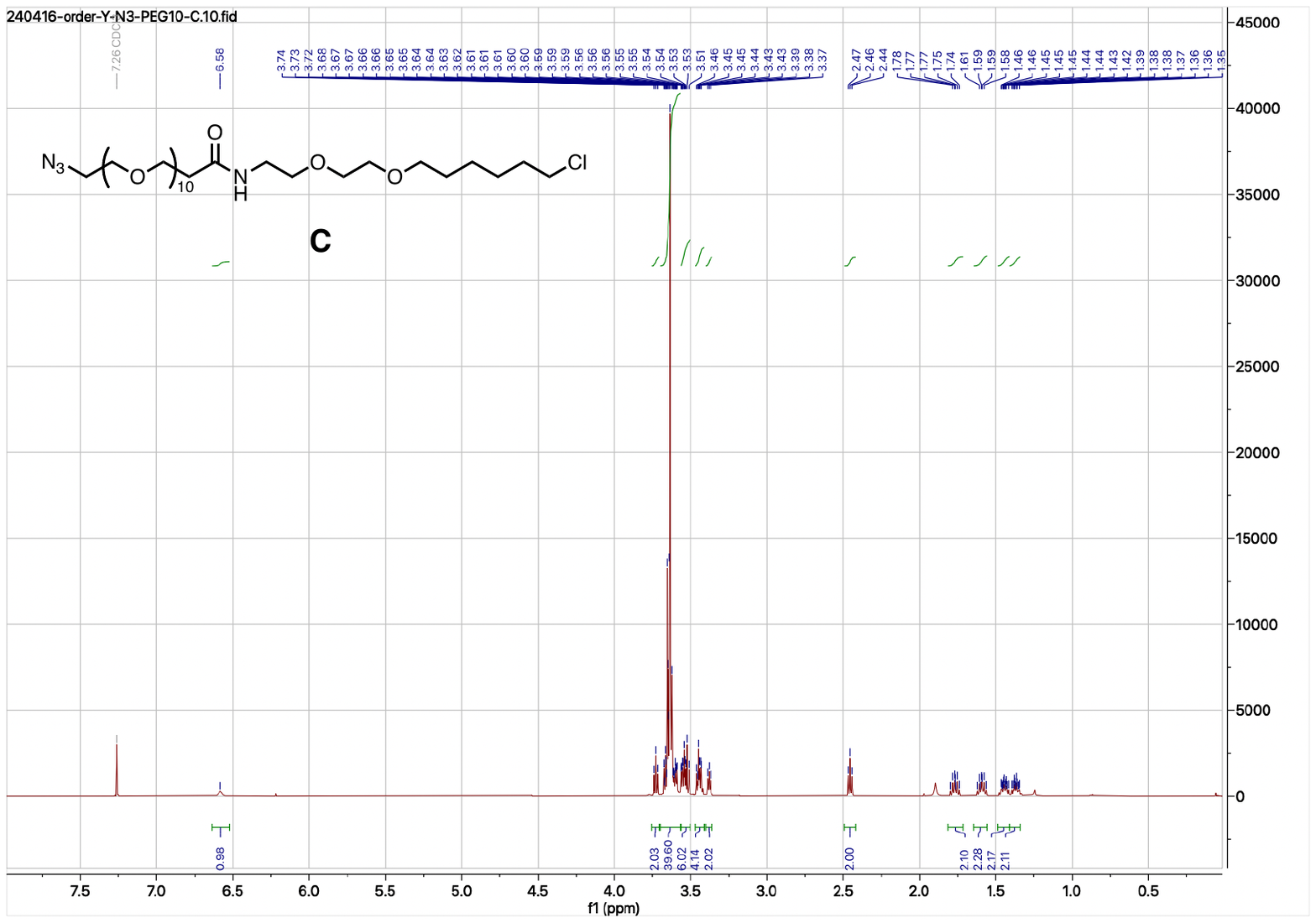}
\end{center}

\newpage
\noindent ${}^{13}$C NMR of compound \textbf{C}

\begin{center}
\includegraphics[width=\linewidth]{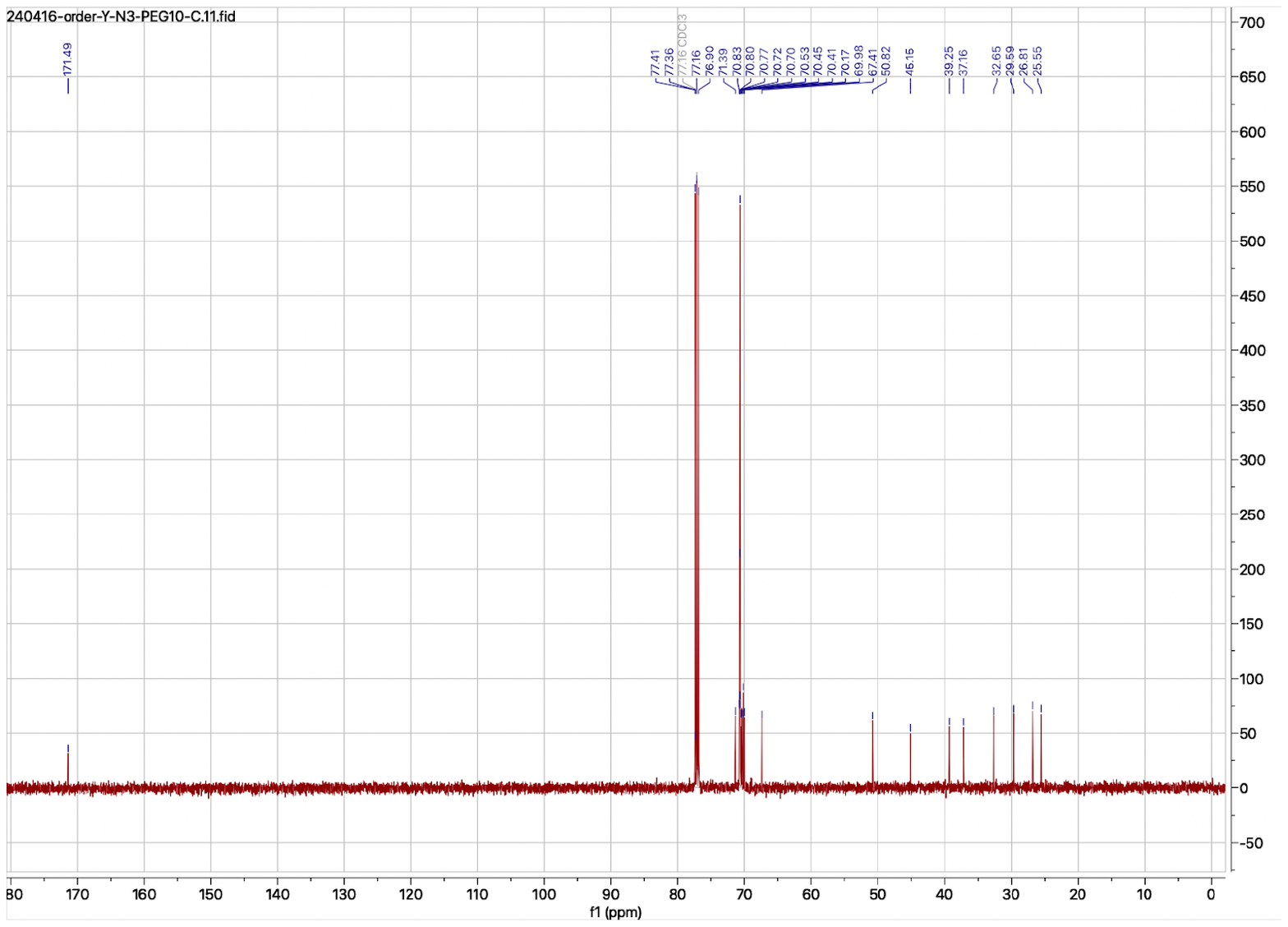}
\end{center}

\newpage
\begingroup
\definecolor{proteinHeader}{HTML}{3B3B3B}
\definecolor{proteinCyan}{HTML}{00B0F0}
\definecolor{proteinOrange}{HTML}{E97132}
\definecolor{proteinBlue}{HTML}{0070C0}
\definecolor{proteinRed}{HTML}{FF0000}
\definecolor{proteinGray}{HTML}{808080}
\definecolor{proteinPurple}{HTML}{A02B93}
\definecolor{proteinGreen}{HTML}{3A7D22}
\setlength{\tabcolsep}{2pt}
\renewcommand{\arraystretch}{0.92}
\newcommand{\proteinmetafont}{\fontsize{6.4}{7.0}\selectfont}
\newcommand{\proteinseq}[1]{\begingroup\renewcommand{\arraystretch}{1.1}\fontsize{5.8}{7.0}\selectfont\begin{tabular}[t]{@{}l@{}}#1\end{tabular}\endgroup}

\noindent{\raggedright\textbf{Supplementary Table 1. DNA and amino acid sequences of HaloTag-LgBiT. The His affinity tag is shown in cyan, HaloTag in orange, and LgBiT in blue.}\par}

\begin{adjustbox}{max width=\textwidth,center}
\begin{tabular}{|>{\raggedright\arraybackslash}p{17mm}|>{\raggedright\arraybackslash}p{14mm}|>{\raggedright\arraybackslash}p{88mm}|>{\raggedright\arraybackslash}p{36mm}|}
\hline
\cellcolor{proteinHeader}\textcolor{white}{\textbf{ID}} & \cellcolor{proteinHeader}\textcolor{white}{\textbf{Vector}} & \cellcolor{proteinHeader}\textcolor{white}{\textbf{DNA sequence}} & \cellcolor{proteinHeader}\textcolor{white}{\textbf{Amino acid sequence}} \\
\hline
\end{tabular}
\end{adjustbox}

\vspace{1.5mm}

\newpage
\noindent{\raggedright\textbf{Supplementary Table 2. DNA and amino acid sequences of HaloTag-SmBiT and HaloTag-DarkBiT~\textbf{1}. The His affinity tag is shown in cyan, HaloTag in orange, SmBiT in red, and DarkBiT~\textbf{1} in grey.}\par}

\begin{adjustbox}{max width=\textwidth,center}
\begin{tabular}{|>{\raggedright\arraybackslash}p{17mm}|>{\raggedright\arraybackslash}p{14mm}|>{\raggedright\arraybackslash}p{88mm}|>{\raggedright\arraybackslash}p{36mm}|}
\hline
\cellcolor{proteinHeader}\textcolor{white}{\textbf{ID}} & \cellcolor{proteinHeader}\textcolor{white}{\textbf{Vector}} & \cellcolor{proteinHeader}\textcolor{white}{\textbf{DNA sequence}} & \cellcolor{proteinHeader}\textcolor{white}{\textbf{Amino acid sequence}} \\
\hline
\end{tabular}
\end{adjustbox}

\vspace{1.5mm}

\begin{adjustbox}{max width=\textwidth,center}
\begin{tabular}{|>{\raggedright\arraybackslash}p{17mm}|>{\raggedright\arraybackslash}p{14mm}|>{\raggedright\arraybackslash}p{88mm}|>{\raggedright\arraybackslash}p{36mm}|}
\hline
\cellcolor{proteinHeader}\textcolor{white}{\textbf{ID}} & \cellcolor{proteinHeader}\textcolor{white}{\textbf{Vector}} & \cellcolor{proteinHeader}\textcolor{white}{\textbf{DNA sequence}} & \cellcolor{proteinHeader}\textcolor{white}{\textbf{Amino acid sequence}} \\
\hline
\end{tabular}
\end{adjustbox}

\vspace{1.5mm}

\begin{adjustbox}{max width=\textwidth,center}
\begin{tabular}{|>{\raggedright\arraybackslash}p{17mm}|>{\raggedright\arraybackslash}p{14mm}|>{\raggedright\arraybackslash}p{88mm}|>{\raggedright\arraybackslash}p{36mm}|}
\hline
\cellcolor{proteinHeader}\textcolor{white}{\textbf{ID}} & \cellcolor{proteinHeader}\textcolor{white}{\textbf{Vector}} & \cellcolor{proteinHeader}\textcolor{white}{\textbf{DNA sequence}} & \cellcolor{proteinHeader}\textcolor{white}{\textbf{Amino acid sequence}} \\
\hline
\end{tabular}
\end{adjustbox}

\vspace{1.5mm}

\begin{adjustbox}{max width=\textwidth,center}
\begin{tabular}{|>{\raggedright\arraybackslash}p{17mm}|>{\raggedright\arraybackslash}p{14mm}|>{\raggedright\arraybackslash}p{88mm}|>{\raggedright\arraybackslash}p{36mm}|}
\hline
\cellcolor{proteinHeader}\textcolor{white}{\textbf{ID}} & \cellcolor{proteinHeader}\textcolor{white}{\textbf{Vector}} & \cellcolor{proteinHeader}\textcolor{white}{\textbf{DNA sequence}} & \cellcolor{proteinHeader}\textcolor{white}{\textbf{Amino acid sequence}} \\
\hline
\end{tabular}
\end{adjustbox}

\vspace{1.5mm}

\begin{adjustbox}{max width=\textwidth,center}
\begin{tabular}{|>{\raggedright\arraybackslash}p{17mm}|>{\raggedright\arraybackslash}p{14mm}|>{\raggedright\arraybackslash}p{88mm}|>{\raggedright\arraybackslash}p{36mm}|}
\hline
\cellcolor{proteinHeader}\textcolor{white}{\textbf{ID}} & \cellcolor{proteinHeader}\textcolor{white}{\textbf{Vector}} & \cellcolor{proteinHeader}\textcolor{white}{\textbf{DNA sequence}} & \cellcolor{proteinHeader}\textcolor{white}{\textbf{Amino acid sequence}} \\
\hline
\end{tabular}
\end{adjustbox}

\vspace{1.5mm}

\begin{adjustbox}{max width=\textwidth,center}
\begin{tabular}{|>{\raggedright\arraybackslash}p{17mm}|>{\raggedright\arraybackslash}p{14mm}|>{\raggedright\arraybackslash}p{88mm}|>{\raggedright\arraybackslash}p{36mm}|}
\hline
\cellcolor{proteinHeader}\textcolor{white}{\textbf{ID}} & \cellcolor{proteinHeader}\textcolor{white}{\textbf{Vector}} & \cellcolor{proteinHeader}\textcolor{white}{\textbf{DNA sequence}} & \cellcolor{proteinHeader}\textcolor{white}{\textbf{Amino acid sequence}} \\
\hline
\end{tabular}
\end{adjustbox}

\vspace{1.5mm}

\begin{adjustbox}{max width=\textwidth,center}
\begin{tabular}{|>{\raggedright\arraybackslash}p{17mm}|>{\raggedright\arraybackslash}p{14mm}|>{\raggedright\arraybackslash}p{88mm}|>{\raggedright\arraybackslash}p{36mm}|}
\hline
\cellcolor{proteinHeader}\textcolor{white}{\textbf{ID}} & \cellcolor{proteinHeader}\textcolor{white}{\textbf{Vector}} & \cellcolor{proteinHeader}\textcolor{white}{\textbf{DNA sequence}} & \cellcolor{proteinHeader}\textcolor{white}{\textbf{Amino acid sequence}} \\
\hline
\end{tabular}
\end{adjustbox}

\vspace{1.5mm}

\newpage
\noindent{\raggedright\textbf{Supplementary Table 3. DNA and amino acid sequences of the six single-chain HaloTag-LgBiT-SmBiT fusions, one per SmBiT variant. The His affinity tag is shown in cyan, HaloTag in orange, LgBiT in blue, SmBiT in red.}\par}

\begin{adjustbox}{max width=\textwidth,center}
\begin{tabular}{|>{\raggedright\arraybackslash}p{17mm}|>{\raggedright\arraybackslash}p{14mm}|>{\raggedright\arraybackslash}p{88mm}|>{\raggedright\arraybackslash}p{36mm}|}
\hline
\cellcolor{proteinHeader}\textcolor{white}{\textbf{ID}} & \cellcolor{proteinHeader}\textcolor{white}{\textbf{Vector}} & \cellcolor{proteinHeader}\textcolor{white}{\textbf{DNA sequence}} & \cellcolor{proteinHeader}\textcolor{white}{\textbf{Amino acid sequence}} \\
\hline
\end{tabular}
\end{adjustbox}

\vspace{1.5mm}

\begin{adjustbox}{max width=\textwidth,center}
\begin{tabular}{|>{\raggedright\arraybackslash}p{17mm}|>{\raggedright\arraybackslash}p{14mm}|>{\raggedright\arraybackslash}p{88mm}|>{\raggedright\arraybackslash}p{36mm}|}
\hline
\cellcolor{proteinHeader}\textcolor{white}{\textbf{ID}} & \cellcolor{proteinHeader}\textcolor{white}{\textbf{Vector}} & \cellcolor{proteinHeader}\textcolor{white}{\textbf{DNA sequence}} & \cellcolor{proteinHeader}\textcolor{white}{\textbf{Amino acid sequence}} \\
\hline
\end{tabular}
\end{adjustbox}

\vspace{1.5mm}

\begin{adjustbox}{max width=\textwidth,center}
\begin{tabular}{|>{\raggedright\arraybackslash}p{17mm}|>{\raggedright\arraybackslash}p{14mm}|>{\raggedright\arraybackslash}p{88mm}|>{\raggedright\arraybackslash}p{36mm}|}
\hline
\cellcolor{proteinHeader}\textcolor{white}{\textbf{ID}} & \cellcolor{proteinHeader}\textcolor{white}{\textbf{Vector}} & \cellcolor{proteinHeader}\textcolor{white}{\textbf{DNA sequence}} & \cellcolor{proteinHeader}\textcolor{white}{\textbf{Amino acid sequence}} \\
\hline
\end{tabular}
\end{adjustbox}

\vspace{1.5mm}

\begin{adjustbox}{max width=\textwidth,center}
\begin{tabular}{|>{\raggedright\arraybackslash}p{17mm}|>{\raggedright\arraybackslash}p{14mm}|>{\raggedright\arraybackslash}p{88mm}|>{\raggedright\arraybackslash}p{36mm}|}
\hline
\cellcolor{proteinHeader}\textcolor{white}{\textbf{ID}} & \cellcolor{proteinHeader}\textcolor{white}{\textbf{Vector}} & \cellcolor{proteinHeader}\textcolor{white}{\textbf{DNA sequence}} & \cellcolor{proteinHeader}\textcolor{white}{\textbf{Amino acid sequence}} \\
\hline
\end{tabular}
\end{adjustbox}

\vspace{1.5mm}

\begin{adjustbox}{max width=\textwidth,center}
\begin{tabular}{|>{\raggedright\arraybackslash}p{17mm}|>{\raggedright\arraybackslash}p{14mm}|>{\raggedright\arraybackslash}p{88mm}|>{\raggedright\arraybackslash}p{36mm}|}
\hline
\cellcolor{proteinHeader}\textcolor{white}{\textbf{ID}} & \cellcolor{proteinHeader}\textcolor{white}{\textbf{Vector}} & \cellcolor{proteinHeader}\textcolor{white}{\textbf{DNA sequence}} & \cellcolor{proteinHeader}\textcolor{white}{\textbf{Amino acid sequence}} \\
\hline
\end{tabular}
\end{adjustbox}

\vspace{1.5mm}

\begin{adjustbox}{max width=\textwidth,center}
\begin{tabular}{|>{\raggedright\arraybackslash}p{17mm}|>{\raggedright\arraybackslash}p{14mm}|>{\raggedright\arraybackslash}p{88mm}|>{\raggedright\arraybackslash}p{36mm}|}
\hline
\cellcolor{proteinHeader}\textcolor{white}{\textbf{ID}} & \cellcolor{proteinHeader}\textcolor{white}{\textbf{Vector}} & \cellcolor{proteinHeader}\textcolor{white}{\textbf{DNA sequence}} & \cellcolor{proteinHeader}\textcolor{white}{\textbf{Amino acid sequence}} \\
\hline
\end{tabular}
\end{adjustbox}

\vspace{1.5mm}

\newpage
\noindent{\raggedright\textbf{Supplementary Table 4. DNA and amino acid sequences of the HaloTag-TLN1 and LgBiT-HaloTag-ITGB1A constructs. The His affinity tag is shown in cyan, HaloTag in orange, LgBiT in blue, ITGB1A in green, and TLN1 in purple.}\par}

\begin{adjustbox}{max width=\textwidth,center}
\begin{tabular}{|>{\raggedright\arraybackslash}p{17mm}|>{\raggedright\arraybackslash}p{14mm}|>{\raggedright\arraybackslash}p{88mm}|>{\raggedright\arraybackslash}p{36mm}|}
\hline
\cellcolor{proteinHeader}\textcolor{white}{\textbf{ID}} & \cellcolor{proteinHeader}\textcolor{white}{\textbf{Vector}} & \cellcolor{proteinHeader}\textcolor{white}{\textbf{DNA sequence}} & \cellcolor{proteinHeader}\textcolor{white}{\textbf{Amino acid sequence}} \\
\hline
\end{tabular}
\end{adjustbox}

\vspace{1.5mm}

\begin{adjustbox}{max width=\textwidth,center}
\begin{tabular}{|>{\raggedright\arraybackslash}p{17mm}|>{\raggedright\arraybackslash}p{14mm}|>{\raggedright\arraybackslash}p{88mm}|>{\raggedright\arraybackslash}p{36mm}|}
\hline
\cellcolor{proteinHeader}\textcolor{white}{\textbf{ID}} & \cellcolor{proteinHeader}\textcolor{white}{\textbf{Vector}} & \cellcolor{proteinHeader}\textcolor{white}{\textbf{DNA sequence}} & \cellcolor{proteinHeader}\textcolor{white}{\textbf{Amino acid sequence}} \\
\hline
\end{tabular}
\end{adjustbox}

\vspace{1.5mm}

\endgroup

\newpage
\noindent \textbf{Supplementary Table 5. Peptide sequences and modifications. Pep1--Pep3 are the screening peptides in Fig. \ref{fig:6}b. DarkBiT~\textbf{2} is the soluble competitor titrated in Fig. \ref{fig:5}.}

\begin{adjustbox}{max width=\textwidth,center}
\begin{tabular}{|l|p{43mm}|p{50mm}|p{25mm}|} \hline
   \textbf{ID}& \textbf{Amino acid sequence} & \textbf{Modification} & \textbf{Article info.}\\ \hline
   ITGB3 & KFEEERARAKWDT & $N$-terminal myristoylation & Ref.\cite{Shen2013}\\ \hline
   Pep1 (non-binding control) & KAEKEKMNAKADT & $N$-terminal myristoylation & this work\\ \hline
   Pep2 (primary screening hit) & KFEEERMRCKWMT & $N$-terminal myristoylation & this work\\ \hline
   Pep3 (= compound~\textbf{1}) & KFEEERSRCKWMT & $N$-terminal myristoylation & this work\\ \hline
   DarkBiT~\textbf{2} (the free DarkBiT) & VTGYALFEKES &  & Ref.\cite{DarkBiT}\\ \hline

 \end{tabular}
\end{adjustbox}

\noindent \textbf{Supplementary Table 6. DNA sequences of the chemically modified single-stranded DNAs (ssDNAs) used to synthesize the AccuLinkers. The 5′ and 3′ modifications are listed. Cy3 denotes the cyanine 3 dye. Tethering helix denotes the duplex to which the AccuLinker hybridizes. The helix numbering follows Supplementary Fig. \ref{fig:Extended Data Fig. Detail Bricks design2}.}

\begin{adjustbox}{max width=\textwidth,center}
\begin{tabular}{|l|l|p{30mm}|p{30mm}|p{30mm}|} \hline
   \textbf{No.}& \textbf{Oligo\_sequence (5′ to 3′)} & \textbf{5′ modification} & \textbf{3′ modification} & \textbf{Tethering helix }\\ \hline
   1 & GGTGATGATGGTTTGG & Azide & Cy3 & 29\\
   \hline
   2 & CTTCTCCATTTTCCAG & Azide & Cy3 & 45\\ \hline
   3 & GTGATTTGAAAGTTGC & Cy3 & Azide & 67\\ \hline
   4 & CACGTAACAAGATCCC & Azide & Cy3 & 91\\ \hline
   5 & TTTTCTACTCTTCACG & Azide & Cy3 & 175\\ \hline

 \end{tabular}
\end{adjustbox}

\newpage
\noindent \textbf{Supplementary Table 7. Chemical information of synthetic AccuLinkers. 5′ and 3′ indicate 5′ modification and 3′ modification of AccuLinker, respectively. HL, HaloTag ligand; LL, linker length; TH, tethering helix. LL is the maximum distance between the DB01 wall and the tethered HaloTag protein, shown schematically in Supplementary Fig. \ref{fig:Extended Data Fig. AccuLinker Pro}. In the identifier AccuLinker\_n\_X, n is the number of OEG units and X the tethering helix. The helix numbering follows Supplementary Fig. \ref{fig:Extended Data Fig. Detail Bricks design2}.}

\begin{adjustbox}{max width=\textwidth,center}
\begin{tabular}{|p{35mm}|p{45mm}|p{10mm}|p{10mm}|p{12mm}|p{16mm}|p{6mm}|} \hline
   \textbf{ID}& \textbf{Oligo\_sequence (5′ to 3′)} & \textbf{5′} & \textbf{3′} & \textbf{OEGn} & \textbf{LL [nm]} & \textbf{TH}\\ \hline
   \textbf{AccuLinker\_0\_29}   & GGTGATGATGGTTTGG & HL & Cy3            & 0  & 1.0  & 29  \\\hline
\textbf{AccuLinker\_4\_29}   & GGTGATGATGGTTTGG & HL & Cy3            & 4  & 4.1  & 29  \\\hline
\textbf{AccuLinker\_14\_29}  & GGTGATGATGGTTTGG & HL & Cy3            & 14 & 7.6  & 29  \\\hline
\textbf{AccuLinker\_17\_29}  & GGTGATGATGGTTTGG & HL & Cy3            & 17 & 9.0  & 29  \\\hline
\textbf{AccuLinker\_22\_29}  & GGTGATGATGGTTTGG & HL & Cy3            & 22 & 10.7 & 29  \\\hline
\textbf{AccuLinker\_27\_29}  & GGTGATGATGGTTTGG & HL & Cy3            & 27 & 12.5 & 29  \\\hline
\textbf{AccuLinker\_32\_29}  & GGTGATGATGGTTTGG & HL & Cy3            & 32 & 14.2 & 29  \\\hline
\textbf{AccuLinker\_0\_45}   & CTTCTCCATTTTCCAG & HL & Cy3            & 0  & 1.0  & 45  \\\hline
\textbf{AccuLinker\_4\_45}   & CTTCTCCATTTTCCAG & HL & Cy3            & 4  & 4.1  & 45  \\\hline
\textbf{AccuLinker\_14\_45}  & CTTCTCCATTTTCCAG & HL & Cy3            & 14 & 7.6  & 45  \\\hline
\textbf{AccuLinker\_17\_45}  & CTTCTCCATTTTCCAG & HL & Cy3            & 17 & 9.0  & 45  \\\hline
\textbf{AccuLinker\_22\_45}  & CTTCTCCATTTTCCAG & HL & Cy3            & 22 & 10.7 & 45  \\\hline
\textbf{AccuLinker\_27\_45}  & CTTCTCCATTTTCCAG & HL & Cy3            & 27 & 12.5 & 45  \\\hline
\textbf{AccuLinker\_32\_45}  & CTTCTCCATTTTCCAG & HL & Cy3            & 32 & 14.2 & 45  \\\hline
\end{tabular}
\end{adjustbox}

\begin{adjustbox}{max width=\textwidth,center}
\begin{tabular}{|p{35mm}|p{45mm}|p{10mm}|p{10mm}|p{12mm}|p{16mm}|p{6mm}|} \hline
   \textbf{ID}& \textbf{Oligo\_sequence (5′ to 3′)} & \textbf{5′} & \textbf{3′} & \textbf{OEGn} & \textbf{LL [nm]} & \textbf{TH}\\ \hline
\textbf{AccuLinker\_0\_67}   & GTGATTTGAAAGTTGC & Cy3 & HL & 0  & 1.0  & 67  \\\hline
\textbf{AccuLinker\_4\_67}   & GTGATTTGAAAGTTGC & Cy3            & HL & 4  & 4.1  & 67  \\\hline
\textbf{AccuLinker\_14\_67}  & GTGATTTGAAAGTTGC & Cy3            & HL & 14 & 7.6  & 67  \\\hline
\textbf{AccuLinker\_17\_67}  & GTGATTTGAAAGTTGC & Cy3            & HL & 17 & 9.0  & 67  \\\hline
\textbf{AccuLinker\_22\_67}  & GTGATTTGAAAGTTGC & Cy3            & HL & 22 & 10.7 & 67  \\\hline
\textbf{AccuLinker\_27\_67}  & GTGATTTGAAAGTTGC & Cy3            & HL & 27 & 12.5 & 67  \\\hline
\textbf{AccuLinker\_32\_67}  & GTGATTTGAAAGTTGC & Cy3            & HL & 32 & 14.2 & 67  \\\hline
\textbf{AccuLinker\_0\_91}   & CACGTAACAAGATCCC & HL & Cy3            & 0  & 1.0  & 91  \\\hline
\textbf{AccuLinker\_4\_91}   & CACGTAACAAGATCCC & HL & Cy3            & 4  & 4.1  & 91  \\\hline
\textbf{AccuLinker\_14\_91}  & CACGTAACAAGATCCC & HL & Cy3            & 14 & 7.6  & 91  \\\hline
\textbf{AccuLinker\_17\_91}  & CACGTAACAAGATCCC & HL & Cy3            & 17 & 9.0  & 91  \\\hline
\textbf{AccuLinker\_22\_91}  & CACGTAACAAGATCCC & HL & Cy3            & 22 & 10.7 & 91  \\\hline
\textbf{AccuLinker\_27\_91}  & CACGTAACAAGATCCC & HL & Cy3            & 27 & 12.5 & 91  \\\hline
\textbf{AccuLinker\_32\_91}  & CACGTAACAAGATCCC & HL & Cy3            & 32 & 14.2 & 91  \\\hline
\textbf{AccuLinker\_0\_175}  & TTTTCTACTCTTCACG & HL & Cy3            & 0  & 1.0  & 175 \\\hline
\textbf{AccuLinker\_4\_175}  & TTTTCTACTCTTCACG & HL & Cy3            & 4  & 4.1  & 175 \\\hline
\textbf{AccuLinker\_14\_175} & TTTTCTACTCTTCACG & HL & Cy3            & 14 & 7.6  & 175 \\\hline
\textbf{AccuLinker\_17\_175} & TTTTCTACTCTTCACG & HL & Cy3            & 17 & 9.0  & 175 \\\hline
\textbf{AccuLinker\_22\_175} & TTTTCTACTCTTCACG & HL & Cy3            & 22 & 10.7 & 175 \\\hline
\textbf{AccuLinker\_27\_175} & TTTTCTACTCTTCACG & HL & Cy3            & 27 & 12.5 & 175 \\\hline
\textbf{AccuLinker\_32\_175} & TTTTCTACTCTTCACG & HL & Cy3            & 32 & 14.2 & 175\\\hline
\end{tabular}
\end{adjustbox}

\newpage
\noindent \textbf{Supplementary Table 8. Sequences of ssDNA overhangs for protein tethering in DB01. The helix numbering follows Supplementary Fig. \ref{fig:Extended Data Fig. Detail Bricks design2}. The 16-nt sequence shown in red is a protein tethering sequence, complementary to the ssDNA end of the AccuLinker, whereas the sequence shown in black is a DB01-folding sequence. Blue text in between is a spacer within the DB01 strand, shown in Supplementary Fig. \ref{fig:Extended Data Fig. Detail Bricks design2}b,c. The supplier part codes denote a C3 (propanediol) spacer (iSpC3), a triethylene glycol spacer (iSp9) and a hexaethylene glycol spacer (iSp18). The side, back, and front positions are illustrated schematically in Supplementary Fig. \ref{fig:Extended Data Fig. Detail Bricks design2}b.}

\begin{adjustbox}{max width=\textwidth,center}
\begin{tabular}{|l|p{120mm}|p{25mm}|} \hline
   \textbf{helix} & \textbf{Oligo\_sequence (5′ to 3′)} & \textbf{DNA Position}\\ \hline
   \textbf{29} & \textcolor{red}{CCAAACCATCATCACC}\textcolor{blue}{(iSp9)}\textcolor{black}{GCTAATCGATTATAGTTCTAACC TGATCGTATACATGGCTACGCCTTAACGC} & side\\ \hline
   \textbf{45} & \textcolor{red}{CTGGAAAATGGAGAAG}\textcolor{blue}{(iSp18)}\textcolor{black}{GAACCGGTCCTTGATACGATTA ATGTATTCTAAGGAAGCCAAGACCGGTCTA} & back\\ \hline
   \textbf{67} & \textcolor{black}{ATATTGAGACGTTAGCAGACGGTATG}\textcolor{blue}{(iSpC3)}\textcolor{red}{GCAACTTTCAAA  TCAC} & front\\ \hline
   \textbf{91} & \textcolor{red}{GGGATCTTGTTACGTG}\textcolor{blue}{(iSp18)}\textcolor{black}{CAGTTGTGCCTATAGTAGGCGC  GGTT} & back\\ \hline
   \textbf{175} & \textcolor{red}{CGTGAAGAGTAGAAAA}\textcolor{blue}{(iSp9)}\textcolor{black}{TGCACCTCCGCTCACACCAAGT GCTTCCAGCTACCACACTGTATGTCTGGCA} & side\\ \hline
 \end{tabular}
\end{adjustbox}

\newpage
\noindent \textbf{Supplementary Table 9. Sequences of short ssDNAs for folding DB01. The sequence IDs of the five ssDNAs that were chemically modified for tethering proteins are highlighted in red. Detailed information on the sequences and spacers is available in Supplementary Table 8.}

\begingroup
\fontsize{6.5}{7.0}\selectfont
\setlength{\tabcolsep}{2pt}
\renewcommand{\arraystretch}{1.2}
\setlength{\LTpre}{3pt}
\setlength{\LTpost}{0pt}
\begin{longtable}{|>{\centering\arraybackslash}m{8mm}|>{\raggedright\arraybackslash}m{\dimexpr\textwidth-8mm-4\tabcolsep-3\arrayrulewidth\relax}|}
\hline
\textbf{ID} & \textbf{Oligo\_sequence (5′ to 3′)} \\ \hline
\endfirsthead
\hline
\endhead
   \textbf{1} & TTTCAACGCGGTTATC \\ \hline
\textbf{2} & TTTTCAGTCGGCCGTCAACATACAGCATATTT \\ \hline
\textbf{3} & TTTTTGGCTAACAAGCCACTCGTTGAATATTT \\ \hline
\textbf{4} & TTTCCTTACGAACGATATGTTCATGCGGTTTT \\ \hline
\textbf{5} & TTTTAGACCGAGTATCCAAGGATAATGTCTTT \\ \hline
\textbf{6} & TTTTCCGAGAGAGCCATACCAGCAGCTAATTT \\ \hline
\textbf{7} & TTTCTATGCTGTTCAGTATATCTGACGGTTTT \\ \hline
\textbf{8} & TTTAGCTAGGAGTCCTGCTGGCCGTATCCTTT \\
\hline
\end{longtable}
\endgroup
\newpage

\renewcommand{\figurename}{Supplementary Fig.}

\noindent Unless stated otherwise, the following symbols are used throughout the Supplementary Figures: grey sphere, HaloTag; black wavy line, oligo(ethylene glycol) (OEG) chain; dark red helix, single-stranded DNA; yellow star, Cy3.

\begin{figure}[htbp]
\centering
\makebox[\linewidth][c]{\includegraphics[
  width=1.35\linewidth,
  height=0.80\textheight,
  keepaspectratio
]{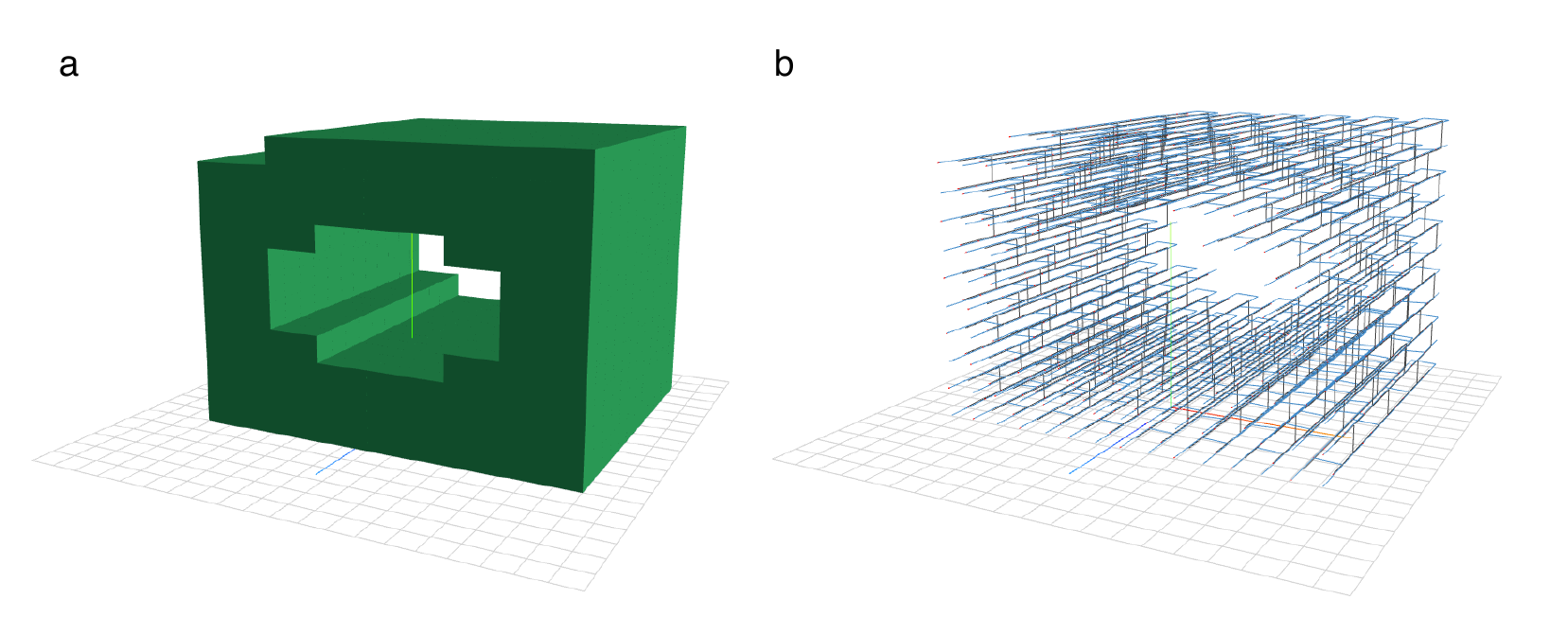}}
\caption{NanoBricks design of DB01. a. Voxelized image of DB01. b. Image of the voxelized DB01 converted into a set of DNA strands. The images were created with NanoBricks\cite{DNA_Bricks_Nature}. }
\label{fig:Extended Data Fig. Bricks design}
\end{figure}

\begin{figure}[htbp]
\centering
\makebox[\linewidth][c]{\includegraphics[
  width=1.35\linewidth,
  height=0.80\textheight,
  keepaspectratio
]{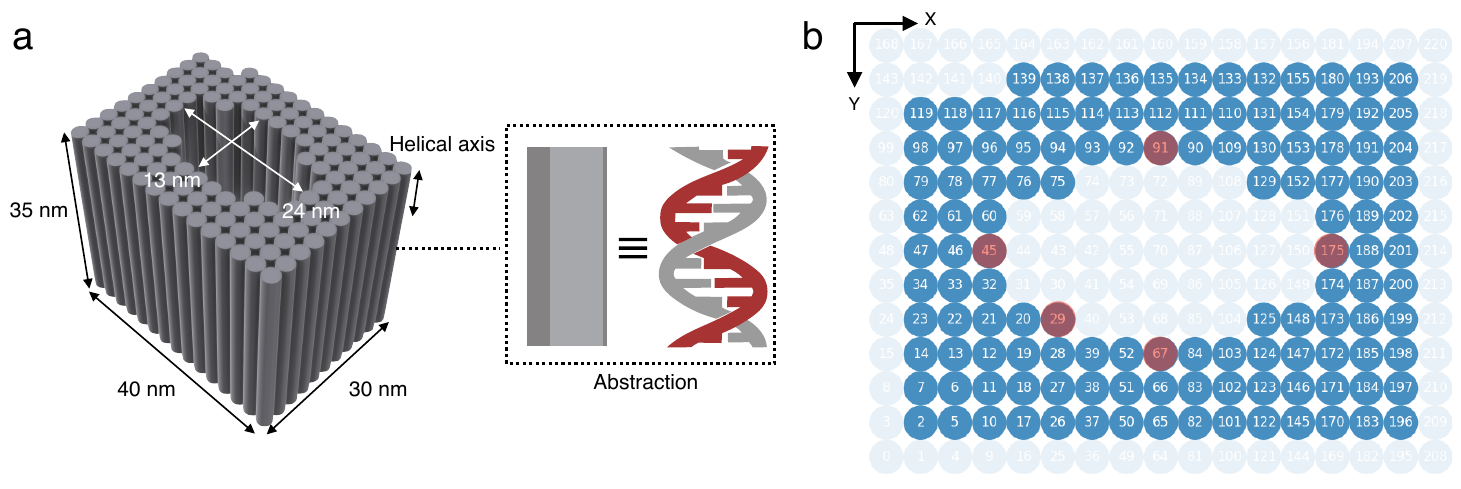}}
\caption{Overview of structural features of the DNA-brick structure DB01. a. DB01 is built by folding 705 short single-stranded DNA oligonucleotides into a 125-helix, 40$\times$30$\times$35 nm${}^3$ cuboid (W $\times$ D $\times$ H, corresponding to X, Y and Z) containing an elliptical through-hole cavity that spans 24 nm in X and 13 nm in Y over the full height. DNA duplexes are abstracted to cylinders, indicated in grey. b. The X-Y cross-section looking down the Z+ direction. The five helices highlighted in red carry single-stranded DNA overhangs for protein tethering, 17.6 nm from the top face along Z. }
\label{fig:Extended Data Fig. Detail Bricks design}
\end{figure}

\begin{figure}[htbp]
\centering
\makebox[\linewidth][c]{\includegraphics[
  width=1.35\linewidth,
  height=0.80\textheight,
  keepaspectratio
]{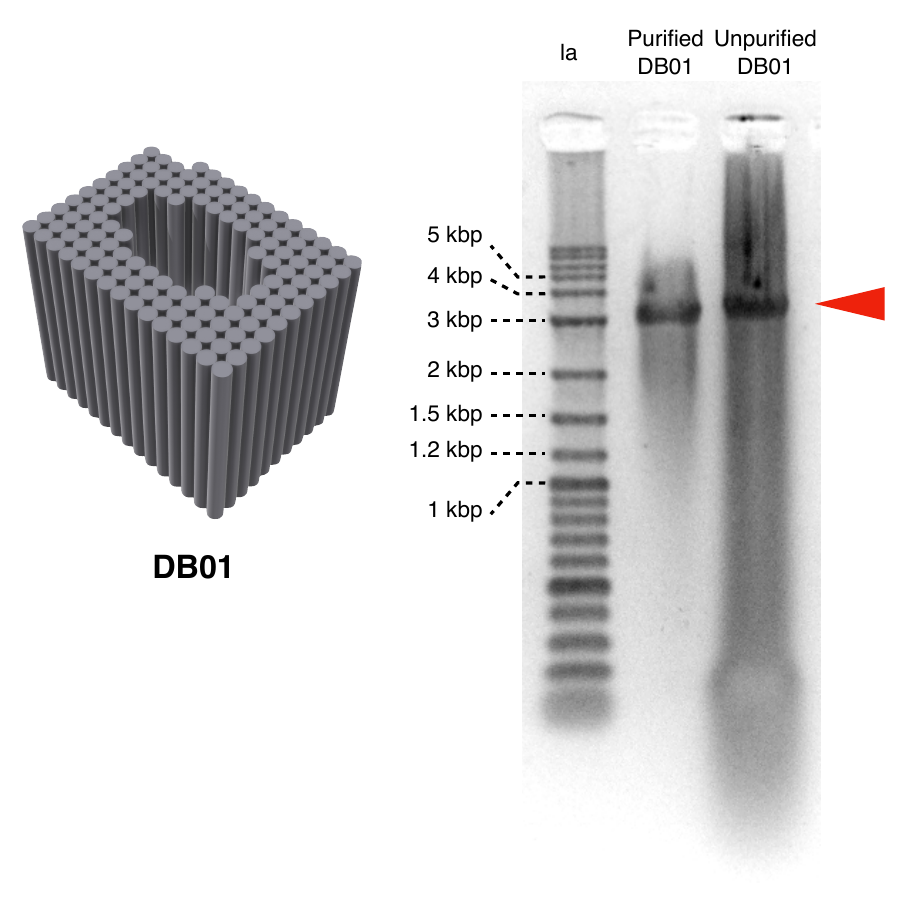}}
\caption{Design and gel analysis of DB01. Left, cylinder model of the DB01 structure, in which each cylinder is a DNA duplex. Right, 1.5\% agarose gel of purified and unpurified DB01, stained for DNA with SYBR Gold; la, DNA ladder with the indicated sizes. The red arrow marks the DB01 band.}
\label{fig:Extended Data Fig. DB01}
\end{figure}

\begin{figure}[htbp]
\centering
\makebox[\linewidth][c]{\includegraphics[
  width=1.35\linewidth,
  height=0.80\textheight,
  keepaspectratio
]{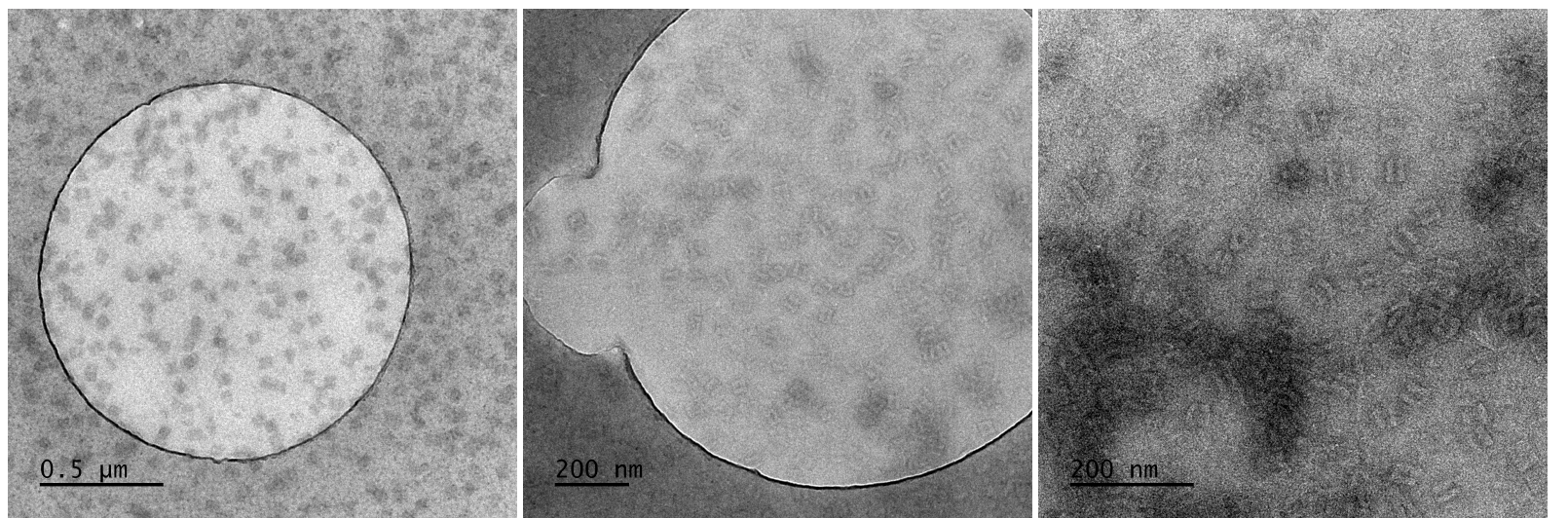}}
\caption{Negative-stain TEM analysis of DB01, purified from the band indicated by the red arrow in Supplementary Fig.~\ref{fig:Extended Data Fig. DB01}. Left, overview field; centre and right, higher-magnification fields. Scale bars, 0.5~\textmu m (left) and 200~nm (centre, right).}
\label{fig:Extended Data Fig. TEM.}
\end{figure}

\begin{figure}[htbp]
\centering
\makebox[\linewidth][c]{\includegraphics[
  width=1.35\linewidth,
  height=0.68\textheight,
  keepaspectratio
]{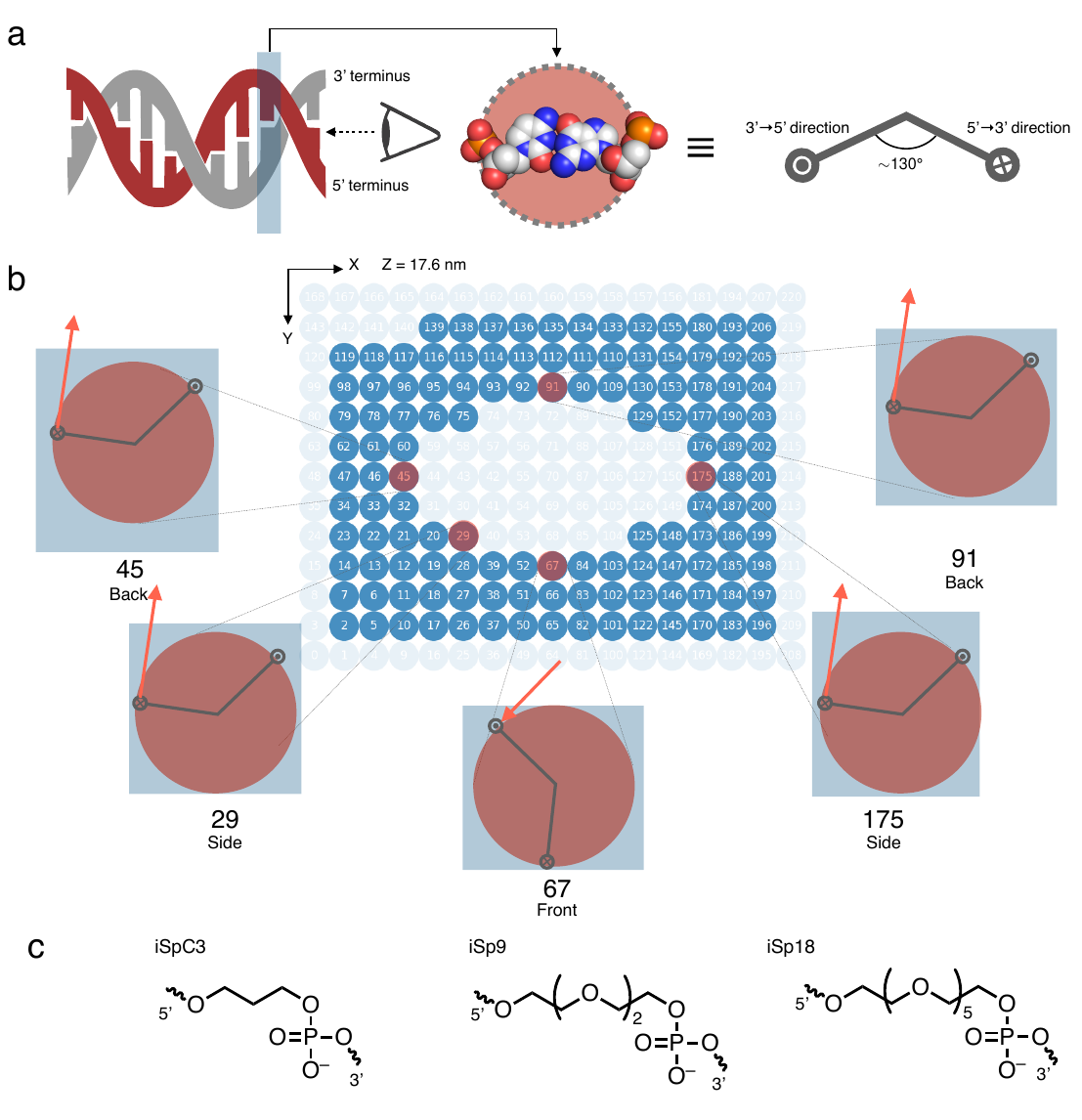}}
\caption{Helix information and chemical modification of the protein tethering helices. a. Schematic illustration of the phase of the DNA bases within the double helix at a particular Z-slice. $\odot$ indicates the 5′ to 3′ direction from the back to front. $\otimes$ indicates the 5′ to 3′ direction from the front to back. b. The phase of the DNA bases within the protein tethering helices, highlighted by red circles, at Z = 17.6 nm. Red arrows indicate the single-stranded DNAs for the protein tethering. In helix 67, a single-stranded DNA extends from the front of the helix towards the inner cavity. In helices 45 and 91, a single-stranded DNA extends from the back of the helix towards the inner cavity. In helices 29 and 175, a single-stranded DNA extends from the side of helix towards the inner cavity. To extend the single-stranded DNA for protein tethering (complementary to the ssDNA end of the AccuLinker) from the front surface of the helix towards the inner cavity, a chemical spacer of appropriate length was incorporated. Specifically, an iSpC3 spacer (three carbons, ca. 0.4--0.5 nm) was used for helix 67, an iSp18 spacer (hexaethylene glycol, ca. 2 nm) for helices 45 and 91, and an iSp9 spacer (triethylene glycol, ca. 1 nm) for helices 29 and 175. c. Chemical structures of the iSpC3, iSp18 and iSp9 spacers. }
\label{fig:Extended Data Fig. Detail Bricks design2}
\end{figure}

\begin{figure}[htbp]
\centering
\makebox[\linewidth][c]{\includegraphics[
  width=1.35\linewidth,
  height=0.80\textheight,
  keepaspectratio
]{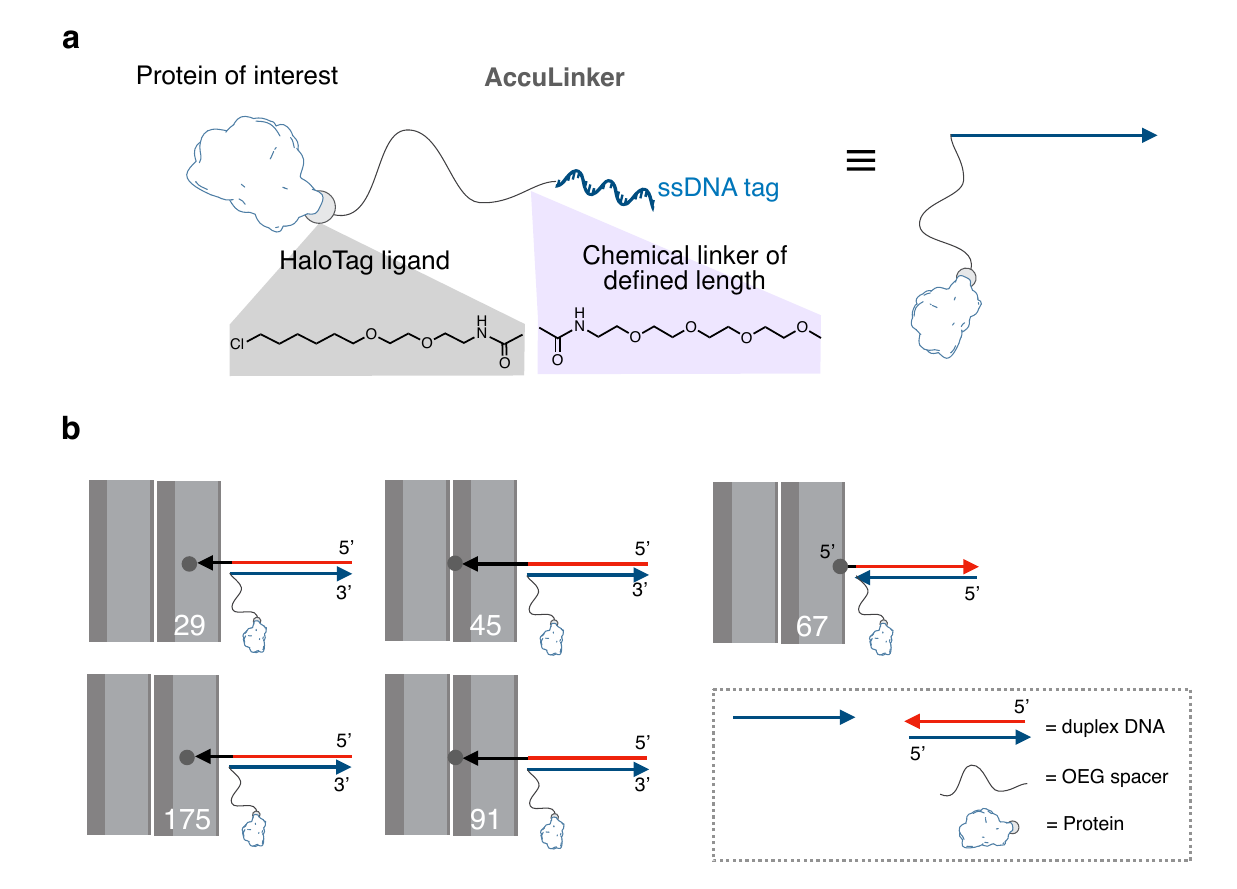}}
\caption{AccuLinker and protein confinement in DB01. a. Schematic illustration of AccuLinker. Each AccuLinker is a chemical linker of defined length, carrying a specific tag at one end that binds the protein of interest and a single-stranded DNA (ssDNA) at the other. b. Schematic image of protein confinement. The red and blue arrows, shown in the 5′-to-3′ direction, indicate the 16-nt ssDNA. The ssDNA of the AccuLinker (blue arrow) binds to pre-installed ssDNA (red arrow) overhangs on the inner surface of DB01. DNA duplexes of DB01 are abstracted to cylinders, indicated in grey. The numbers correspond to the helix numbers in Supplementary Fig.~\ref{fig:Extended Data Fig. Detail Bricks design2}.}
\label{fig:Extended Data Fig. AccuLinker and protein tethering}
\end{figure}

\begin{figure}[htbp]
\centering
\makebox[\linewidth][c]{\includegraphics[
  width=1.35\linewidth,
  height=0.80\textheight,
  keepaspectratio
]{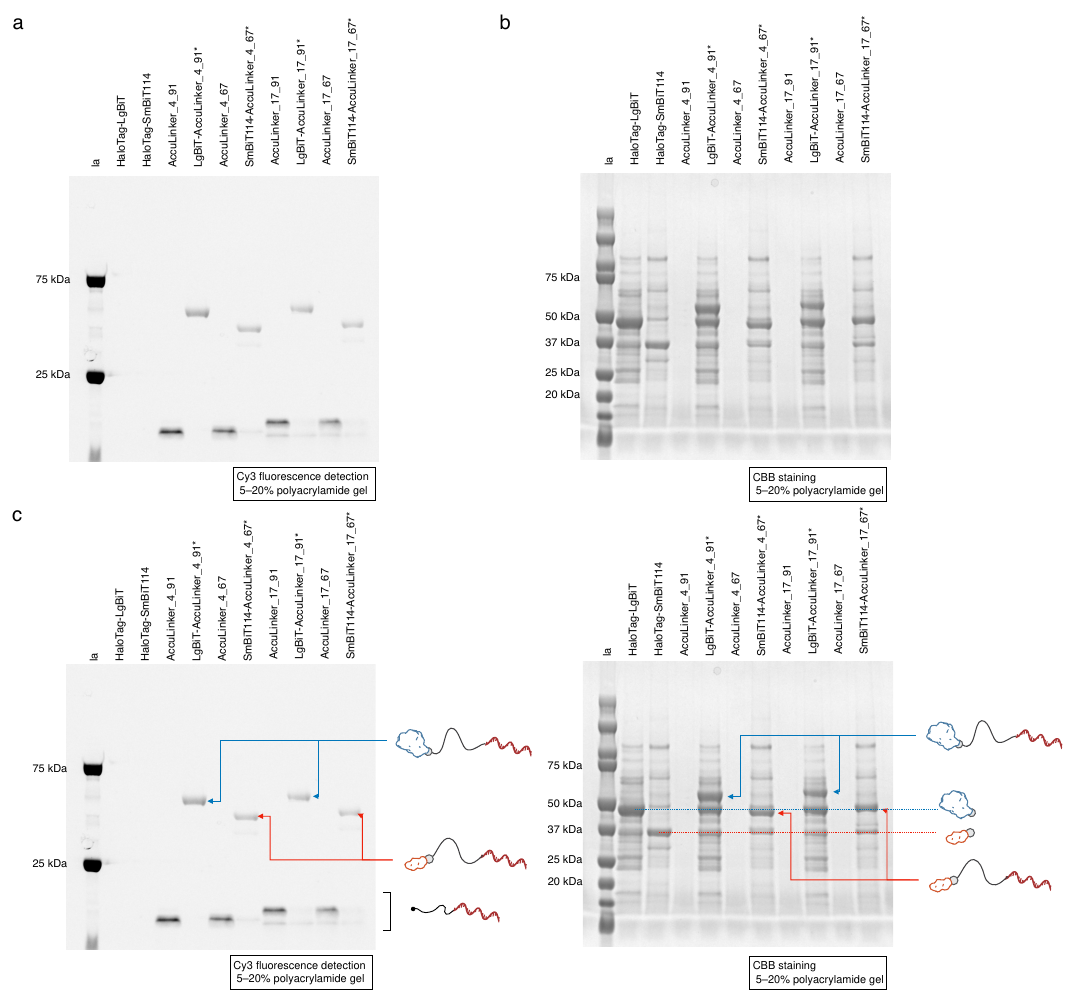}}
\caption{HaloTag-mediated conjugation of AccuLinker to HaloTag-LgBiT and HaloTag-SmBiT114. a. Cy3 fluorescence image of the PAGE gel. b. Coomassie Brilliant Blue (CBB) staining image of the PAGE gel. c. Molecular annotation of the bands in a and b. la, molecular-weight marker. The gel electrophoresis was conducted on a 5--20\% gradient polyacrylamide gel. The gel shown is representative of independent preparations. ${}^*$The samples are unpurified reaction mixtures. AccuLinker\_n\_X: n indicates the number of OEG units and X indicates the helix number for the protein tethering. The numbering corresponds to the helix numbers in Supplementary Fig.~\ref{fig:Extended Data Fig. Detail Bricks design2}.}
\label{fig:Extended Data Fig. Protein conju1}
\end{figure}

\begin{figure}[htbp]
\centering
\makebox[\linewidth][c]{\includegraphics[
  width=1.35\linewidth,
  height=0.80\textheight,
  keepaspectratio
]{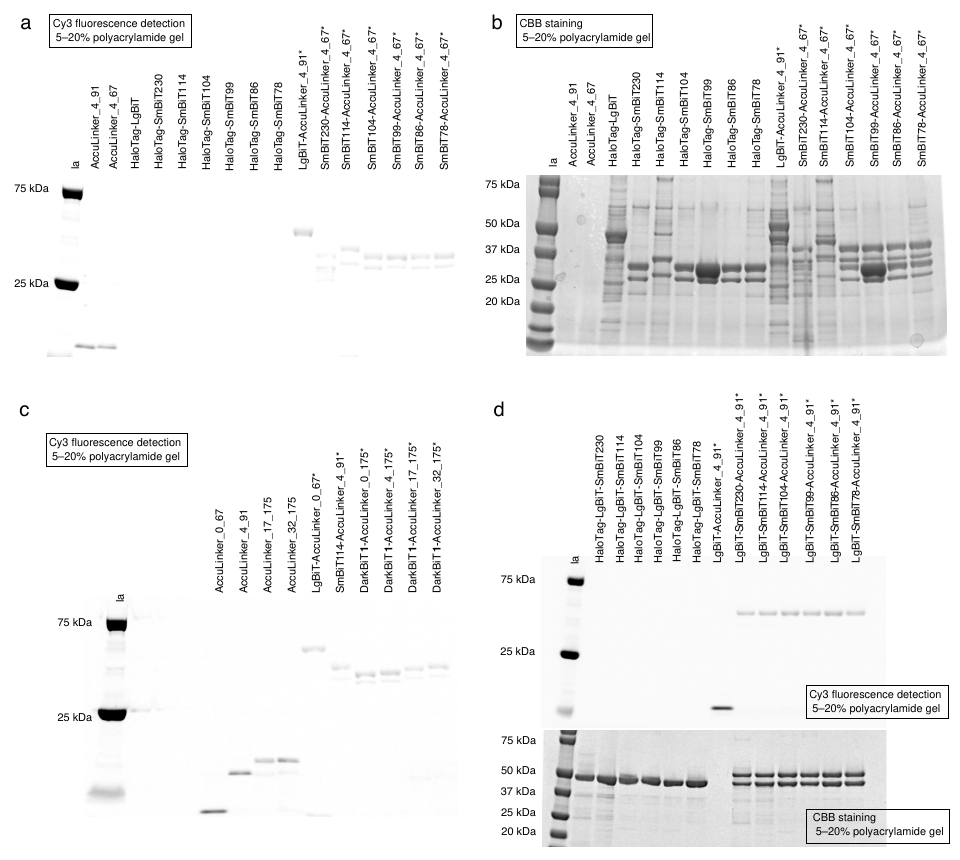}}
\caption{HaloTag-mediated conjugation of AccuLinker to a series of HaloTag-fused LgBiT, SmBiT, DarkBiT~\textbf{1} and single-chain LgBiT-SmBiT constructs. a. Cy3 fluorescence image of the LgBiT and SmBiT gel. b. CBB staining of the same gel. c. Cy3 fluorescence image of the DarkBiT~\textbf{1} gel. d. Cy3 fluorescence (upper) and CBB staining (lower) of the gel for the six single-chain HaloTag-LgBiT-SmBiT fusions. la, molecular-weight marker. The gel electrophoresis was conducted on a 5--20\% gradient polyacrylamide gel. The gel shown is representative of independent preparations. ${}^*$The samples are unpurified reaction mixtures. AccuLinker\_n\_X: n indicates the number of OEG units and X indicates the helix number for the protein tethering. The numbering corresponds to the helix numbers in Supplementary Fig.~\ref{fig:Extended Data Fig. Detail Bricks design2}.}
\label{fig:Extended Data Fig. Protein conju2}
\end{figure}

\begin{figure}[htbp]
\centering
\makebox[\linewidth][c]{\includegraphics[
  width=1.35\linewidth,
  height=0.80\textheight,
  keepaspectratio
]{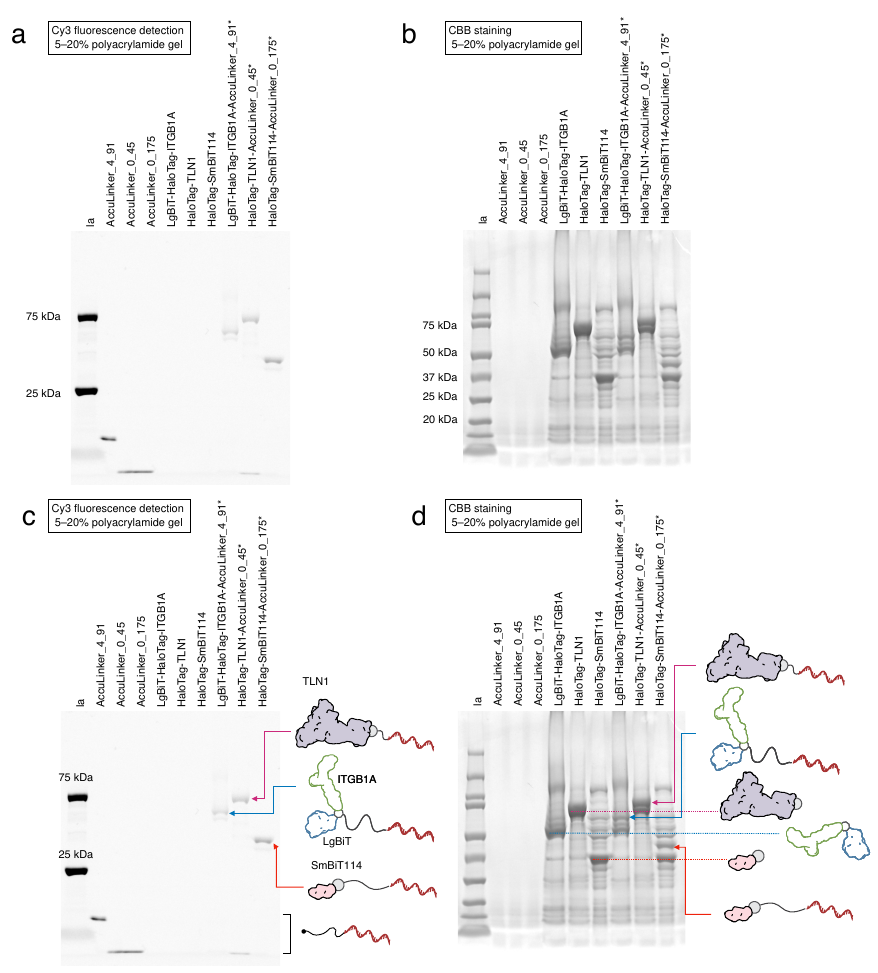}}
\caption{
HaloTag-mediated conjugation of AccuLinker to the ITGB1A, TLN1 and SmBiT114 constructs used in the modulator screen. a. Cy3 fluorescence image of the PAGE gel. b. CBB staining image of the PAGE gel. c, d. Molecular annotation of the bands in a and b, respectively. la, molecular-weight marker. The gel electrophoresis was conducted on a 5--20\% gradient polyacrylamide gel. The gel shown is representative of independent preparations. ${}^*$The samples are unpurified reaction mixtures. AccuLinker\_n\_X: n indicates the number of OEG units and X indicates the helix number for the protein tethering. The numbering corresponds to the helix numbers in Supplementary Fig.~\ref{fig:Extended Data Fig. Detail Bricks design2}.}
\label{fig:Extended Data Fig. Protein conju3}
\end{figure}

\begin{figure}[htbp]
\centering
\makebox[\linewidth][c]{\includegraphics[
  width=1.35\linewidth,
  height=0.80\textheight,
  keepaspectratio
]{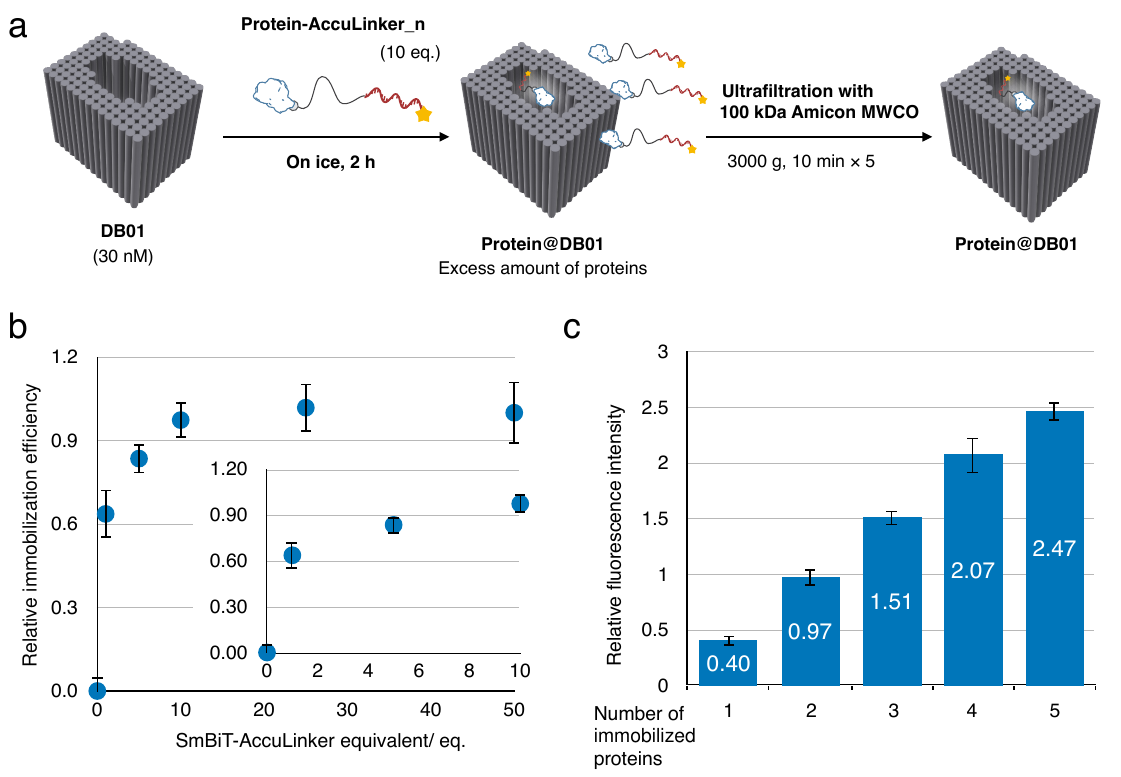}}
\caption{
   Protein tethering into DB01. a. Schematic procedure for tethering protein--AccuLinker\_n conjugates to DB01. b. Relative immobilization efficiency at different equivalents of SmBiT114--AccuLinker conjugate, reacted on ice for 2 hours. Efficiency is the Cy3 fluorescence per unit DB01 concentration, normalized to its saturating value. Immobilization was quantitative at 10 or more equivalents. Data are mean $\pm$ s.d. ($n = 3$). c. Cy3 fluorescence of DB01 carrying one to five LgBiT molecules, set by the sequence specificity of the AccuLinker overhangs. Values in the bars are mean relative fluorescence intensities, which scale linearly with the number immobilized. Data are mean $\pm$ s.d. ($n = 3$).}
   \label{fig:Extended Data Fig. Protein tethering}
\end{figure}

\begin{figure}[htbp]
\centering
\makebox[\linewidth][c]{\includegraphics[
  width=1.35\linewidth,
  height=0.80\textheight,
  keepaspectratio
]{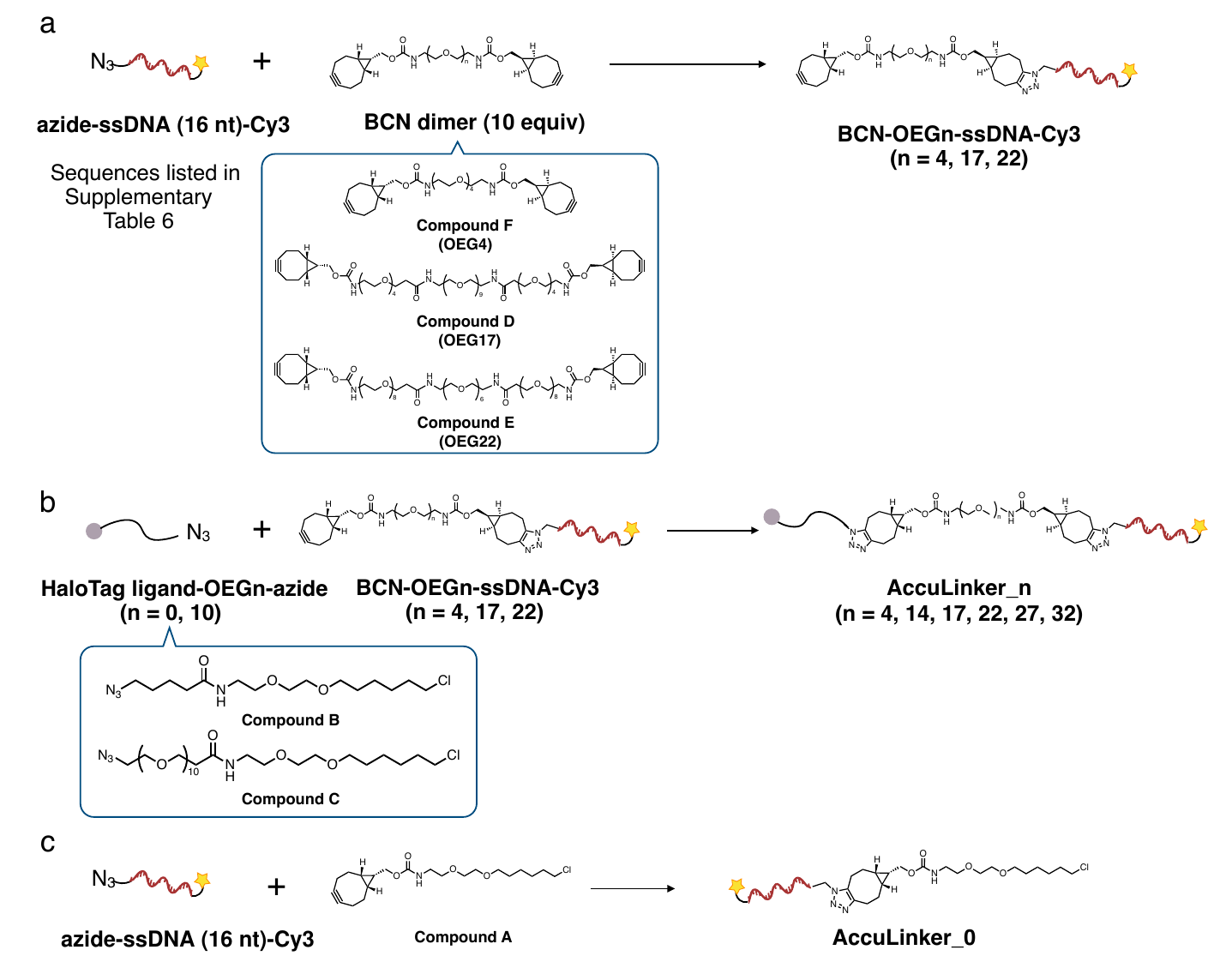}}
\caption{
   Synthetic scheme of AccuLinker\_n. a. Synthetic scheme of BCN-OEGn-ssDNA-Cy3. The sequences of the azide-ssDNA-Cy3 used are listed in \textbf{Supplementary Table 6}. BCN dimers (the synthesized compounds \textbf{D} and \textbf{E}, or the commercial compound \textbf{F}; Section~\ref{sec:Synthesis of molecules}) were used for the first click reaction. b. Synthetic scheme of AccuLinker\_n. The synthesized BCN-OEGn-ssDNA-Cy3 conjugates were reacted with compound \textbf{B} or \textbf{C} in a second click reaction to prepare the desired AccuLinker\_n constructs of various lengths. c. Synthetic scheme of AccuLinker\_0.
}
\label{fig:Extended Data Fig. Synthetic plan}
\end{figure}

\begin{figure}[htbp]
\centering
\makebox[\linewidth][c]{\includegraphics[
  width=1.35\linewidth,
  height=0.80\textheight,
  keepaspectratio
]{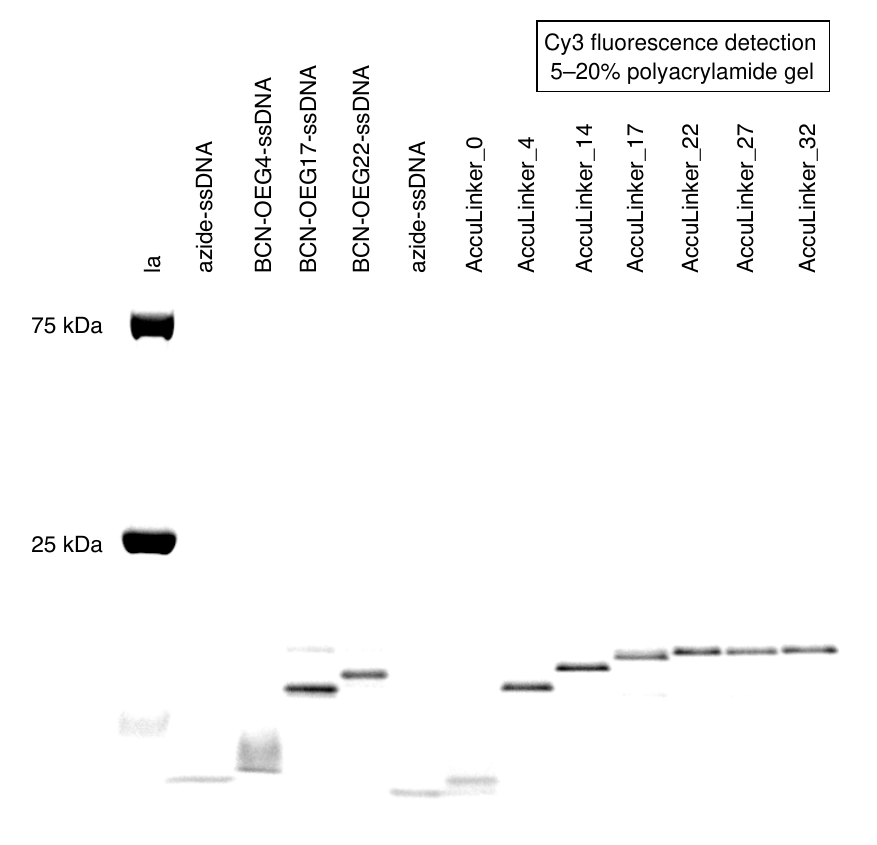}}
\caption{
   Gel-electrophoresis image of synthesized BCN-OEGn-ssDNA and AccuLinker\_n. Gel mobility analysis of the unreacted azide-modified ssDNA, synthesized BCN-OEGn-ssDNAs (n = 4, 17, and 22) and synthesized AccuLinker\_n (n = 0, 4, 14, 17, 22, 27, and 32). Each intermediate yields two AccuLinkers, because compound \textbf{B} adds no OEG unit and compound \textbf{C} adds ten. AccuLinker\_0 follows the separate route in Supplementary Fig.~\ref{fig:Extended Data Fig. Synthetic plan}c. The sequence of ssDNA is GTGATTTGAAAGTTGC, and the synthesized AccuLinkers correspond to the AccuLinker\_n\_67 series listed in Supplementary Table 7. The ssDNA carried a Cy3 label, so the purity and size of each product are read from its Cy3 fluorescence in the gel. The gel electrophoresis was conducted on a 5--20\% gradient polyacrylamide gel. The gel shown is representative of independent preparations.}
 \label{fig:Extended Data Fig. AccuLinker Evi}
\end{figure}

\begin{figure}[htbp]
\centering
\makebox[\linewidth][c]{\includegraphics[
  width=1.35\linewidth,
  height=0.80\textheight,
  keepaspectratio
]{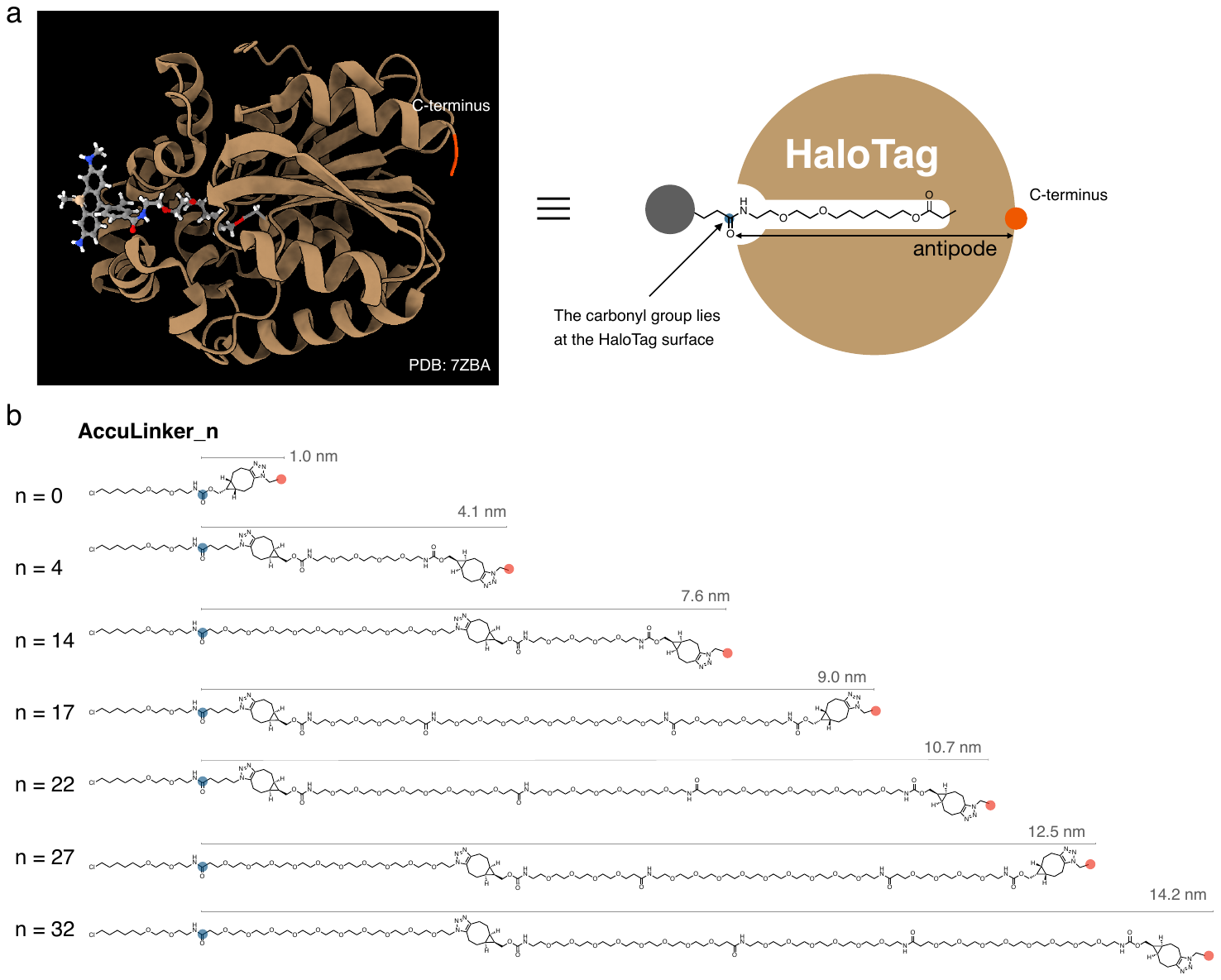}}
\caption{
   Synthetic AccuLinker\_n. a. X-ray structural analysis of HaloTag protein-conjugated with specific ligand (Protein Data Bank (PDB) entry 7ZBA). b. Chemical structure of AccuLinker\_n (n = 0, 4, 14, 17, 22, 27, 32). The linker length of AccuLinker\_n is estimated from an OEG unit length of 0.35 nm and a BCN-azide conjugate length of 1.0 nm, both calculated using Material Studio.}
   \label{fig:Extended Data Fig. AccuLinker Pro}
\end{figure}

\begin{figure}[htbp]
\centering
\makebox[\linewidth][c]{\includegraphics[
  width=1.35\linewidth,
  height=0.80\textheight,
  keepaspectratio
]{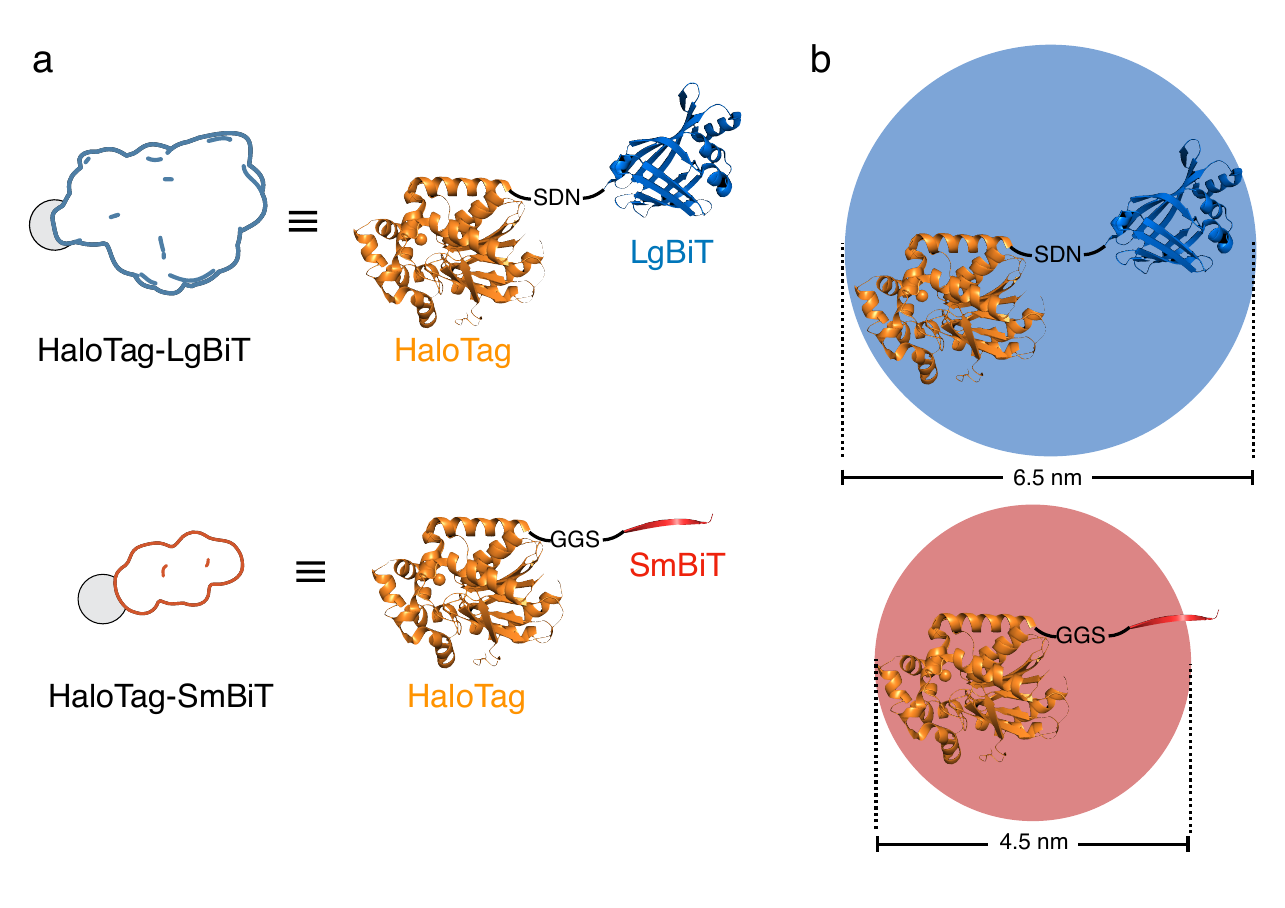}}
\caption{
   Sizes of HaloTag-fused proteins. a. Schematic illustration of HaloTag-fused LgBiT or SmBiT with a three-amino-acids spacer. b. HaloTag-fused LgBiT and SmBiT are approximated as spherical particles 6.5 and 4.5 nm in diameter, respectively for theoretical simulation. Detailed sequence information of these proteins and the other proteins was summarized in \textbf{Supplementary Tables 1--4}.}
   \label{fig:Extended Data Fig. Size of LgBiT and SmBiT}
\end{figure}

\begin{figure}[htbp]
\centering
\makebox[\linewidth][c]{\includegraphics[
  width=1.35\linewidth,
  height=0.80\textheight,
  keepaspectratio
]{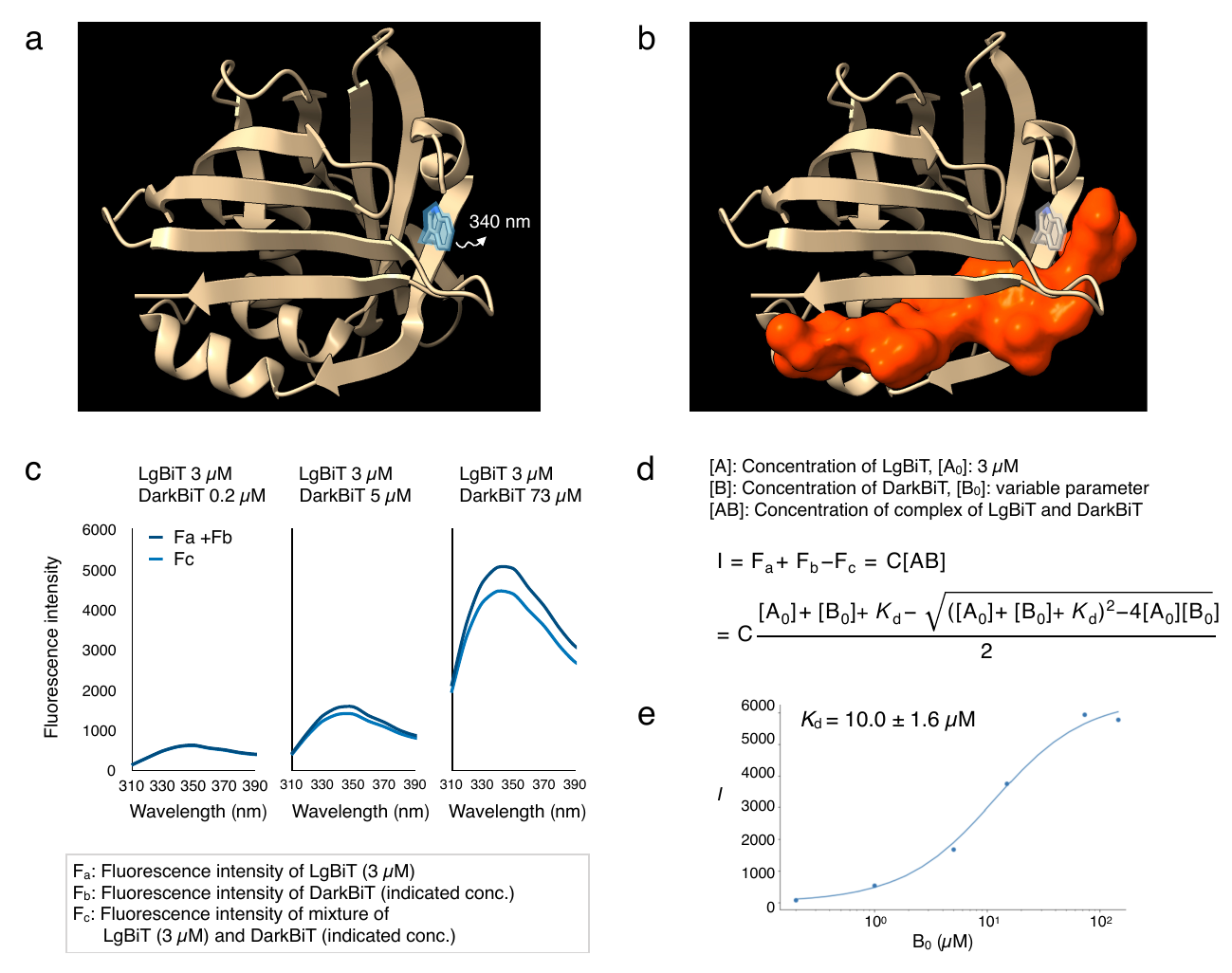}}
\caption{
   Binding analysis of DarkBiT~\textbf{1} (labelled DarkBiT) to LgBiT. a. LgBiT from the LgBiT--SmBiT complex structure (PDB 7SNX), with SmBiT hidden. The tryptophan residue emitting at 340 nm is highlighted in blue. b. The same structure with SmBiT shown in orange. DarkBiT~\textbf{1} competes for this site, and peptide binding quenches the tryptophan fluorescence. c. Fluorescence spectra at three DarkBiT~\textbf{1} concentrations. The gap between $F_a+F_b$ and $F_c$ is the quenching $I$, which grows with concentration. d. Symbol definitions and the binding equation used to obtain the LgBiT--DarkBiT~\textbf{1} dissociation constant ($K_{\mathrm{d}}$). e. Fitted curve for DarkBiT~\textbf{1} binding to LgBiT. DarkBiT~\textbf{1} binds to LgBiT with $K_{\mathrm{d}} = 10.0 \pm 1.6$~\textmu M (standard error of the fit).}
   \label{fig:Extended Data Fig. Trp analysis}
\end{figure}

\begin{figure}[htbp]
\centering
\makebox[\linewidth][c]{\includegraphics[
  width=1.35\linewidth,
  height=0.80\textheight,
  keepaspectratio
]{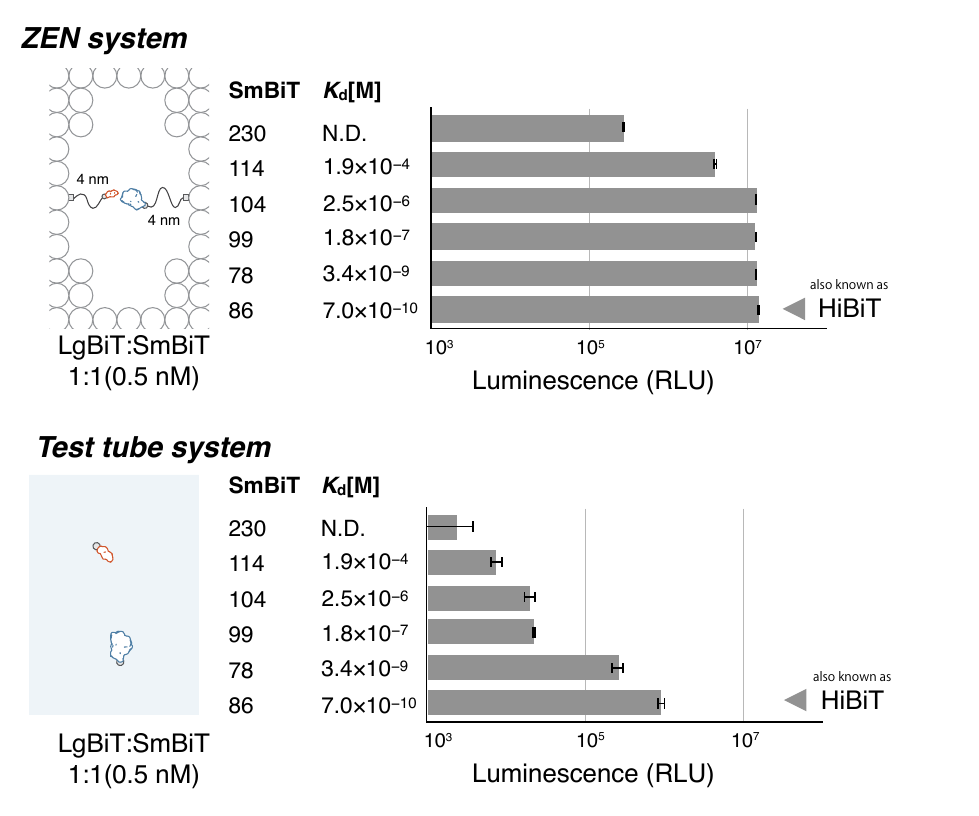}}
\caption{
   Comparison of LgBiT-SmBiT complexation in the Zonal Engineered Nano (ZEN) system and the test tube system. LgBiT-SmBiT complexation was evaluated by measuring the complexation-dependent luminescence. In the ZEN system, LgBiT and SmBiT were tethered with AccuLinker\_4\_91 and AccuLinker\_4\_67, respectively. In both systems, LgBiT and SmBiT were present at 0.5 nM each. The SmBiT numbering and the reported dissociation constants ($K_{\mathrm{d}}$) follow the original NanoBiT report\cite{LgBiT-SmBiT}. The red arrow marks SmBiT86, which is also known as HiBiT. The luminescence axis is logarithmic. Data are mean $\pm$ s.d. ($n = 3$).}
   \label{fig:local conc}
\end{figure}

\begin{figure}[htbp]
\centering
\makebox[\linewidth][c]{\includegraphics[
  width=1.35\linewidth,
  height=0.80\textheight,
  keepaspectratio
]{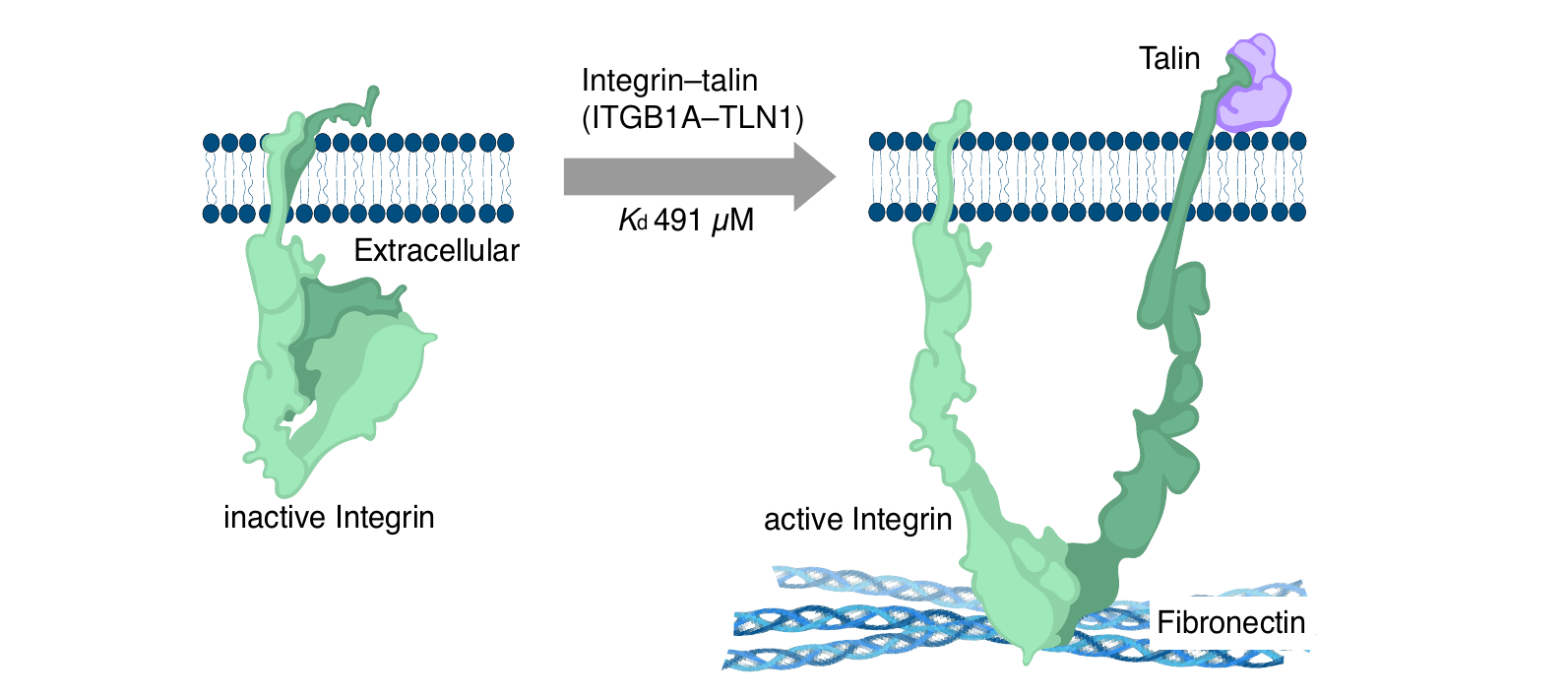}}
\caption{
   Molecular mechanism of integrin activation\cite{Integrin_activation_endgame}. Binding of talin to the intracellular domain of integrin promotes a structural change in the integrin. This structural change allows the integrin to bind extracellular-matrix proteins such as fibronectin.}
   \label{fig:Talin-Integrin}
\end{figure}

\begin{figure}[htbp]
\centering
\makebox[\linewidth][c]{\includegraphics[
  width=1.35\linewidth,
  height=0.80\textheight,
  keepaspectratio
]{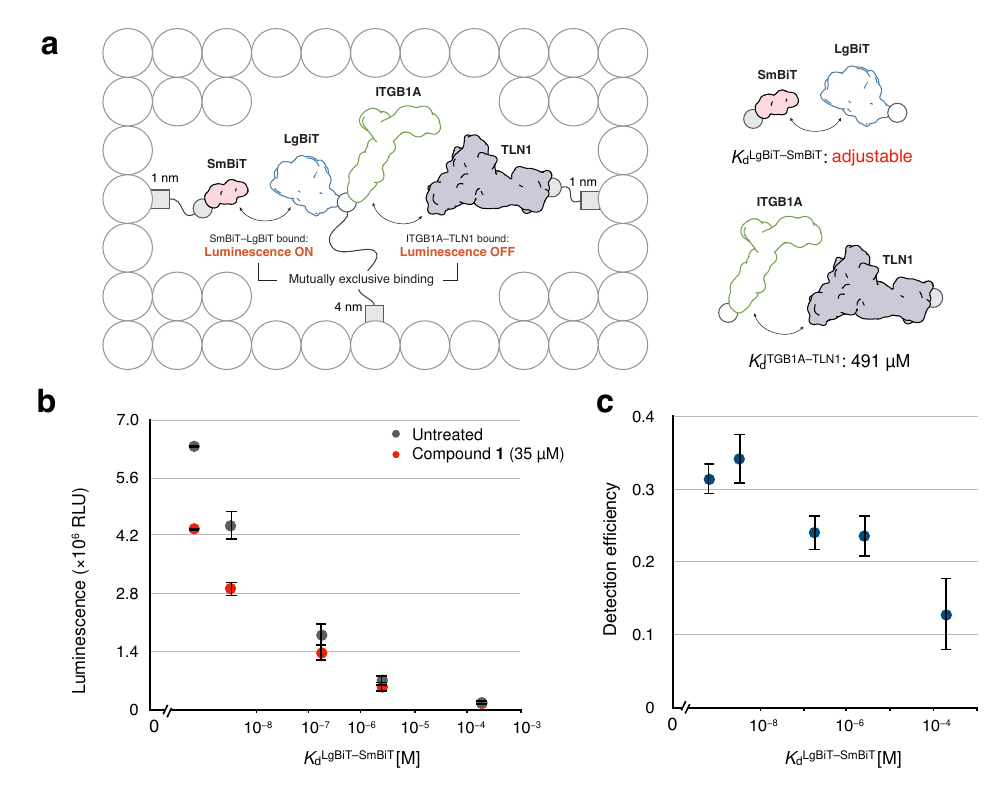}}
\caption{
   Adjustment of PPI modulator sensitivity. a. Protein placement in DB01 for detection of a  PPI modulator. HaloTag-fused SmBiT, HaloTag-fused TLN1 and LgBiT-HaloTag-ITGB1A are tethered into DB01 with AccuLinker\_0\_175, AccuLinker\_0\_45, and AccuLinker\_4\_91, respectively. Treatment with a PPI modulator (compound~\textbf{1}) switches the binding partners, resulting in the dissociation of LgBiT from SmBiT. b. Changes in luminescence of LgBiT--SmBiT in the presence or absence of compound~\textbf{1}, shown against their dissociation constants ($K_{\mathrm{d}}{}^{\mathrm{LgBiT\text{--}SmBiT}}$) on the $x$ axis. Red data points represent samples treated with 35 \textmu M of compound~\textbf{1}, while grey points indicate untreated samples. Data are mean $\pm$ s.d. ($n = 3$). c. Fractional decrease in luminescence caused by treatment with compound~\textbf{1}. The $x$ axis is the LgBiT--SmBiT dissociation constant. The $y$ axis shows $1-(\mathrm{treated}/\mathrm{untreated})$, so larger values indicate a stronger switching response. The response varies systematically with the LgBiT--SmBiT affinity, so the switch threshold can be set by the choice of SmBiT variant. Data are mean $\pm$ s.d. ($n = 3$).}
   \label{fig:switch sensitivity}
\end{figure}

\begin{figure}[htbp]
\centering
\makebox[\linewidth][c]{\includegraphics[
  width=1.35\linewidth,
  keepaspectratio
]{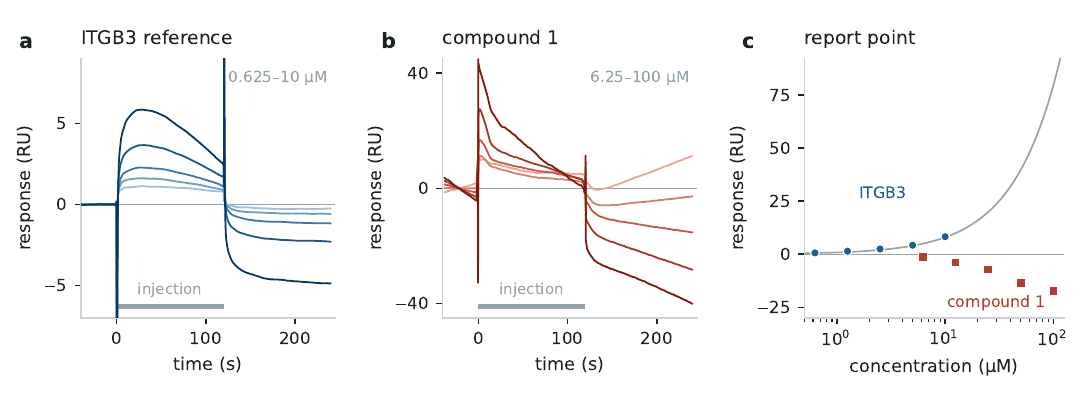}}
\caption{
   Compound~\textbf{1} shows no binding to talin-1 by surface plasmon resonance.
   \textbf{a}, Sensorgrams, response against time, for the ITGB3 reference peptide at 0.625--10~\textmu M over captured talin-1 FERM domain. \textbf{b}, The same for compound~\textbf{1} at 6.25--100~\textmu M on the same surface. The response scale is wider than in \textbf{a}, and the baseline drifts across the series. \textbf{c}, Response against concentration, measured as the mean over 15--25~s minus the mean over 0.5--2~s, which removes the refractive-index step. The line fits the reference points and is extended to show what compound~\textbf{1} would give at the same affinity. Compound~\textbf{1} gives no positive response, and its negative values follow the size of the refractive-index step at the start of each injection.}
   \label{fig:SPR}
\end{figure}

\begin{figure}[htbp]
\centering
\includegraphics[
  width=\linewidth,
  height=0.68\textheight,
  keepaspectratio
]{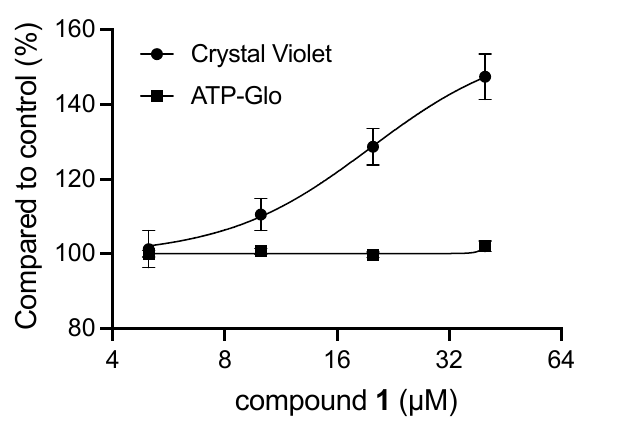}
\caption{
   Cell proliferation and cell biomass analyses. Crystal violet staining (circles) and the CellTiter-Glo 3D ATP signal (squares, labelled ATP-Glo in the figure) are plotted against compound~\textbf{1} concentration, each as a percentage of the vehicle control.
   To test the effect of compound~\textbf{1} on integrin-dependent cell adhesion, U87MG cells were cultured for 5 days with DMEM containing 1\% FBS on fibronectin-coated dishes (10 \textmu g/mL). Cell proliferation was quantified using the CellTiter-Glo 3D assay, which estimates viable cell number based on intracellular ATP content. In parallel, crystal violet staining was used to assess cellular biomass, providing an integrated measure of adherent cell mass that is influenced by both the number of adherent cells and their size. Data are mean $\pm$ s.d. ($n = 6$ replicate wells).}
 \label{fig:adhesion_biomass}
\end{figure}

\clearpage

\section{Bayesian estimation of the SmBiT230
\texorpdfstring{\(K_{\mathrm{d}}\)}{Kd}}
\label{sec:mcmc_kd_digest}

The competitive DarkBiT~\textbf{2} titration series place the six SmBiT variants
on a common occupancy scale, and anchoring that scale to published affinities
gives \(K_{\mathrm{d}}^{\mathrm{SmBiT230}}=9.60\ [4.24,\,22.04]\,\mathrm{mM}\).
The median is insensitive to the calibration and likelihood-weighting choices
examined, staying between 8.82 and 9.98 mM, and the fitted effective
concentration agrees to within a factor of 1.3 with the value computed
independently from cavity geometry. This section sets out the model and its
calibration, the estimate and its diagnostics, the sensitivity analyses, and the
choices held fixed.

\subsection{Analysis scope}

The SmBiT230 \(K_{\mathrm{d}}\) was estimated from six DarkBiT~\textbf{2} competition titrations, using the
finite-capacity \({}_2F_1\) response model
derived and parameterized in the supplementary information of the companion
paper\cite{CompanionPaper}. The DarkBiT~\textbf{2} affinity is not an input to
this response model, so the reported estimate does not depend on it. The original
report estimates it to be comparable to that of the parent SmBiT
peptide\cite{DarkBiT}. The present section
describes the external-affinity calibration, the resulting SmBiT230 estimate,
and the analyses used to assess its dependence on that calibration. The full
response function, prior specification for the curve-shape and observation
parameters, implementation, and posterior predictive checks are given
there\cite{CompanionPaper}.

Published SmBiT affinity measurements provide the external calibration needed
to place the fitted occupancy ratios on a conventional \(K_{\mathrm{d}}\)
scale. The reported SmBiT230 value is therefore a calibrated model-based
estimate whose absolute scale is set by that calibration.

The fitted data comprised six series: SmBiT86, SmBiT78, SmBiT99, SmBiT104, SmBiT114 and
SmBiT230. Each contained 12 DarkBiT~\textbf{2} concentrations measured in triplicate,
giving \(6\times12\times3=216\) replicate-level observations. The
linker-length titration data were not included in these \(K_{\mathrm{d}}\)
analyses.

\subsection{Primary external-affinity calibration}
\label{subsec:external_calibration_primary}

For each SmBiT series \(i\), the response model uses
\begin{equation}
  \rho_i=\frac{c_{\mathrm{eff}}}{K_{\mathrm{d},i}},
  \label{eq:experimental_rho}
\end{equation}
where \(c_{\mathrm{eff}}\) is a shared effective-concentration scale. The
titration curves determine \(\rho_i\) more directly than they determine
\(c_{\mathrm{eff}}\) and \(K_{\mathrm{d},i}\) separately. External affinity
information is therefore required to calibrate the absolute \(K_{\mathrm{d}}\)
scale. The geometric meaning of \(c_{\mathrm{eff}}\), and its relation to the
intrinsic \(K_{\mathrm{d}}\) through the two-state occupancy odds, are derived in
Section~\ref{SI_detailed_derivation}
(Eqs.~\eqref{eq:SI_ceff_K0_definitions} and \eqref{eq:SI_local_odds}).

The spatial-overlap calculation underlying Fig.~4b provides a position-based
effective-concentration scale under its simplified geometric encounter
kernel. That geometric value was not imposed as the
\(c_{\mathrm{eff}}\) of the present response model, because the simplified
geometric kernel does not
explicitly resolve every orientational, conformational, or molecular-shape
contribution. Such contributions can in principle be incorporated into a
higher-resolution encounter kernel. Here their net effect is absorbed into
the fitted response-model scale.

In the primary analysis, the published SmBiT99 and SmBiT104 dissociation
constants were fixed at \(180\,\mathrm{nM}\) and \(1.3\,\mu\mathrm{M}\),
respectively\cite{LgBiT-SmBiT}. The two reported SmBiT114 values,
\(190\,\mu\mathrm{M}\) and \(440\,\mu\mathrm{M}\), were obtained by different
measurement procedures\cite{LgBiT-SmBiT} and were not treated as error-free
replicates. Instead, they were modeled as external observations of one latent
SmBiT114 dissociation constant:
\begin{align}
  \log_{10}\!\left(\frac{K_{\mathrm{d},114}^{(k)}}{\mathrm M}\right)
  &\sim
  \mathcal{N}\!\left[
    \log_{10}\!\left(\frac{K_{\mathrm{d},114}}{\mathrm M}\right),
    \sigma_{\mathrm{ext}}^2
  \right],
  \qquad k=1,2, \\
  \sigma_{\mathrm{ext}}
  &\sim \operatorname{HalfNormal}(0.30\,\mathrm{dex}).
  \label{eq:mcmc_digest_external_114}
\end{align}
This formulation treats the discrepancy between the two values as uncertainty
between measurement procedures while retaining a single underlying SmBiT114
affinity.

The shared \(c_{\mathrm{eff}}\) and the SmBiT86, SmBiT78, SmBiT114, and
SmBiT230 dissociation constants were estimated simultaneously.

Each titration series was normalized to the luminescence of the corresponding
single-chain HaloTag-LgBiT-SmBiT fusion (Supplementary Table~3), which reports
the intact-luciferase activity of that SmBiT variant and was measured separately
in the absence of competitor. The reference value was the mean of three
technical replicates, each background-subtracted before averaging. A single
common factor was then applied to all series so that the mean zero-competitor
response of SmBiT86 equalled one. SmBiT86 binds most strongly and therefore has
the bound fraction closest to one in the absence of competitor. Every titration
point was measured in triplicate, and all replicates are plotted in Fig.~5c,d.

The finite-capacity response model, its fixed parameters \(M=24\) and
\(\beta_\nu=-0.5\), and the series-specific signal scale \(s_i\) are all defined
in the supplementary information of the companion paper\cite{CompanionPaper},
which is the source for every equation used here. The predicted response
is \(s_i\) times a bound fraction that equals \(\rho_i/(1+\rho_i)\) at zero
competitor, so fixing \(s_i\) does not force the zero-competitor response to
one: series with small \(\rho_i\) begin below full signal. For SmBiT114 and
SmBiT230, \(s_i\) was fixed at one as an explicit model constraint, because over
the measured range a change in \(K_{\mathrm{d},i}\) for these two series appears
mainly as a change in response amplitude, leaving \(s_i\) and
\(K_{\mathrm{d},i}\) not jointly identifiable. The signal scales for the other four
series were estimated.

\subsection{Priors, sampling, and primary estimate}

For each freely estimated dissociation constant,
\begin{equation}
  \log_{10}\!\left(\frac{K_{\mathrm{d},i}}{\mathrm M}\right)
  \sim \operatorname{Uniform}(-12,-1).
\end{equation}
The shared effective concentration was given the prior
\begin{equation}
  \log_{10}\!\left(\frac{c_{\mathrm{eff}}}{\mathrm M}\right)
  \sim \mathcal{N}(-5,1.5^2)
  \quad\text{truncated to }[-12,2].
\end{equation}
The primary fit used a shape-weighted objective based on the
replicate-level Student-\(t_4\) observation model:
\begin{equation}
  \log\widetilde p(D\mid\theta)
  =
  \sum_{ij}\omega_{ij}\log p(y_{ij}\mid\theta).
  \label{eq:mcmc_digest_weighted_objective}
\end{equation}
The weights increase the influence of the transition and high-concentration
regions on the simultaneous curve-shape fit. The group definitions and values
are given in the deposited code. The weighting defines a weighted
pseudo-posterior rather than an additional generative observation model.

The weighted objective was sampled with the No-U-Turn Sampler in Turing.jl
using four chains, 2,000 warmup iterations per chain, 4,000 retained samples
per chain, and a target acceptance probability of 0.95. This gave 16,000
retained samples. The random seed was 20260733.

The primary weighted pseudo-posterior estimate for SmBiT230 was
\begin{equation}
  K_{\mathrm{d}}^{\mathrm{SmBiT230}}
  =
  9.60\ [4.24,\,22.04]\,\mathrm{mM}.
  \label{eq:mcmc_digest_kd230_main}
\end{equation}
The central value is the weighted pseudo-posterior median, and the brackets
give the 5th and 95th percentiles, defining a 90\% weighted pseudo-posterior interval.
Sampling was well behaved: the maximum monitored \(\widehat R\) was 1.0013, the
minimum effective sample size was 3,704, and no divergent transitions occurred.
The corresponding latent SmBiT114 estimate was
\(285.5\,\mu\mathrm{M}\)
\((90\%\ \mathrm{weighted\mbox{-}posterior\ interval}:
128.3\text{--}637.4\,\mu\mathrm{M})\),
and the shared effective concentration was
\(0.978\,\mathrm{mM}\)
\((0.437\text{--}2.21\,\mathrm{mM})\).
The weighted pseudo-posterior median of \(\sigma_{\mathrm{ext}}\) was \(0.271\,\mathrm{dex}\)
\((0.130\text{--}0.566\,\mathrm{dex})\).

\renewcommand{\tablename}{Supplementary Table}
\renewcommand{\thetable}{\arabic{table}}
\setcounter{table}{9}

\begin{table}[!htbp]
\centering
\small
\begin{adjustbox}{max width=\linewidth}
\begin{tabular}{@{}lllll@{}}
\toprule
Series & Literature \(K_{\mathrm{d}}\) & Use in the primary fit & Median & 90\% interval \\
\midrule
SmBiT86  & \(0.70\,\mathrm{nM}\)          & free; literature value withheld & \(0.699\,\mathrm{nM}\)        & \(0.450\text{--}1.03\,\mathrm{nM}\) \\
SmBiT78  & \(3.4\,\mathrm{nM}\)           & free; literature value withheld & \(3.60\,\mathrm{nM}\)         & \(2.34\text{--}5.47\,\mathrm{nM}\) \\
SmBiT99  & \(180\,\mathrm{nM}\)           & exact anchor                    & not estimated                 & --- \\
SmBiT104 & \(1.3\,\mu\mathrm{M}\)         & exact anchor                    & not estimated                 & --- \\
SmBiT114 & \(190/440\,\mu\mathrm{M}\)     & external-observation model      & \(286\,\mu\mathrm{M}\)        & \(128\text{--}637\,\mu\mathrm{M}\) \\
SmBiT230 & none                           & free                            & \(9.60\,\mathrm{mM}\)         & \(4.24\text{--}22.04\,\mathrm{mM}\) \\
\bottomrule
\end{tabular}
\end{adjustbox}
\caption{Literature affinities and how each entered the primary fit. The
SmBiT86 and SmBiT78 literature values were withheld from fitting and used only
for post-fit comparison, so their estimates test the calibrated model against
affinities it was not given. Both literature values fall inside the reported
intervals. SmBiT99 and SmBiT104 were exact calibration anchors, and the two
SmBiT114 measurements entered an external-observation model. Literature values
are from the NanoBiT characterization\cite{LgBiT-SmBiT}. Estimates are weighted
pseudo-posterior medians with 5th--95th percentile intervals.}
\label{tab:external_validation}
\end{table}

\subsection{Stability across calibration and weighting choices}

The primary analysis was compared with four alternative treatments of the
SmBiT114 information. The first retained the exact SmBiT99 and SmBiT104
constraints but omitted external SmBiT114 calibration. Two further analyses
placed a Normal and a Student-\(t_4\) prior on
\(\log_{10}(K_{\mathrm{d},114}/\mathrm M)\), centred at the geometric mean of
the two reported values (\(289\,\mu\mathrm{M}\)) with scale
\(0.182\,\mathrm{dex}\). The final analysis retained the external-observation
formulation but increased the Half-Normal prior scale for
\(\sigma_{\mathrm{ext}}\) from \(0.30\) to \(0.60\,\mathrm{dex}\).

\begin{table}[!htbp]
\centering
\small
\setlength{\tabcolsep}{4pt}
\begin{adjustbox}{max width=\linewidth}
\begin{tabular}{@{}>{\raggedright\arraybackslash}p{0.39\textwidth}
                p{0.24\textwidth}p{0.24\textwidth}@{}}
\toprule
SmBiT114 treatment & SmBiT114 \(K_{\mathrm{d}}\)
& SmBiT230 \(K_{\mathrm{d}}\) \\
\midrule
No external SmBiT114 data; SmBiT99/104 exact constraints retained
& \(262\ [31.7,\,2112]\,\mu\mathrm M\)
& \(8.82\ [1.04,\,71.26]\,\mathrm{mM}\) \\
Direct Normal prior
& \(287\ [145,\,559]\,\mu\mathrm M\)
& \(9.65\ [4.79,\,19.31]\,\mathrm{mM}\) \\
Direct Student-\(t_4\) prior
& \(283\ [125,\,633]\,\mu\mathrm M\)
& \(9.52\ [4.10,\,21.78]\,\mathrm{mM}\) \\
\textbf{Two external observations; Half-Normal
\((0.30\,\mathrm{dex})\) s.d. prior (primary)}
& \(286\ [128,\,637]\,\mu\mathrm M\)
& \(9.60\ [4.24,\,22.04]\,\mathrm{mM}\) \\
Two external observations; Half-Normal
\((0.60\,\mathrm{dex})\) s.d. prior
& \(284\ [94.2,\,841]\,\mu\mathrm M\)
& \(9.52\ [3.13,\,28.47]\,\mathrm{mM}\) \\
\bottomrule
\end{tabular}
\end{adjustbox}
\caption{Sensitivity of the conditional SmBiT230 \(K_{\mathrm{d}}\) estimate
to the treatment of the SmBiT114 external information. Values are
weighted pseudo-posterior medians with 5th--95th percentile intervals. All
analyses retain the exact SmBiT99 and SmBiT104 literature constraints. The
analysis without SmBiT114 external calibration is an identifiability reference,
not an alternative primary estimate.}
\label{tab:mcmc_kd230_digest_summary}
\end{table}

The entries in this comparison table are weighted pseudo-posterior medians and 90\%
weighted pseudo-posterior intervals obtained with the same shape-weighted objective.
These are the intervals reported as uncertainty intervals in the main text.
These intervals summarize a weighted pseudo-posterior and should not be
interpreted as nominal credible intervals for the unweighted Student-\(t_4\)
observation model.

Across the five calibration treatments, the SmBiT230 median remained within
\(8.82\)--\(9.65\,\mathrm{mM}\). Broadening the prior for between-method
variation changed the median by less than 1\% while widening the interval.
Omitting the SmBiT114 external calibration produced a similar median but a
substantially broader interval. In that identifiability reference, the
posterior correlations of \(\log_{10}c_{\mathrm{eff}}\) with
\(\log_{10}K_{\mathrm{d},114}\) and \(\log_{10}K_{\mathrm{d},230}\) were 0.9986
and 0.9977, respectively. Thus the central estimate was stable across the
examined calibration treatments, whereas the interval width reflected the
uncertainty in the absolute affinity scale.

As a separate likelihood-weighting sensitivity analysis, the shape weighting
was disabled and mean-one observation weights were used. The
highest-concentration 25\% of observations contributed 25\%, 33\%, 40\%, or
50\% of the total likelihood weight. Under the 25\% setting every observation
carried equal weight. Across these four settings, the SmBiT230 pseudo-posterior
median spanned 9.25 to 9.98 mM, with 90\% interval bounds from 3.77 to
23.53 mM. The equal-weight result was
\(9.98\ [4.35,\,23.53]\,\mathrm{mM}\). The central
estimate was therefore insensitive to this weighting choice at the scale
relevant to the reported affinity, although interval width remained dependent
on the calibration and weighting assumptions.

Across all nine analysis variants examined here, five calibration treatments and
four likelihood-weighting settings, the SmBiT230 median stayed between 8.82 and
9.98 mM, within 9\% of the reported value. The calibration governs the width of
the interval rather than its centre.

\subsection{Calibration-free identifiability analysis}
\label{subsec:calibration_free_identifiability}

To separate what the titration data determine from what the calibration
supplies, the staged finite-capacity \({}_2F_1\) analysis was repeated with all
six dissociation constants treated as latent parameters. No literature
\(K_{\mathrm{d}}\) was fixed and no external-affinity observation entered the
likelihood. Each \(\log_{10}(K_{\mathrm{d},i}/\mathrm M)\) was assigned a
\(\operatorname{Uniform}(-14,2)\) prior, and the shared
\(\log_{10}(c_{\mathrm{eff}}/\mathrm M)\) a Normal prior with standard
deviation 1.5 truncated to \([-12,2]\). Prior means of \(-6\), \(-5\) and
\(-4\) were run to expose the dependence of the absolute scale on that choice.
All remaining settings matched the primary analysis, including \(M=24\),
\(\beta_\nu=-0.5\), and the signal scales \(s_i\) fixed at one for SmBiT114 and
SmBiT230. Each condition used four chains, 2,000 warmup iterations per chain,
4,000 retained samples per chain, and a target acceptance probability of 0.95.
Every posterior lies at least \(1.39\,\mathrm{dex}\) inside the \(\operatorname{Uniform}\)
bounds, so the \(K_{\mathrm{d}}\) priors contribute no gradient along the
direction described below.

All three fits met the prespecified diagnostics (maximum \(\hat R\le1.003\),
minimum bulk effective sample size \(\ge2{,}585\), minimum tail effective
sample size \(\ge1{,}128\), no divergent transitions). Convergence did not
confer an absolute affinity scale. The posterior correlation between
\(\log_{10}K_{\mathrm{d}}\) and \(\log_{10}c_{\mathrm{eff}}\) exceeded 0.999
for both SmBiT114 and SmBiT230, and moving the prior mean from \(-6\) to
\(-4\) moved the \(c_{\mathrm{eff}}\) median from 43.4 to
708\,\(\mu\)M and the SmBiT230 median from 0.429 to 6.99\,mM, while
\(K_{\mathrm{d}}/c_{\mathrm{eff}}\) stayed at 9.84--9.86
(Table~\ref{tab:calibration_free_prior_sensitivity}).

The response model is invariant under
\begin{equation}
  \log_{10}c_{\mathrm{eff}}'=\log_{10}c_{\mathrm{eff}}+\delta,\qquad
  \log_{10}K_{\mathrm{d},i}'=\log_{10}K_{\mathrm{d},i}+\delta,\qquad
  \log_{10}\nu_0'=\log_{10}\nu_0-\beta_\nu\delta ,
  \label{eq:SI_scale_invariance}
\end{equation}
so the likelihood constrains ratios, \(K_{\mathrm{d},i}/c_{\mathrm{eff}}\) and
\(K_{\mathrm{d},i}/K_{\mathrm{d},j}\), rather than either factor alone. The
identifiable combinations are
\(\eta_i=\log_{10}(K_{\mathrm{d},i}/c_{\mathrm{eff}})\) and
\(\omega=\log_{10}\nu_0+\beta_\nu\log_{10}c_{\mathrm{eff}}\). Where the
absolute scale settles is set jointly by the \(c_{\mathrm{eff}}\), \(\nu_0\)
and \(K_{\mathrm{d}}\) priors and their bounds, not by the
\(c_{\mathrm{eff}}\) prior alone: the \(\log_{10}\nu_0\) medians of
\(-1.316\), \(-1.193\) and \(-1.080\) move by \(0.11\text{--}0.12\,\mathrm{dex}\), whereas
Eq.~\eqref{eq:SI_scale_invariance} alone would move them by \(0.29\text{--}0.32\,\mathrm{dex}\).

What the data do determine transfers unchanged to the calibrated fit. The
SmBiT230-to-SmBiT114 ratio was \(33.61\ [28.16,\,40.50]\) without any
literature input and \(33.56\ [28.23,\,40.61]\) with it, a difference of
0.15\%. Combining that ratio with an independently computed effective
concentration therefore yields an affinity without any affinity standard: the
position-based \(c_{\mathrm{eff}}^{\mathrm{spatial}}=1.20\,\mathrm{mM}\) of
Section~\ref{subsec:geometric_agreement} gives
\(K_{\mathrm{d}}^{\mathrm{SmBiT230}}=11.8\,\mathrm{mM}\) and
\(K_{\mathrm{d}}^{\mathrm{SmBiT114}}=352\,\mu\mathrm{M}\), within a factor of
1.23 of the calibrated values in both cases. Extending this route requires
only a more accurate \(c_{\mathrm{eff}}\), which follows from geometry.

The calibration-free fits are therefore reported as an identifiability and
prior-sensitivity reference. The reported estimate remains the calibrated,
explicitly conditional
\(K_{\mathrm{d}}^{\mathrm{SmBiT230}}=9.60\ [4.24,\,22.04]\,\mathrm{mM}\).
Machine-readable tables, the complete per-series results, and the sampler
diagnostics are provided under \texttt{docs/calibration\_free\_tables/} in the
deposited analysis code (\url{https://github.com/FujitaG/zen-kd-estimation}).

\begin{table}[!htbp]
\centering
\small
\setlength{\tabcolsep}{4pt}
\begin{adjustbox}{max width=\linewidth}
\begin{tabular}{@{}lrrrr@{}}
\toprule
Prior mean & SmBiT114 \(K_{\mathrm{d}}\) (\(\mu\)M)
& SmBiT230 \(K_{\mathrm{d}}\) (mM)
& SmBiT114 \(K_{\mathrm{d}}/c_{\mathrm{eff}}\)
& SmBiT230 \(K_{\mathrm{d}}/c_{\mathrm{eff}}\) \\
\midrule
\(-6\) & 12.6 [0.215, 1{,}045] & 0.429 [0.00722, 35.3] & 0.293 [0.262, 0.329] & 9.84 [8.63, 11.43] \\
\(-5\) & 47.4 [0.710, 5{,}108] & 1.60 [0.0243, 170]    & 0.294 [0.262, 0.329] & 9.86 [8.62, 11.45] \\
\(-4\) & 206 [2.57, 28{,}034]  & 6.99 [0.0859, 964]    & 0.294 [0.262, 0.328] & 9.86 [8.64, 11.46] \\
\bottomrule
\end{tabular}
\end{adjustbox}
\caption{Dependence of the calibration-free absolute scale on the
\(\log_{10}c_{\mathrm{eff}}\) prior mean. Intervals are 90\%
weighted pseudo-posterior intervals. The \(K_{\mathrm{d}}\) values are deliberately
uncalibrated and are not independent affinity measurements. The
\(K_{\mathrm{d}}/c_{\mathrm{eff}}\) ratios are the identifiable quantities.}
\label{tab:calibration_free_prior_sensitivity}
\end{table}

\begin{table}[!htbp]
\centering
\small
\begin{adjustbox}{max width=\linewidth}
\begin{tabular}{@{}lrr@{}}
\toprule
Series & Median & 90\% interval \\
\midrule
SmBiT86  & 0.0139\,nM        & 0.000246--0.611\,nM \\
SmBiT78  & 0.0750\,nM        & 0.00134--3.50\,nM \\
SmBiT99  & 3.58\,nM          & 0.0619--165\,nM \\
SmBiT104 & 0.116\,\(\mu\)M   & 0.00146--14.98\,\(\mu\)M \\
SmBiT114 & 47.4\,\(\mu\)M    & 0.710--5{,}108\,\(\mu\)M \\
SmBiT230 & 1.60\,mM          & 0.0243--170\,mM \\
\bottomrule
\end{tabular}
\end{adjustbox}
\caption{Calibration-free estimates for all six series at the reference
condition (\(\log_{10}c_{\mathrm{eff}}\) prior mean \(-5\)). Values lie on an
arbitrary reference scale set by the priors and are not independent absolute
affinities. Their ratios are the identifiable content.}
\label{tab:calibration_free_all_series}
\end{table}

\subsection{Agreement with the geometric effective concentration}
\label{subsec:geometric_agreement}

The position-based spatial-overlap calculation gave an effective-concentration
scale of \(c_{\mathrm{eff}}^{\mathrm{spatial}}=1.20\,\mathrm{mM}\). This value
was not fixed as the shared response-model scale in the primary affinity
analysis, which instead sampled \(c_{\mathrm{eff}}\) jointly with the unknown
dissociation constants. The fitted median of \(0.978\,\mathrm{mM}\) agrees with
the spatial estimate to within a factor of 1.3, although this agreement was not
imposed.

The spatial implementation resolves tether geometry, steric exclusion, and
positional overlap at its present resolution. A higher-resolution treatment
could be developed to represent linker bending stiffness, detailed molecular
shape, conformational flexibility, cavity fluctuations, and
orientation-dependent reactivity, and to assess how uncertainty in these
effects influences future affinity inference.

\subsection{Model choices held fixed}

Three response-model choices remained fixed across the primary fit and the
calibration and likelihood-weighting sensitivity analyses above. The reported
estimate and intervals are conditional on these choices.

For SmBiT114 and SmBiT230, the signal scale \(s_i\) was fixed at one because
\(s_i\) and \(K_{\mathrm{d},i}\) are not jointly identifiable over the measured
range. The signal scales for the other variants were estimated under a weakly
informative prior\cite{CompanionPaper}.

Finite local capacity is physically motivated by the limited occupancy of a
single cavity. Its numerical value \(M=24\) was chosen after exploratory
comparisons among candidate integer values and then held fixed during the
reported Markov chain Monte Carlo (MCMC) sampling. Because \(M\) remained fixed, uncertainty from this exploratory
selection is not included in the reported intervals.

The empirical coefficient \(\beta_\nu\), which describes the affinity dependence
of curve shape, is not microscopically derived. It was fixed at \(-0.5\) after
exploratory model checks to stabilize that trend and was not sampled during the
reported MCMC\cite{CompanionPaper}.

\section{Accessible volume, state occupancy, and the definition of \texorpdfstring{$K_{\mathrm{d}}$}{Kd}}
\label{SI_detailed_derivation}

This section explains why the accessible-volume expression in the main text does not imply that the intrinsic dissociation constant changes with confinement. The apparent contradiction arises from applying a bulk concentration identity to state-conditioned microscopic volumes. In a homogeneous bulk solution, the concentrations of A, B, and AB are defined in the same physical volume, and their ratio already contains the equilibrium balance between unbound and bound populations. After mutually isolated local states are resolved, that population balance must instead be retained explicitly.

Throughout this section, concentrations are number concentrations. Conversion to molar units introduces only the corresponding constant factor. We also use an idealized state-counting model to expose the relevant probability term. In this minimal notation, geometry-independent internal and solvent contributions are coarse-grained into an effective non-spatial binding term. Retaining them as separate partition-function factors gives the same result.

\subsection{A mutually isolated pair is a two-state system}

Consider one mutually isolated compartment containing one specified A--B pair. The compartment has two mutually exclusive states,
\begin{equation}
  \mathrm{A}+\mathrm{B}
  \rightleftharpoons
  \mathrm{AB},
\end{equation}
which we call the unbound and bound states, respectively. Let $\overline v_A$ and $\overline v_B$ be the volumes accessible to the unbound partners, and let $\overline v_{AB}$ be the volume accessible to the complex. In a minimal state-counting representation, the statistical weights of the two states are
\begin{equation}
  q_{\mathrm{u}}
  = \frac{\overline v_A\overline v_B}{v_0^2},
  \qquad
  q_{\mathrm{b}}
  = \frac{\overline v_{AB}}{v_0}\exp(\beta\epsilon),
  \label{eq:SI_local_state_weights}
\end{equation}
where $v_0$ is a microscopic reference volume, $\epsilon>0$ denotes the geometry-independent binding-energy gain in this minimal model, and $\beta=(k_{\mathrm B}T)^{-1}$. Additional non-spatial state factors can be retained explicitly or absorbed into the corresponding effective Boltzmann factor without changing the argument. We define the geometric effective concentration and the geometry-independent concentration scale $K_0$ as
\begin{equation}
  C_{\mathrm{eff}}
  := \frac{\overline v_{AB}}
  {\overline v_A\overline v_B},
  \qquad
  K_0
  := \frac{1}{v_0\exp(\beta\epsilon)}.
  \label{eq:SI_ceff_K0_definitions}
\end{equation}
Both quantities are number concentrations, as is every concentration in this section. Calibrating the minimal state-counting model to its homogeneous bulk limit identifies $K_0$ with the intrinsic dissociation constant $K_{\mathrm{d}}$, as derived in Section~\ref{sec:SI_bulk_calibration}. Thus $K_0=K_{\mathrm{d}}$, and Eqs.~\eqref{eq:SI_local_state_weights} and \eqref{eq:SI_ceff_K0_definitions} give
\begin{equation}
  \frac{q_{\mathrm b}}{q_{\mathrm u}}
  = \frac{C_{\mathrm{eff}}}{K_0}
  = \frac{C_{\mathrm{eff}}}{K_{\mathrm{d}}}.
  \label{eq:SI_local_weight_ratio}
\end{equation}

Expressing every concentration in molar units divides each by the Avogadro constant and leaves these ratios unchanged, so the same relations hold for the molar $c_{\mathrm{eff}}=C_{\mathrm{eff}}/N_{\mathrm{A}}$ and $K_{\mathrm{d}}$ reported elsewhere in this work.

Let $p$ be the equilibrium probability of the bound state. The normalized two-state probabilities are $P_{\mathrm b}=p$ and $P_{\mathrm u}=1-p$, and therefore
\begin{equation}
  \frac{p}{1-p}
  = \frac{q_{\mathrm b}}{q_{\mathrm u}}
  = \frac{C_{\mathrm{eff}}}{K_{\mathrm{d}}}.
  \label{eq:SI_local_odds}
\end{equation}
Equivalently,
\begin{equation}
  \boxed{
  K_{\mathrm{d}}
  = \frac{\overline v_{AB}}
  {\overline v_A\overline v_B}
  \frac{1-p}{p}}
  \label{eq:SI_corrected_volume_Kd}
\end{equation}
and
\begin{equation}
  p=\frac{C_{\mathrm{eff}}}
  {K_{\mathrm{d}}+C_{\mathrm{eff}}}.
  \label{eq:SI_local_binding_fraction}
\end{equation}
Equation~\eqref{eq:SI_corrected_volume_Kd} is the complete form of the schematic accessible-volume expression in the main text. The factor $(1-p)/p$ is the missing balance between the unbound and bound states. Changing accessible volume changes $C_{\mathrm{eff}}$ and therefore changes $p$ according to Eq.~\eqref{eq:SI_local_binding_fraction}; it does not require $K_{\mathrm{d}}$ to change.

The unbound probability in Eq.~\eqref{eq:SI_corrected_volume_Kd} is a joint-state probability. Within an isolated compartment, A and B are unbound together or are converted together into AB. It would therefore be incorrect to replace $1-p$ by the product of two independent marginal probabilities, $(1-p)^2$. Mutual isolation removes cross-pair encounters and creates this A--B pairing constraint, so the factorization underlying the ordinary bulk product $[\mathrm A][\mathrm B]$ no longer applies at the single-pair level.

\subsection{The occupancy factor from an ensemble of isolated compartments}

The same result follows by considering $N$ mutually isolated, equivalent compartments. If $n$ compartments are bound, the statistical weight of that population state is
\begin{equation}
  W_n
  = \binom{N}{n}
  q_{\mathrm u}^{N-n}q_{\mathrm b}^{n}.
  \label{eq:SI_isolated_population_weight}
\end{equation}
The binomial coefficient counts the ways of choosing which compartments are bound. Equivalently, it removes the overcounting among the $n$ bound and $N-n$ unbound compartments. Its contribution can be written as the combinatorial entropy
\begin{equation}
  S_{\mathrm{comb}}(n)
  = k_{\mathrm B}\log\binom{N}{n}.
  \label{eq:SI_combinatorial_entropy}
\end{equation}
For large $N$, with $p=n/N$, the change in this entropy upon increasing the number of bound compartments is
\begin{equation}
  \frac{1}{k_{\mathrm B}}
  \frac{\partial S_{\mathrm{comb}}}{\partial n}
  \simeq \log\frac{N-n}{n}
  = -\log\frac{p}{1-p}.
  \label{eq:SI_occupancy_entropy}
\end{equation}
This is the origin of the bound--unbound odds term in an entropy derivation. Without taking the large-$N$ derivative, the corresponding exact finite-system statement is obtained directly from adjacent population states:
\begin{equation}
  \frac{W_{n+1}}{W_n}
  = \frac{N-n}{n+1}
  \frac{C_{\mathrm{eff}}}{K_{\mathrm{d}}}.
  \label{eq:SI_isolated_adjacent_ratio}
\end{equation}
Thus, the combinatorial state count supplies precisely the occupancy information that disappears if molecule number is cancelled against a species-specific occupied volume. The normalized distribution generated by Eq.~\eqref{eq:SI_isolated_population_weight} is binomial, with mean bound fraction $\langle n/N\rangle=p$ given by Eq.~\eqref{eq:SI_local_binding_fraction}. For a large ensemble, the adjacent-state ratio crosses unity near the most probable $n$, and Eq.~\eqref{eq:SI_isolated_adjacent_ratio} reduces to Eq.~\eqref{eq:SI_local_odds} with $p\simeq n/N$.

The substitution $V_X=N_X\overline v_X$ makes
\begin{equation}
  \frac{N_X}{V_X}=\frac{1}{\overline v_X}
\end{equation}
a density conditioned on state $X$ already being present. Because both numerator and denominator change with occupancy, recovering the bulk activity additionally requires the probability that state $X$ occurs. Equation~\eqref{eq:SI_corrected_volume_Kd} supplies that state probability.

\subsection{Where the same probability balance appears in bulk}
\label{sec:SI_bulk_calibration}

In a homogeneous bulk system, the equilibrium composition encodes the probability balance. Let the conserved molecular totals be $N_A^0$ and $N_B^0$, and let $j$ complexes be present. The particle numbers in state $j$ are
\begin{equation}
  n_A=N_A^0-j,
  \qquad
  n_B=N_B^0-j,
  \qquad
  n_{AB}=j.
\end{equation}
For an ideal, homogeneous solution of volume $V$, the equilibrium probability $P_j$ is proportional to the state weight
\begin{equation}
  \Omega_j
  \propto
  \frac{1}{n_A!n_B!j!}
  \left(\frac{V}{v_0}\right)^{n_A+n_B+j}
  \exp(\beta j\epsilon).
  \label{eq:SI_bulk_state_weight}
\end{equation}
The factorials are the combinatorial terms associated with indistinguishable bulk molecules. Taking the ratio of adjacent states gives the exact finite-system expression
\begin{equation}
  \frac{P_{j+1}}{P_j}
  = \frac{\Omega_{j+1}}{\Omega_j}
  = v_0\exp(\beta\epsilon)
  \frac{n_A n_B}{V(j+1)}
  = \frac{1}{K_0}
  \frac{[\mathrm A][\mathrm B]}
  {[\mathrm{AB}]+1/V},
  \label{eq:SI_bulk_adjacent_ratio}
\end{equation}
where $[\mathrm A]=n_A/V$, $[\mathrm B]=n_B/V$, and $[\mathrm{AB}]=j/V$.

The bulk equilibrium distribution has a maximum near some $j=j_*$. For a finite discrete system, the adjacent ratios on the two sides of the maximum bracket unity, attaining it only in the degenerate case of two equally probable neighbouring states. In the macroscopic limit, however, the distribution is sharply peaked and
\begin{equation}
  \frac{P_{j_*+1}}{P_{j_*}}\longrightarrow 1.
\end{equation}
Equation~\eqref{eq:SI_bulk_adjacent_ratio} then yields
\begin{equation}
  K_0
  \simeq \frac{[\mathrm A][\mathrm B]}
  {[\mathrm{AB}]+1/V}
  \longrightarrow
  \frac{[\mathrm A][\mathrm B]}{[\mathrm{AB}]}
  \equiv K_{\mathrm{d}}.
  \label{eq:SI_bulk_limit_Kd}
\end{equation}
The last step is the thermodynamic limit at fixed concentrations and establishes the promised identification
\begin{equation}
  K_{\mathrm{d}}=K_0
  =\frac{1}{v_0\exp(\beta\epsilon)}.
  \label{eq:SI_bulk_calibrated_Kd}
\end{equation}
This relation is therefore a consequence of calibrating the minimal statistical model to the conventional bulk definition, rather than an independent assumption about $K_{\mathrm{d}}$. The relative size of the finite-particle correction is $1/(V[\mathrm{AB}])=1/j$. The probability term is not discarded as an arbitrary constant. Its composition-dependent part becomes the concentration ratio, while the remaining local slope of the equilibrium probability distribution, $\log(P_{j+1}/P_j)$, approaches zero at the bulk peak.

The different appearances of the same information can now be summarized directly. A mutually isolated pair has only two states, so equilibrium is represented by their explicit odds $p/(1-p)$. A bulk system has many possible values of $j$, so equilibrium is represented primarily by the location of a sharply peaked distribution, which is, equivalently, the equilibrium concentration ratio. The standard bulk formula already contains the bound--unbound balance and should not be reapplied to volumes conditioned separately on A, B, and AB occupancy.

\subsection{Interpretation and scope}

The thermodynamically general statement is written in terms of activities rather than raw concentrations. In a homogeneous ideal solution, activities reduce to concentration factors and the familiar expression $K_{\mathrm{d}}=[\mathrm A][\mathrm B]/[\mathrm{AB}]$ follows. In a mutually isolated microscopic system, correlations and pairing prevent the unbound joint state from being represented by the product of two independently averaged local concentrations. The explicit state odds in Eq.~\eqref{eq:SI_corrected_volume_Kd}, or an equivalent correlated activity, must be used instead.

Accordingly, accessible volume controls occupancy through $C_{\mathrm{eff}}$, whereas $K_{\mathrm{d}}$ describes the intrinsic A--B binding thermodynamics under the volume-only assumption used here: geometry changes translational accessibility while the binding energy and the internal state factors of A, B, and AB remain fixed. An extended state-weight treatment could represent confinement-induced strain, surface interactions, electrostatic changes, or conformational selection through geometry-dependent free-energy factors and, where necessary, additional conformational states. These contributions are outside the present treatment and could change the intrinsic or apparent affinity. Within the volume-only treatment, the apparent volume dependence of $K_{\mathrm{d}}$ is not a physical prediction but a consequence of omitting the bound--unbound occupancy factor when transferring a bulk concentration expression to resolved local states.

\end{document}